\documentclass[aps,prbl,showpacs,preprintnumbers,reprint,floatfix,superscriptaddress,amsmath,amssymb]{revtex4-2}

\usepackage{graphicx}
\usepackage[linkcolor=blue,citecolor=blue,filecolor=blue,colorlinks=true,urlcolor=blue]{hyperref}

\usepackage{bm}

\usepackage{upgreek}
\usepackage{color}
\usepackage{upgreek}
\usepackage{natbib}

\usepackage{amsmath,amsthm,amsfonts,amssymb,amscd}
\usepackage{lastpage}
\usepackage{enumerate}
\usepackage{fancyhdr}
\usepackage{mathrsfs}
\usepackage{xcolor}
\usepackage{graphicx}
\usepackage{listings}
\usepackage{hyperref}
\usepackage{braket}
\usepackage{amsmath}
\usepackage{mathtools}
\DeclareMathAlphabet{\pazocal}{OMS}{zplm}{m}{n}

\newcommand{\eps}{\varepsilon}
\newcommand{\mpar}{m_\parallel^*}
\newcommand{\ve}[1]{\mathbf #1}
\newcommand{\innp}[3]{\left< #1 | #2| #3 \right>}

\newcommand{\rhoint}{ \tilde{\rho} }
\newcommand{\Hint}{ \tilde{H_i} }
\newcommand{\trph}[1]{\mathrm{Tr}_\mathrm{ph} \left\{ #1 \right\} }
\newcommand{\dpr}[2]{\mathbf{#1}\cdot\mathbf{#2}}
\newcommand{\phpr}[2]{ \left(#1|#2\right)_{Q_z} }
\newcommand{\irpr}[3]{ \left(#1|#2\right)_{#3,\mathrm{IR}} }
\newcommand{\iipr}[4]{ \left(#1|#2\right)^{\mathrm{II}}_{#3,#4} }
\newcommand{\iiprc}[4]{ \left(#1|#2\right)^{\mathrm{II*}}_{#3,#4} }
\newcommand{\aprod}[3]{ \left(#1 | #2\right)_\mathrm{alloy}(#3) }
\newcommand{\paz}{\frac{\partial}{\partial z}}

\newcommand{\pth}[1]{\left( #1 \right)}
\newcommand{\sth}[1]{\left[ #1 \right]}

\newcommand{\pa}[2]{\frac{\partial #1}{\partial #2}}

\def \FigureWidth{3.3in}

\begin{document}

\title{Numerically efficient density-matrix technique for modeling electronic transport in midinfrared quantum cascade lasers}

\author{S. Soleimanikahnoj}\email{soleimanikah@wisc.edu}
\affiliation{Department of Electrical and Computer Engineering, University
of Wisconsin-Madison, 1415 Engineering Dr., Madison, Wisconsin 53706, USA}

\author{O. Jonasson}\email{michelle.king@wisc.edu}
\affiliation{Department of Electrical and Computer Engineering, University
of Wisconsin-Madison, 1415 Engineering Dr., Madison, Wisconsin 53706, USA}

\author{F. Karimi}
\affiliation{Department of Electrical and Computer Engineering, University
of Wisconsin-Madison, 1415 Engineering Dr., Madison, Wisconsin 53706, USA}

\author{I. Knezevic}\email{knezevic@engr.wisc.edu}
\affiliation{Department of Electrical and Computer Engineering, University
of Wisconsin-Madison, 1415 Engineering Dr., Madison, Wisconsin 53706, USA}

\date{\today}

\begin{abstract}
We present a numerically efficient density-matrix model applicable to midinfrared quantum cascade lasers. The model is based on a Markovian master equation for the density matrix that includes in-plane dynamics, preserves positivity of the density matrix and does not rely on phenomenologically introduced dephasing times. Nonparabolicity in the bandstructure is accounted for with a three-band $\dpr{k}{p}$ model, which includes the conduction, light-hole, and spin-orbit split-off bands. We compare the model to experimental results for QCLs based on lattice-matched as well as strain-balanced InGaAs/InAlAs heterostructures grown on InP. We find that our density-matrix model is in quantitative agreement with experiment up to threshold and is capable of reproducing results obtained using the more computationally expensive nonequilibrium Green's function formalism. We compare our density-matrix model to a semiclassical model where off-diagonal elements of the density matrix are ignored. We find that the semiclassical model overestimates the threshold current density by 29\% for a 8.5-$\upmu$m-QCL based on a lattice-matched heterostructure and 40\% for a 4.6-$\upmu$m QCL based on a strain-balanced heterostructure, demonstrating the need to include off diagonal density matrix elements for accurate description of midinfrared QCLs.
\end{abstract}

\maketitle

\section{Introduction}
\label{sec:intro}

Quantum cascade lasers (QCLs) are unipolar light-emitting sources in the midinfrared and terahertz parts of the electromagnetic spectrum ~\cite{faist_book_2013,vitiello_OE_2015}. The gain medium in QCLs is a vertically-grown semiconductor heterostructure made of alternating layers of quantum wells and barriers. A population inversion between quasibound quantum states is achieved by a careful design of the layer structure and application of electrical bias. In this work, we focus on midinfrared QCLs based on a InGaAs/InAlAs heterostructure grown on InP substrates. QCLs based on this material system have achieved room-temperature, continuous-wave, high-power operation in the midinfrared range ~\cite{capasso_OE_2010,yao_NATP_2012,razeghi_OME_2013}, and have found applications in the fields of remote sensing~\cite{curl_CPL_2010} and metrology ~\cite{bartalini_MST_2014}.

Numerical simulations play an essential role in the development of QCLs and can provide detailed insights into the microscopic details of QCL operation ~\cite{dupont_PRB_2010,jirauschek_APR_2014}. Electron transport in quantum cascade lasers in both midinfared and terahertz regimes has been simulated via semiclassical (rate equations~\cite{indjin_JAP_2002,indjin_APL_2002,mircetic_JAP_2005,wang2018rate} and Monte Carlo~\cite{iotti2001,callebaut_APL_2004,gao_JAP_2007,shi_JAP_2014,borowik2017monte}) and quantum techniques (density matrix~\cite{willenberg_PRB_2003,kumar_PRB_2009,weber_PRB_2009,dupont_PRB_2010,terazzi_NJP_2010,burnett2018robust,pan2017density,demic2017infinite,
jirauschek2017density,riesch2018performance,riesch2019analyzing,Demic2019_AIPAdvances,Jirauschek2019_MaxwellBloch_AdvancedTheoSimulation} and nonequilibrium Green's functions (NEGF)~\cite{lee_PRB_2002,bugajski_PSS_2014,wacker_JSTQE_2013,lindskog_APL_2014,kolek_JSTQE_2015,haldas2019implementation,kolek2019implementation,grange2019room}). Semiclassical models are attractive due to their low computational costs, allowing inclusion of advanced phenomena such as nonequilibrium phonons \cite{shi_JAP_2014}. However, semiclassical models involve the assumption that off-diagonal density-matrix (DM) elements (also referred to as the coherences) are much smaller than the diagonal ones (which correspond to level occupations). This approximation has been shown to fail in QCLs working in the terahertz~\cite{callebaut_JAP_2005,kumar_PRB_2009}, as well as in the midinfrared range ~\cite{jonasson_PHT_2016}.

NEGF simulations accurately capture quantum transport in QCLs \cite{lee_PRB_2002,wacker_JSTQE_2013}, but are computationally demanding. Density-matrix models offer a compromise between speed an accuracy. They are capable of describing partially coherent transport and require considerably less computational resources than NEGF. Previous theoretical work on QCLs using density-matrix approaches features various approximations: simplified treatment of in-plane dynamics, where either a thermal distribution (Maxwell or Fermi-Dirac) is assumed~\cite{kumar_PRB_2009,lindskog_APL_2014} or the density matrix is assumed to be separable into the cross-plain and in-plane directions (same in-plane distribution assumed for all subbands) \cite{iotti_RPP_2005}, both of which may pose a problem because QCLs operate far from equilibrium, so in-plane distributions can deviate strongly from a thermal distribution and in-plane distributions can vary drastically between subbands \cite{shi_JAP_2014,lindskog_APL_2014,jonasson_JCEL_2016}; semiclassical treatment within a subregion of a device (typically a single stage separated by thick injection barriers), while coupling between adjacent subregions is treated quantum mechanically using phenomological dephasing times~\cite{kumar_PRB_2009,lindskog_APL_2014,terazzi_NJP_2010,callebaut_JAP_2005}; a Redfield equation of motion for the density matrix that does not require the adoption of an artificially localized basis, which is important in THz QCLs ~\cite{pan2017density}; a completely positive equation of motion for the density matrix with model dissipative terms of the Lindblad form and empirical parameters \cite{Jirauschek2019_MaxwellBloch_AdvancedTheoSimulation}.

In this paper, we present a numerically efficient density-matrix model applicable to electronic transport in midinfrared and terahertz QCLs. The model is based on a Markovian master equation for the density matrix that is of the so-called Lindblad form \cite{Lindblad1976,breuer_2002,knezevic_JCEL_2013,Jirauschek2019_MaxwellBloch_AdvancedTheoSimulation,Demic2019_AIPAdvances} and thus preserves the positivity of the density matrix. The model supplants the need for a phenomenological treatment of dissipation or dephasing: electronic interaction with phonons, impurities, interface roughness, and random alloy disorder are described through the dissipator term derived directly from microscopic Hamiltonians. Full in-plane dynamics are included without assuming a thermal distribution or separation of the density matrix into in-plane and cross-plane components. Nonparabolicity in the band structure is treated using a three-band $\dpr{k}{p}$ model. We note that the model is ideal for QCLs in which a finite number of reasonably well localized states governs electronic transport, and is therefore better suited for midinfrared than terahertz structures.

This paper is organized as follows. Section~\ref{sec:theory} contains an overview of the theory. We derive a Markovian master equation that governs the time evolutions of the single-electron density matrix and the linear response to a harmonic electric field. In Sec.~\ref{sec:numerical}, we discuss details regarding the numerical solution of the master equation and derive an iterative scheme to efficiently solve for the steady-state and linear-response density matrices. In Sec.~\ref{sec:results}, we verify the model by simulating midinfrared QCLs based on InGaAs/InAlAs heterostructures grown on InP. We consider an 8.5~$\upmu$m-QCL based on a lattice-matched heterostructure and a 4.6-$\upmu$m-QCL based on a strain-balanced heterostructure. We compare our results to experiment, as well as to theoretical results based on NEGF and to a semiclassical model where only diagonal values of the density matrix are considered. Concluding remarks are given in Sec.~\ref{sec:conclusion}.

\section{Theory}
\label{sec:theory}

Electrons interacting with an environment can be described using a Hamiltonian on the form
\begin{align}
  \label{eq:interaction}
  H = H_0 + H_{i} + H_{ph} ,
\end{align}
where $H_0$ is the unperturbed electron Hamiltonian (in our case a semiconductor superlattice under an applied electric field), $H_{ph}$ is the Hamiltonian of the environment (e.g., the free-phonon Hamiltonian), and $H_i$ contains the interaction terms (e.g., electron--phonon interaction or scattering due to interface roughness). For transport calculations, we will use the eigenstates of $H_0$ as a basis, which are calculated using a three-band $\dpr{k}{p}$ Hamiltonian described in  Appendix ~\ref{sec:electronic}. We assume that the device area in the $x$-$y$ plane perpendicular to the transport ($z$) direction is macroscopic with respect to the length of a single period and that the system behaves as translationally invariant in this plane. The eigenstates of $H_0$ are labelled as $\left |n,\mathbf k\right>=\left | n\right>\otimes \left| \mathbf k \right>$, where the discrete index $n$ quantizes states in the transport direction ($z$) and $\mathbf k$ is the wave vector associated with free motion in the $x$-$y$ plane.


Owing to in-plane translational invariance, the reduced density matrix of electrons is diagonal in $\mathbf k$ and only depends on the magnitude $k=|\mathbf k|$. The central quantities of interest are then the matrix elements $\rho_{n,m}^{E_k}$, which for $n\neq m$ represent the coherence between eigenstates $\ket n$ and $\ket m$, at the in-plane energy $E_k$ while, $\rho_{n,n}^{E_k}$ represents the occupation of the $z$-direction eigenstate $\ket N$ at in-plane energy $E_k$.

When working with periodic systems, it is convenient to work with relative coordinates: $f_{N,M}^{E_k}\equiv\rho_{N,N+M}^{E_k}$, with $M=0$ terms giving occupations and terms with $|M|>0$ represent coherences between states $\ket{N}$ and $\ket{N+M}$. Relative coordinates are convenient because $f_{N,M}^{E_k}$ is periodic in $N$, with a period of $N_s$, where $N_s$ is the number of relevant eigenstates in a single period. We also have $f_{N,M}^{E_k}\rightarrow 0$ as $|M|\rightarrow \infty$, providing an obvious truncation scheme $|M|\leq N_c$, where $N_c$ is a number we refer to as the coherence cutoff. Henceforth, we will assume that $f$ is simply $\rho$, conveniently written in relative coordinates; the relative (absolute) coordinates will be denoted by capital (lowercase) indices.

In the interaction picture (with $\hbar=1$), the equation of motion for the density matrix can be written as
\begin{subequations}
\begin{align}
  \label{eq:EOM_density_a}
  \pa{ \rhoint }{t}&=-i \left[ \Hint(t),\rhoint(t) \right], \\
  \label{eq:EOM_density_b}
  \rhoint(t)&=\rhoint(0)-i \int_{0}^{t}\left[ \Hint(s),\rhoint(s) \right] ds ,
\end{align}
\end{subequations}
where $\tilde{A}(t)=e^{i(H_0+H_\mathrm{ph})t} A e^{-i(H_0+H_\mathrm{ph})t}$ denotes operator $A$ in the interaction picture. We will assume that the electron--phonon interaction is weak enough to warrant the assumption that the phonon distribution is negligibly affected by the electrons and write the density matrix as a product $\rhoint(t)=\tilde \rho_e(t)\otimes \tilde \rho_\mathrm{ph}$, where $\tilde \rho_e(t)$ is the electron density matrix and $\tilde \rho_\mathrm{ph}$ is the thermal equilibrium density matrix for the phonons. This approximation is called the \emph{Born approximation}. Inserting Eq.~\eqref{eq:EOM_density_a} into Eq.~\eqref{eq:EOM_density_b} and tracing over the phonon degrees of freedom gives
\begin{align}
\label{eq:EOM_traced}
\begin{split}
  \pa{\tilde\rho_e}{t}&=
  -i \trph{ \left[ \Hint(t),\tilde\rho_e(0)\otimes \tilde \rho_\mathrm{ph} \right] } \\
  &-\int_0^t ds
  \trph{ \left[ \Hint(t), \left[ \Hint(s),\rhoint_e(s)\otimes \rhoint_\mathrm{ph}\right]   \right] }.
\end{split}
\end{align}
The first term on the RHS is zero for interaction Hamiltonians that are linear in creation/annihilation operators ~\cite{knezevic_JCEL_2013}, which is the case for the electron--phonon interaction Hamiltonians considered in this work. An alternative justification for its omission is that we can assume that, after a sufficiently long time, the density matrix has no memory of its initial state $\tilde \rho_e(0)$. Next we assume that the time evolution of the density matrix is memoryless (Markov approximation), and replace $\tilde \rho_e(s)$ in the integrand of Eq.~\eqref{eq:EOM_traced} by the density matrix at the current time $\tilde \rho_e(t)$, giving
\begin{align}
  \label{eq:markov}
  \pa{\tilde\rho_e}{t}=-\int_0^t ds
  \trph{ \left[ \Hint(t), \left[ \Hint(s),\rhoint_e(t)\otimes \rhoint_\mathrm{ph}\right]   \right] }.
\end{align}
Equation~\eqref{eq:markov} is the Redfield equation and is not a completely positive map (therefore the density matrix is not guaranteed to remain positive during evolution). Making a change of variables $s\rightarrow t-s$ and extending the range of integration to infinity (in doing so, we assume environmental correlations decay fast with increasing $|t-s|$) gives
\begin{align}
  \label{eq:redfield}
  \pa{\tilde\rho_e}{t}=-\int_0^\infty ds
  \trph{ \left[ \Hint(t), \left[ \Hint(t-s),\rhoint_e(t)\otimes \rhoint_\mathrm{ph}\right]   \right] }.
\end{align}
The net energy change in the electron system due to the two $\tilde H_i$ is zero ~\cite{breuer_2002}. The energy change has two components: The in-plane energy change and the transport-direction energy change. Since we assumed that the density matrix is translationally invariant in the in-plane direction, i.e., $\langle n \mathbf{k}|\rho |m \mathbf{k}'\rangle = \rho_{n,m}^{E_k}\delta(\mathbf{k}-\mathbf{k'})$, we will see [Eq.~\eqref{eq:completeness} after integration of $\ve k_2$, $\ve k_3$ and $\ve k_4$] that the in-plane energy change is zero. Also, because in the transport direction the energy levels are discrete, in the limit $t\rightarrow \infty$ we can use the rotating-wave approximation, which states the terms resulting in non-zero energy changes average out to zero, and shows that the transport-direction energy change vanishes as well (see Eq. 3.132 in Ref.~\cite{breuer_2002}) Therefore, it will be possible to write Eq.~\eqref{eq:redfield} in the Lindblad form, i.e., the equation is completely positive and trace preserving. We remove the remaining time dependence by switching back to the Schr{\"o}dinger picture:
\begin{align}
  \label{eq:master_schrod}
  &\pa{\rho_e}{t}=-i\left[H_0,\rho_e(t)\right]
  - \int_0^{\infty} ds
  \mathrm{Tr}_\mathrm{ph} \\ \nonumber
   &\left\{ \left[H_i, \left[ e^{-i(H_0+H_\mathrm{ph})s} H_i e^{i(H_0+H_\mathrm{ph})s},
   \rho_e(t)\otimes \rho_\mathrm{ph}\right]   \right] \right\},
\end{align}
The above term can be written as a sum of unitary time evolution and a dissipative term that includes interaction with phonons,
\begin{align}
  \label{eq:diss_def}
  \pa{\rho_e}{t}=-i\left[H_0,\rho_e(t)\right]
  +D[\rho_e(t)],
\end{align}
where the superoperator $D$ describes dissipation and is often referred to as the dissipator. Appendix~\ref{sec:phonon} is devoted to the derivation of the dissipator for electron-phonon interaction  while interactions with other scattering mechanisms are covered in Appendix~\ref{sec:elastic}. \textcolor{blue}{Again, the dissipator was derived according to a procedure that guarantees the Lindblad form (see more in \cite{breuer_2002,KarimiDissertation2017,Karimi_PRB_2016,jonasson_JCEL_2016}). However, once can readily convince oneself that this is true for any particular mechanism, as shown in Appendix \ref{sec:lindblad} on the example of electron--phonon interaction.}

Equation~(\ref{eq:diss_def}) written  for the matrix elements $f_{N,M}^{E_k}$ in relative coordinate has the form
\begin{align}
  \label{eq:EOM}
  \frac{\partial f_{N,M}^{E_k}}{\partial t} = -i\frac{\Delta_{N,N+M}}{\hbar}f_{N,M}^{E_k}
  + \mathcal D_{N,M}^{E_k},
\end{align}
where $\Delta_{N,N+M}$ is the energy spacing between states $N$ and $M$. For brevity, we use $\mathcal{D}$ to denote $Df$ and will henceforth call this whole term (rather than the superoperator) the dissipator.
\begin{align}
  \nonumber
  \mathcal D_{N,M}^{E_k}&=
   -\sum_{n,m,g,\pm} \Gamma_{NMnmE_k}^{\mathrm{out},g,\pm}f_{N,n}^{E_k} \\  \label{eq:diss_elements}
  &+\sum_{n,m,g,\pm}\Gamma_{NMnmE_k}^{\mathrm{in},g,\pm}f_{N+n,M+m-n}^{E_k\mp E_{g}+\Delta_{N+M,N+M+m}}+\mathrm{h.c.}.
\end{align}
Here, $\Gamma_{NMnmE_k}^{\mathrm{in},g,\pm}$ and $\Gamma_{NMnmE_k}^{\mathrm{out},g,\pm}$ are the quantum-mechanical generalizations of semiclassical scattering rates ~\cite{gao_JAP_2007,jirauschek_APR_2014}. They are given in Appendix \ref{app:rates}. In this context, h.c. should be understood as the matrix elements of the Hermitian conjugate (it is not simply the complex conjugate because indices need to be swapped as well). In this work we will refer to them as scattering rates (or simply rates). The $g$ index represents the different scattering mechanisms, $+(-)$ refers to absorption (emission), and $E_{g}$ is the energy associated with scattering mechanism $g$ (i.e., phonon energy). For elastic scattering mechanisms, such as ionized-impurity scattering, we have $E_{g}=0$.


In this work, we are interested in the steady-state solution of Eq.~\eqref{eq:EOM}. Once the steady-state density matrix is known, experimentally measurable quantities can easily be calculated. Detailed information regarding the different observables, such as the current density and charge density, are given in Section.~\ref{sec:ss_observ}. In Section~\ref{sec:gain}, we derive an expression for the optical gain by considering the linear response of the density matrix to an optical field, treating it as a small perturbation. This section concludes with Sec.~\ref{sec:semiclassical}, where we provide a comparison with semiclassical model based on the Pauli master equation \cite{fischetti1999}, where only the diagonal values of the density matrix are included.

\subsection{Steady-State Observables}
\label{sec:ss_observ}
Once the steady-state density matrix elements have been calculated from Eq.~\eqref{eq:EOM}, the expectation value of any observable $A$ can be calculated using
\begin{align}
  \label{eq:expect}
  \left< A \right>= \sum_{N=1}^{N_s}\sum_{M=-N_c}^{N_c}\int dE_k \innp{N}{A}{N+M}f_{N,M}^{E_k},
\end{align}
Here, $\ket{N}$ represents an eigenstate of $H_0$ in Dirac notation. As before, $N_s$ is the number of eigenstates per period and $N_c$ is a coherence cutoff, denoting how much coupling there si between the states in the neighboring stages (see further discussion in Sec.~\ref{subsec:steadystate}). The observable $A$ can depend on position $z$, any derivative of position $\frac{\partial^n}{\partial z^n}$, as well as in-plane energy. An important steady-state observable is the drift velocity $v$, which is used to calculate the expectation value of current density $J=q n_{3D} \left< v \right>$, where $n_{3D}$ is the average 3D electron density in the device,
\begin{align}
  \label{eq:n3d}
  n_{3D}=\frac{1}{L_P}\int_0^{L_P} N_D(z) dz ,
\end{align}
with $L_P$ the period length of the QCL and $N_D(z)$ the doping density. Matrix elements of the drift velocity operator are obtained from the time derivative of the position operator using
\begin{align}
  \label{eq:drift_op}
\begin{split}
  &\innp{n}{v}{m} = \left<n \bigg| \frac{dz}{dt} \bigg|m\right>
  \innp{n}{\sth{H,z}}{m}  \\
  &=\frac{i}{\hbar}(E_n-E_m) \innp{n}{z}{m}=\frac{i}{\hbar}(E_n-E_m)z_{n,m} ,
\end{split}
\end{align}
where we have used $\innp{n}{\sth{H_i,z}}{m}=0$ because the interaction potentials in this work only contain the position operator $z$, but not momentum $p_z$. The steady-state current density can be written as
\begin{align}
  \label{eq:expect}
  J = q n_{3D}\sum_{N=1}^{N_s}\sum_{M=-N_c}^{N_c}\int dE_k v_{N,N+M}f_{N,M}^{E_k}.
\end{align}
In this work, electron--electron interaction is treated at a mean-field level by obtaining a self-consistent solution with Poisson's equation (see further discussion in Sec.~\ref{subsec:steadystate}). We therefore need to calculate the spatially resolved electron density $n_\mathrm{el}(z)$ using
\begin{align}
\label{eq:electron_dens}
\begin{split}
  n_\mathrm{el}(z)=& n_{3D} \sum_{N=1}^{N_s}\sum_{M=-N_c}^{N_c}\int dE_k \\
                 & \times \sum_b \sth{ \psi_N^{(b)}(z)}^*\psi_{N+M}^{(b)}(z) f_{N,M}^{E_k},
\end{split}
\end{align}
Here, $\psi_N^{(b)}(z)$ is the $N$th electron eigenfunction in band $b$. The sum involving $b$ is over the 3 bands considered in the $\dpr{k}{p}$ model described in Appendix ~\ref{sec:electronic}. We define the occupation of state $N$ using
\begin{align}
  \label{eq:occdef}
  f_N=\int dE_k f_{N,0}^{E_k}
\end{align}
and the average coherence between states $N$ and $M$ as
\begin{align}
  \label{eq:coherencedef}
  \rho_{N,M}=\int dE_k \rho_{N,M}^{E_k}=\int dE_k f_{N,M-N}^{E_k} \ .
\end{align}
The magnitude of the coherences relative to the occupations give us an idea how important is it to include off-diagonal elements of the density matrix and when a semiclassical treatment is sufficient.

In order to quantify the amount of electron heating that takes place under operating conditions, we define the expectation value of in-plane energy using
\begin{align}
  \label{eq:inplane_av}
  \left< E_{k} \right > = \sum_{N=1}^{N_s}\int dE_k E_k f_{N,0}^{E_k}
\end{align}
and the expectation value of in-plane energy of a single subband using
\begin{align}
  \label{eq:inplane}
  \left< E_{k} \right >_N = \frac{\int dE_k E_k f_{N,0}^{E_k}}{\int dE_k f_{N,0}^{E_k}} .
\end{align}
For low bias, we have $\left< E_{k} \right >_N\simeq k_BT$, where $T$ is the lattice temperature. However, as will be shown in Section~\ref{sec:inplane}, for electric fields around and above threshold, $\left< E_{k} \right >_N$ can greatly exceed $k_BT$ at lattice temperature and can differ substantially between subbands. It is possible to define the electron temperature of state $N$ using $T_{\mathrm{N}}=\left< E_{k} \right >_N /k_B$. The electron temperature can be a useful quantity because it is an input parameter for simplified DM models, where a thermal in-plane distribution is assumed ~\cite{terazzi_NJP_2010,lindskog_APL_2014}. However, we note that the in-plane energy distribution under typical operational electric fields can deviate strongly from a Maxwellian. The term electron temperature should therefore only be interpreted as the temperature equivalent of the average in-plane energy.

\subsection{Optical Gain}
\label{sec:gain}
The $z$-component of an electric field propagating in the $x$-direction, with frequency $\omega$ and amplitude $\mathcal E_0$, can be written as a plane wave of the form~\cite{haug_2004}
\begin{align}
  \label{eq:planewave}
  \mathcal E_z(x,\omega)=\mathcal E_0e^{i n_B \omega x/c }e^{-\alpha(\omega) x/2},
\end{align}
where $c$ is the speed of light in vacuum, $n_B$ is the background index of refraction (which we assume is frequency independent in the frequency range of interest in this work), and $\alpha(\omega)$ is the absorption coefficient. We note that the electric field in typical QCL waveguides (operating in a TM mode) also has a nonzero $x$-component ~\cite{jirauschek_APR_2014}. However, only the dipole matrix element in the $z$ direction are nonzero, so the $x$-component is not amplified. Optical gain is defined as $g(\omega)=-\alpha(\omega)$ and can be obtained from the complex susceptibility using ~\cite{haug_2004}
\begin{align}
  \label{eq:gainsus}
    g(\omega)&=-\frac{\omega}{c n_B} \operatorname{Im}[\chi(\omega)].
\end{align}
In order to calculate the complex susceptibility, we follow the approach in Ref.~\cite{lee_PRB_2002}. We treat the optical field as a small perturbation and consider the linear response in the frequency domain. We assume the wavelength of the optical field is long and ignore the position dependence of the electric field across the QCL. We can then write the perturbation potential as
\begin{align}
  \label{eq:Vomega}
  V_\omega(z)=q \mathcal E_0 z e^{-i\omega t} \ .
\end{align}
Using linear-response theory, the density matrix can be written as
\begin{align}
  \label{eq:linearres}
  \rho(t) = \rho_0 + \rho_\omega e^{-i\omega t},
\end{align}
where $\rho_0$ is the steady-state solution for $\mathcal E_0=0$. The equation of motion for $\rho(t)$ can be written as
\begin{align}
\label{eq:EOM_omg}
\begin{split}
  i\hbar \pa{\rho}{t}&=  \sth{H_0+V_\omega,\rho_0+\rho_\omega} +
                      \mathcal D[\rho_0+\rho_\omega] \\
                     &=\sth{H_0,\rho_0}+\mathcal D[\rho_0]+\sth{H_\omega,\rho_\omega}
                   + \sth{H_0,\rho_\omega} \\
                   &+\sth{V_\omega,\rho_0}
                   +\mathcal D[\rho_\omega] \\
                   &\simeq \sth{H_0,\rho_\omega} +\sth{V_\omega,\rho_0},
\end{split}
\end{align}
where the first two terms in the second line of Eq.~\eqref{eq:EOM_omg} add up to zero and the third term is proportional to $e^{-2i\omega t}$ (second harmonic). The second-harmonic term is a product of two small quantities ($V_\omega$ and $\rho_\omega$) and is ignored. Combining Eqs.~\eqref{eq:Vomega}-\eqref{eq:EOM_omg}  and dividing both sides by $e^{-i \omega t}$ gives an equation for $\rho_\omega$
\begin{align}
  \label{eq:eqomega}
  \hbar\omega \rho_\omega=\sth{H_0,\rho_\omega}+\mathcal D[\rho_\omega]+e\mathcal E_0\sth{z,\rho_0} \ .
\end{align}
In the same way we defined the matrix elements $f_{NM}^{E_k}$, we can define the matrix elements of $\rho_\omega$ in relative coordinates as $f_{NM}^{E_k}(\omega)=\rho_{N,N+M}^{E_k}(\omega)$. Equation~\eqref{eq:eqomega} gives the following equation for the matrix elements
\begin{align}
  \label{eq:rho_omg_mat}
  f_{N,M}^{E_k}(\omega)\sth{i\frac{\Delta_{N,N+M}}{\hbar}-i\omega}=A_{N,M}^{E_k}
  +\mathcal D_{N,M}^{E_k}(\omega),
\end{align}
where $\mathcal D_{N,M}^{E_k}(\omega)$ are  matrix elements of the dissipator acting on the linear-response density matrix and
\begin{align}
  \label{eq:Adef}
    A_{N,M}^{E_k}=-i\frac{e\mathcal E_0}{\hbar} \sum_n z_{N,N+n}f_{N+n,M-n}^{E_k} + \mathrm{h.c.},
\end{align}
where $z_{n,m}$ is a matrix elements of the $z$-position operator. Using Eqs.~\eqref{eq:rho_omg_mat} and \eqref{eq:Adef}, we can calculate $f_{NM}^{E_k}(\omega)$ for relevant values of $\omega$. Note that the dissipator term depends on many of the matrix elements $f_{N,M}^{E_k}(\omega)$, so solving for $f_{N,M}^{E_k}(\omega)$ is not as trivial as it might appear from Eq.~\eqref{eq:rho_omg_mat}. Our numerical method for solving Eq.~\eqref{eq:rho_omg_mat} is given in Sec.~\ref{sec:numerical}.

Once $f_{NM}^{E_k}(\omega)$ is known, we can calculate the field-induced polarization using
\begin{align}
  \label{eq:poldef}
  p(\omega) = \mathrm{Tr} \sth{d\rho_\omega}=e \int dE_k \sum_{N,M} z_{N,N+M} f_{N,M}^{E_k}(\omega)\ ,
\end{align}
where $d$ is the dipole operator $d=qz$. Note that the density matrix is normalized to one, so $p(\omega)$ gives the induced polarization per electron. To get the polarization per unit volume, we use $\mathcal P(\omega)=n_{3D}p(\omega)$, where $n_{3D}$ is the average electron density. The polarization is then used to calculate the complex susceptibility $\chi(\omega)$ using~\cite{haug_2004,lee_PRB_2002}
\begin{align}
  \label{eq:suscept}
  \chi(\omega) = \frac{\mathcal P(\omega)}{\eps_0 \mathcal E_0} .
\end{align}
Alternatively, the complex susceptibility can be calculated using the induced current density, which is the time derivative of the polarization $J(\omega)=-i\omega \mathcal P$, giving
\begin{align}
  \label{eq:suscept_j}
  \chi(\omega) = \frac{i}{\omega}\frac{J(\omega)}{\eps_0 \mathcal E_0} .
\end{align}
We have calculated the susceptibility using both Eq.~\eqref{eq:suscept} and Eq.~\eqref{eq:suscept_j}, and both expressions give the same result (within numerical accuracy). Once the complex susceptibility has been calculated, the gain is obtained using Eq.~\eqref{eq:gainsus}.

\subsection{Comparison With Semiclassical Models}
\label{sec:semiclassical}
Equation~\eqref{eq:EOM} can be thought of as a generalization of the semiclassical Pauli master equation, wherein the rates are calculated using Fermi's golden rule~\cite{gao_JAP_2007,jirauschek_APR_2014}. This fact can be seen by looking at the diagonal terms $(M=0)$ and limiting the out-scattering sum to $n=0$, and the in-scattering sum to $n=m$, giving
\begin{align}
  \label{eq:paulidef}
\begin{split}
  \pa{f_{N,0}^{E_k}}{t}=&-2\sum_{m,g,\pm}\Gamma_{N,0,0,m,E_k}^{\mathrm{out},g,\pm}f_{N,0}^{E_k} \\
  &+ 2\sum_{n,g,\pm} \Gamma_{N,0,n,n,E_k}^{\mathrm{in},g,\pm} f_{N+n,0}^{E_k+\Delta_{N,N+n} \mp E_s} \ .
\end{split}
\end{align}
By defining
\begin{align}
\begin{aligned}
    \label{eq:semi_rate_def}
  f_{N}^{E_k}&=f_{N,0}^{E_k}, \\
  \tilde \Gamma_{N,n,E_k}^{\mathrm{out},g,\pm}&=2\Gamma_{N,0,0,n,E_k}^{\mathrm{out},g,\pm}, \\
  \tilde \Gamma_{N,n,E_k}^{\mathrm{in},g,\pm}&=2\Gamma_{N,0,n,n,E_k}^{\mathrm{in},g,\pm},
\end{aligned}
\end{align}
we can write Eq.~\eqref{eq:paulidef}
\begin{align}
\label{eq:pauli2}
\begin{split}
  \pa{f_N^{E_k}}{t}=&-\sum_{n,g,\pm} \tilde \Gamma_{N,n,E_k}^{\mathrm{out},g,\pm}f_N^{E_k} \\
  &+ \sum_{n,g,\pm} \tilde \Gamma_{N,n,E_k}^{\mathrm{in},g,\pm} f_{N+n}^{E_k+\Delta_{N,N+n} \mp E_s} \ ,
\end{split}
\end{align}
which is equivalent to the Boltzmann equation for the occupation of state $n$ with in-plane energy $E_k$ (see, for example, Eq. ~(129) in Ref.~\cite{jirauschek_APR_2014}). The factor of $2$ in front of the rates comes from the addition of the Hermitian conjugate (which is identical for the diagonal of the density matrix). Because of the similarities between the semiclassical Boltzmann equation in~\eqref{eq:pauli2} and the quantum-mechanical results in Eq.~\eqref{eq:EOM}, we can easily compare results obtained using both models and investigate when semiclassical models give satisfactory results and when off-diagonal density matrix elements must be included.

When calculating current density using semiclassical models, we cannot directly use Eq.~\eqref{eq:expect} because matrix elements of the velocity operator given in Eq.~\eqref{eq:drift_op} are zero for $n=m$ (in other words, all the current is given by off-diagonal matrix elements). In order to calculate drift velocity, we first define
\begin{align}
  \label{eq:totalout}
  \tilde \Gamma_{NnE_k}^{\mathrm{out}}=\sum_{g,\pm} \tilde \Gamma_{N,n,E_k}^{\mathrm{out},g,\pm},
\end{align}
which gives the total rate at which an electron in subband $N$ with in-plane energy $E_k$ is scattered to subband $N+n$. We then define the semiclassical drift velocity as
\begin{align}
  \label{eq:vclass}
  v_c= \int d E_k \sum_{N=1}^{N_s} \sum_{n=-N_c}^{N_c}(z_{N}-z_{N+n})f_N^{E_k}\Gamma_{NnE_k}^{\mathrm{out}},
\end{align}
with $z_N=\innp{\psi_N}{z}{\psi_N}$. The current density is then calculated using $J_c=q n_{3D}v_c$. Equation~\eqref{eq:vclass} can be interpreted as summing up the flux of electrons from position $z_{N}$ to $z_{N+n}$. This intuitive expression for $v_c$ can be derived from Eq.~\eqref{eq:expect} by approximating the off-diagonal density matrix elements using diagonal ones.~\cite{calecki_JPC_1984}

If the optical frequency $\omega$ is much larger than typical scattering rates (due to phonons, impurities etc.), the gain can be calculated from the diagonal density matrix elements using~\cite{gelmont_APL_1996,jirauschek_JAP_2009,jirauschek_APR_2014}
\begin{align}
  \nonumber
  g(\omega)&=\frac{q^2\omega \mpar}{\hbar^3c \eps_0 n_B L_P}
  \sum_{N=1}^{N_s} \sum_{M=-N_c}^{N_c} \frac{\Delta_{N,N+M}}{|\Delta_{N,N+M}|}
  |z_{N,N+M}|^2 \\ \label{eq:gain_noscatt}
  &\times \int dE_k ( f_N^{E_k}-f_{N+M}^{E_k})
  \mathcal L_{N,N+M}(E_k,\omega) ,
\end{align}
with the Lorentzian broadening function
\begin{align}
  \label{eq:lorentz}
  \mathcal L_{n,m}(E_k,\omega)= \frac{\gamma_{n,m}(E_k)}
  {\gamma_{n,m}^2(E_k)+\sth{\omega-\Delta_{nm}/\hbar}^2}.
\end{align}
The broadening of a transition $\gamma_{n,m}(E_K)$ is calculated using~\cite{jirauschek_APR_2014}
\begin{align}
  \label{eq:broadef}
  \gamma_{n,m}(E_k)=\frac12 \sth{ \gamma_{n}(E_k)+\gamma_m (E_k) },
\end{align}
where $\gamma_n(E_k)$ is the total intersubband scattering rate from subband $n$ at energy $E_k$, calculated using
\begin{align}
  \label{eq:ratedef}
  \gamma_n(E_k)= \sum_{n\neq0}\tilde \Gamma_{NnE_k}^{\mathrm{out}}.
\end{align}

\section{Numerical Method}
\label{sec:numerical}
\subsection{Steady-State Density Matrix}
\label{subsec:steadystate}
The central quantities of interest are the matrix elements of the steady-state density matrix $f_{N,M}^{E_k}=\rho_{N,N+M}^{E_k}$. Because of periodicity, we have $f_{N,M}^{E_k}=f_{N+N_s,M}^{E_k}$, where $N_s$ is the number of states in a single period. The index $M$ strictly runs from $-\infty$ to $+\infty$ but is truncated according to $|M|\leq N_c$, where $N_c$ is a coherence cutoff. The coherence cutoff needs to be large enough such that overlaps between states $\ket{N}$ and $\ket{N+N_c}$ is small for all $N$. Typical values for $N_c$ are slightly larger than $N_s$, however, the specific choice of $N_c$ is highly system dependent. We also have to discretize the continuous in-plane energy $E_k$ into $N_E$ evenly spaced values. This truncation scheme results in $N_s(2N_c+1)N_E$ matrix elements.

In order to calculate the steady-state matrix elements $f_{N,M}^{E_k}$, we set the time derivative in Eq.~\eqref{eq:EOM} to zero and get
\begin{align}
  \label{eq:steady}
\begin{split}
  &0 = -i\frac{\Delta_{N,N+M}}{\hbar}f_{N,M}^{E_k}
  -\sum_{n,m,g,\pm} \Gamma_{NMnmE_k}^{\mathrm{out},g\pm}f_{N,n}^{E_k}\\
  &+\sum_{n,m,g,\pm}\Gamma_{NMnmE_k}^{\mathrm{in},g,\pm}f_{N+n,M+m-n}^{E_k\mp E_0+\Delta_{N+M,N+M,m}}
  +\mathrm{h.c.} .
\end{split}
\end{align}
Note that the sums in the Eq.~\eqref{eq:steady} run over matrix elements and energies outside the center period (e.g., $N>N_s$ or $N<1$), which are calculated by taking advantage of periodicity. We also encounter terms for which $|M|>N_c$, where we assume $f_{N,M}^{E_k}=0$. The exact values of $N_s$, $N_c$, and $N_E$ are very system specific, for example, in order to simulate the 8.5~$\upmu$m-QCL considered in Sec.~\ref{sec:lm_design}, we use $N_s=8$, $N_c=12$ and $N_E=251$ resulting in $50,200$ matrix elements and for the 4.6-$\upmu$m-QCL studied in section~\ref{sec:sb_design} we use $N_s=12$, $N_c=14$, and $N_E=251$ resulting in $87,348$ matrix elements. This high number of variables makes it impractical to solve Eq.~\eqref{eq:steady} as a matrix equation, so we resorted to iterative methods.

In order to solve Eq.~\eqref{eq:steady} using the Jacobi iterative method,~\cite{saad_iter_2003} we need to write $f_{N,M}^{E_k}$ in terms of all other terms $f_{N',M'}^{E_k'}$ for $N'\neq N$, $M'\neq M$, and $E_k'\neq E_k$. From the dissipator term in~\eqref{eq:steady}, we see that the terms containing  $f_{N,M}^{E_k}$ in the out-scattering part are the $n=M$ terms. As for the in-scattering part, only the elastic terms ($E_0=0$) with $n=m=0$ contain $f_{N,M}^{E_k}$. Note that the Hermitian conjugate part also contains terms proportional to $f_{N,M}^{E_k}$, which are obtained using $f_{N,M}^{E_k}=[f_{N+M,-M}^{E_k}]^*$, which is equivalent to $\rho_{N,M}^{E_k}=[\rho_{M,N}^{E_k}]^*$. It is convenient to define the reduced dissipator $\tilde{\mathcal D}_{N,M}^{E_k}$, where the terms containing $f_{N,M}^{E_k}$ are excluded. Summing up all the terms multiplying $f_{N,M}^{E_k}$  gives
\begin{align}
  \nonumber
  \gamma_{NME_k}= -\sum_{m,g,\pm}
   &\sth{ \Gamma_{N,M,M,m,E_k}^{\mathrm{out},g\pm} +\Gamma_{N+M,-M,-M,m,E_k}^{\mathrm{out},g\pm} } \\
   \label{eq:denom_iter}
   +\sum_{E_0=0,\pm}&\sth{\Gamma_{N,M,0,0,E_k}^{\mathrm{in},g,\pm}+\Gamma_{N+M,-M,0,0,E_k}^{\mathrm{in},g,\pm}},
\end{align}
where $E_0=0$ represents summing over all elastic scattering mechanisms. Using Eq.~\eqref{eq:denom_iter}, we can write the reduced dissipator as
\begin{align}
  \label{eq:reduced_diss}
  \tilde{\mathcal D}_{N,M}^{E_k}&=\mathcal D_{N,M}^{E_k}-\gamma_{NME_k}f_{NM}^{E_k},
\end{align}
which can depend on all matrix elements of the density matrix except for $f_{NM}^{E_k}$. Using Eqs.~\eqref{eq:reduced_diss}-\eqref{eq:denom_iter}, we can write Eq.~\eqref{eq:steady} as
\begin{align}
  \label{eq:iter_basic}
  f_{N,M}^{E_k}=\frac{\tilde{\mathcal D}_{NM}^{E_k}}{ \frac{i}{\hbar}\Delta_{N,N+M}-\gamma_{N,M,E_k}},
\end{align}
where the RHS does not depend on $f_{N,M}^{E_k}$. Equation~\eqref{eq:iter_basic} is then used as a basis for an iterative solution for $f_{NM}^{E_k}$ using a weighted Jacobi scheme~\cite{saad_iter_2003}
\begin{align}
  \label{eq:iter_relation}
  f_{NM}^{E_k,(j)} &= (1-\xi) f_{NM}^{E_k,(j-1)} \\
  &+ \xi\frac{\mathcal D_{NM}^{E_k,(j-1)}-\gamma_{NME_k} f_{NM}^{E_k,(j-1)}  }{ \frac{i}{\hbar}\Delta_{N,N+M}-\gamma_{N,M,E_k}},
\end{align}
where $\xi\in[0,1]$ is a weight. The quantity $\mathcal D_{NM}^{E_k,(j)}$  should be understood as matrix elements of the dissipator acting on the $j$-th iteration of the density matrix. As an initial guess, we assume a diagonal density matrix, where all subbands are equally occupied with a thermal distribution,
\begin{align}
  \label{eq:guess}
  f_{NM}^{E_k,(1)}= A \delta_{M,0} e^{-E_k/(k_BT)},
\end{align}
where $A$ is normalization constant. In the current work, we used $\xi=0.6$. However, we did not find convergence speed to depend strongly on the specific value. However, we found the numerical method to be more stable when $\xi>0.5$. We iterate Eq.~\eqref{eq:iter_relation} until the following convergence criterion is satisfied:
\begin{align}
  \label{eq:convergence}
   \sum_{N,M} \int dE_k | f_{NM}^{E_k,(j)}-f_{NM}^{E_k,(j-1)} |<\delta_\mathrm{iter},
\end{align}
where we used $\delta_\mathrm{iter}=10^{-3}$ in this work. Note that the density matrix is normalized to one, making $\delta_\mathrm{iter}$ unitless. Typically, $50-150$ iterations are needed to reach the convergence criteria.

In order to include corrections due to electron--electron interaction on a mean-field level, the iterative solution method described above is supplemented with a self-consistent solution to  Poisson's equation. The solutions to Poisson's equation is obtained using an outer self-consistency loop (the inner loop is the iterative solution described earlier). At the end of the inner (iterative) loop, corrections to the bandstructure $V_P(z)$ are calculated based on a self-consistent solution with Poisson's equation. If the change in $V_P(z)$ is above a certain threshold, the eigenfunctions and rates are recalculated and the inner iterative loop is repeated with the new eigenfunctions and rates. The process is repeated until $V_P(z)$ converges. To quantify when convergence is reached, we define the largest first-order correction to the eigenfunctions
\begin{align}
  \label{eq:poisson_corr}
  \delta_\mathrm{self}=\max_{m} \left| \int dz |\psi_m(z)|^2
  \left[ V_P^{\mathrm{new}}(z)-V_P^{\mathrm{old}}(z) \right] \right|,
\end{align}
where $|\psi_m(z)|^2$ is the probability density of eigenfunction $m$. In this work we use $\delta_\mathrm{self}=0.1$~meV, which is small compared with a typical energy spacing in QCLs. The number of iterations needed to satisfy the convergence criterion in Eq.~\eqref{eq:poisson_corr} depends on the electron density in the device (higher density means more iterations). For the device considered in Section.~\ref{sec:lm_design}, $2-5$ iterations are needed for the first value of applied electric field. For other values of electric field, the number of iteration needed can be kept to a minimum by sweeping the field and using results from previous fields as the initial guess for the next value of the electric field.

\subsection{Linear-Response Density Matrix}
Once we have calculated the steady-state density matrix using the procedure described in Sec.~\ref{subsec:steadystate}, we can calculate the linear-response density matrix elements $f_{NM}^{E_k}(\omega)$ using Eqs.~\eqref{eq:rho_omg_mat} and \eqref{eq:Adef}. By following the same steps as in Section~\ref{subsec:steadystate}, we get
\begin{align}
  \label{eq:rho_omg_iter}
  &f_{NM}^{E_k,(j)}(\omega)=  (1-\xi) f_{NM}^{E_k,(j-1)}(\omega) \\ \nonumber
  &+\xi\frac{\mathcal D_{NM}^{E_k,(j-1)}(\omega)-\gamma_{NME_k}f_{NM}^{E_k,(j-1)}(\omega)
  +A_{N,M}^{E_k,(j-1)}}{ \frac{i}{\hbar}(\Delta_{N,N+M}-\hbar \omega)
  -\gamma_{NME_k}} .
\end{align}
where the matrix elements $A_{N,M}^{E_k,(j)}$ only depend on the steady-state density matrix and are given by Eq.~\eqref{eq:Adef}. Note that the matrix elements $\mathcal D_{NM}^{E_k,(j)}(\omega)$ cannot be calculated directly using Eq.~\eqref{eq:diss_elements} because Fourier components of the linear-response density matrix are not Hermitian and the resulting dissipator cannot be written as a pair of Hermitian conjugate terms. The calculations will not be shown here, but the result is
\begin{align}
\label{eq:diss_freq_ele}
\begin{split}
  & \mathcal D_{NM}^{E_k}(\omega)=
   -\sum_{n,m,g,\pm} \Gamma_{NMnmE_k}^{\mathrm{out},g\pm}f_{N,n}^{E_k}(\omega) \\
  &+\sum_{n,m,g,\pm}\Gamma_{NMnmE_k}^{\mathrm{in},g,\pm}f_{N+n,M+m-n}^{E_k\mp E_s+\Delta_{N+M,N+M+m}}(\omega) \\
  &-\sum_{n,m,g,\pm} \Gamma_{N+M,-M,n,m,E_k}^{\mathrm{out},g\pm}f_{N+M+n,-n}^{E_k}(\omega) \\
  &+\sum_{n,m,g,\pm}\Gamma_{N+M,-M,n,m,E_k}^{\mathrm{in},g,\pm}f_{N+m,M-m+n}^{E_k\mp E_s+\Delta_{N,N+m}}(\omega),
\end{split}
\end{align}
which would reduce to Eq.~\eqref{eq:diss_elements} if the linear response DM were Hermitian. Equation~\eqref{eq:rho_omg_iter} is iterated until
\begin{align}
  \label{eq:conv_omg}
  |g^{(j)}(\omega)-g^{(j-1)}(\omega)|<10^{-3}~\mathrm{cm}^{-1},
\end{align}
where $g^{(j)}(\omega)$ is the gain calculated from $f_{NM}^{E_k,(j)}(\omega)$, using Eq.~\eqref{eq:suscept}. This process is repeated for multiple values of $\omega$ in the frequency range of interest.

\section{Results}
\label{sec:results}

\subsection{Lattice-matched design}
\label{sec:lm_design}
In order to verify the accuracy of our model, we simulated a midinfrared QCL proposed in Ref.~\cite{bismuto_APL_2010}. We chose this device because it has previously been successfully modeled using both NEGF and simplified density matrix approaches~\cite{lindskog_APL_2014}, which allowed us to compare our results to existing theoretical results. The QCL was designed for wide voltage tuning, emitting around 8.5~$\upmu$m ($146$~meV), with a tuning range of almost $100$~cm$^{-1}$ ($12$~meV). The device is based on the In$_{0.47}$Ga$_{0.53}$As/In$_{0.52}$Al$_{0.48}$As material system, which is lattice-matched to InP, eliminating the effects of strain. With this material composition, In$_{0.47}$Ga$_{0.53}$As acts as potential wells and In$_{0.52}$Al$_{0.48}$As as barriers with approximately a $520$~meV conduction-band discontinuity between materials. All material parameters relating to band structure (e.g., Kane energy) as well as other constants such as phonon energy and interface-roughness parameters are provided in Appendix~\ref{app:matpar}.

\subsubsection{Bandstructure}
\begin{figure}[h]
\centering
\includegraphics[width=\FigureWidth]{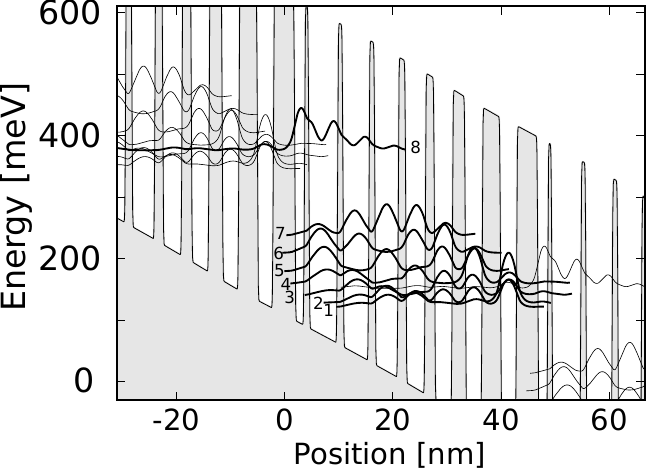}
\caption{
Conduction-band edge (thin curve bounding gray area) and probability densities for the 8 eigenstates used in calculations (bold curves) for a field strength of 50~kV/cm. States belonging to neighboring periods are denoted by thin gray curves. The states are numbered in increasing order of energy, starting with the ground state in the injector. Using this convention, the lasing transition is from state 8 to state 7. The length of one period is 44.9~nm, with a layer structure (in nanometers), starting with the injector barrier (centered at the origin) \textbf{4.0}/1.8/\textbf{0.8}/5.3/\textbf{1.0}/4.8/\textbf{1.1}/4.3/\textbf{1.4}/3.6/\textbf{1.7}/3.3/\textbf{2.4}/\underline{3.1}/ \underline{\textbf{3.4}}/2.9, with barriers denoted bold. Underlined layers are doped to $1.2\times10^{17}$~cm$^{-3}$, resulting in an average charge density of $n_\mathrm{3D}=1.74\times10^{16}$~cm$^{-3}$ over the stage.
}
\label{band_structure}
\end{figure}

\begin{figure}[h!]
\centering
\includegraphics[width=\FigureWidth]{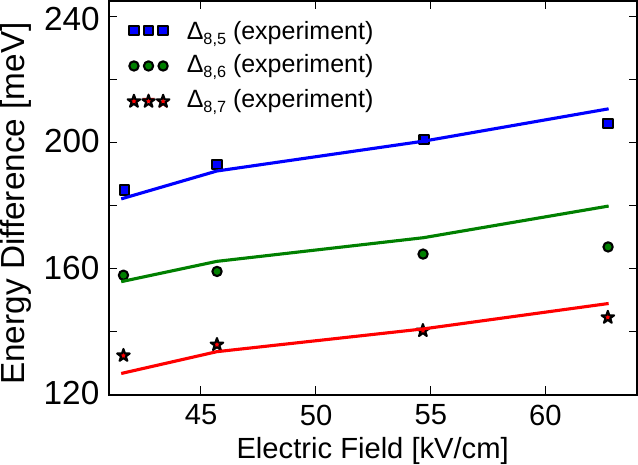}
\caption{
Calculated energy differences $\Delta_{8,7}$ (lowest red curve), $\Delta_{8,6}$ (middle green curve) and $\Delta_{8,5}$ (top blue curve) as a function of field. The symbols show results based on experimentally determined electroluminescence spectra from Ref.~\cite{bismuto_APL_2010}.
}
\label{ediff}
\end{figure}

Figure~\ref{band_structure} shows the conduction-band diagram and the probability density of the relevant eigenfunctions slightly above threshold (50~kV/cm). The figure shows the 8 considered states belonging to a single period in bold and states belonging to neighboring periods are represented as thin curves. The lasing transition is between states 8 and 7, and state 6 is the main extractor state. State number 1 is the energetically lowest state in the injector (often denoted as the ground state).
In order to verify the accuracy of the bandstructure shown in Fig.~\ref{band_structure}, we compare various calculated energy differences $\Delta_{n,m}$ to results obtained from experimentally determined electroluminescence spectra.~\cite{bismuto_APL_2010} Figure~\ref{ediff} shows the calculated energy differences $\Delta_{8,7}$, $\Delta_{8,6}$ and $\Delta_{8,5}$ as a function of electric field (symbols) and a comparison to experimentally obtained values (curves). From Fig.~\ref{ediff}, we can see that for low field strength (41.6 and 45.9~kV/cm), the deviation from experiment is small and the largest error is 3~meV (less than a 2\% relative difference) for the $\Delta_{8,6}$ energy spacing for a field strength of 46~kV/cm. For higher field strengths (54.7 and 62.7~kV/cm), the energy differences are overestimated by up to 13~meV (8\% relative error) for the $\Delta_{8,6}$ energy spacing for a field strength of 62.7~kV/cm. The reason for the discrepancy between experiment and theory at high fields is not clear. A possible reason is the low bandgap of the well material ($\sim 0.74$~meV) compared with the lasing transition at higher fields ($\sim 0.15$~eV), limiting the accuracy of the three-band $\dpr{k}{p}$ model.

\subsubsection{Current Density vs Field}
\label{sec:current_dens_ir}
\begin{figure}[h!]
\centering
\includegraphics[width=\FigureWidth]{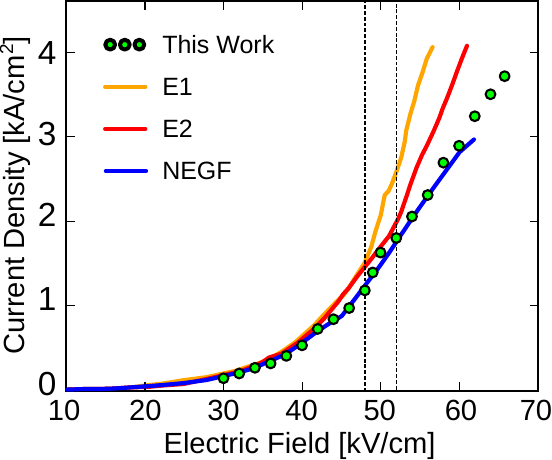}
\caption{
Calculated current density vs field based on this work and comparison to results using NEGF ~\cite{lindskog_APL_2014}. Also shown are experimental results from Ref.~\cite{bismuto_APL_2010} (E1) and a regrown device from Ref.~\cite{lindskog_APL_2014} (E2). Experimentally determined threshold fields of 48~kV/cm (E1) and 52~kV/cm (E2) are denoted by dashed vertical lines. Both theoretical results are given for a lattice temperature of 300~K. Experimental results are provided for pulsed operation, minimizing the effect of lattice heating above the heatsink temperature of 300~K.
}
\label{JE_comparison}
\end{figure}

Figure~\ref{JE_comparison} shows results for current density vs electric field calculated using the density-matrix formalism presented in this work, along with a comparison with two different experiments and theoretical results based on the NEGF formalism. Note that neither theoretical result takes the laser field into account, so we can only expect results to agree with experiment below and around threshold (denoted by dashed vertical lines in Fig.~\ref{JE_comparison}). The two experimental results are based on the same QCL-core design but with different waveguide designs. Experiment 1 (E1) refers to the sample from Ref.~\cite{bismuto_APL_2010}, which is based on a buried-heterostructure waveguide design. Experiment 2 (E2) refers to a sample based on a double-trench waveguide design with higher estimated losses than experiment 1 ~\cite{lindskog_APL_2014}. From Figure~\ref{JE_comparison}, we can see that our results based on the density-matrix formalism shows excellent agreement with the NEGF results for low fields (less than 40~kV/cm) and slightly higher current density for higher fields. Our density-matrix results are in quantitative agreement with experiments up to threshold for both experiments, while the NEGF underestimate current density for fields higher than 40~kV/cm. Above threshold, our density-matrix results underestimate the current density due to the omission of the laser electromagnetic field.

\begin{figure}[h!]
\centering
\includegraphics[width=\FigureWidth]{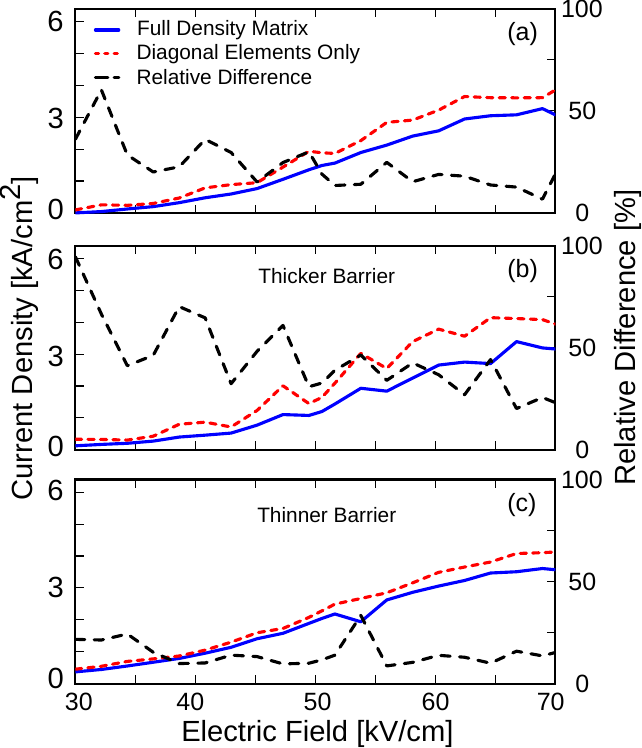}
\caption{
Comparison between the current density obtained using the full density matrix (solid blue curve) and semiclassical results with only diagonal elements included (short-dashed red curve). The relative difference between the two results is also shown (long-dashed black curve). (a) Results for the same device considered in Fig.~\ref{JE_comparison}. (b) Results for a device with a thicker injection barrier ($5$~nm). (c) Results for a device with a thinner injection barrier ($3$~nm).
}
\label{JE_classical}
\end{figure}

In order to get an idea how important the inclusion of off-diagonal density-matrix elements are, Fig.~\ref{JE_classical} shows a comparison between the current density obtained using the full density matrix and results obtained using only the diagonal matrix elements (semiclassical results). The importance of including off-diagonal matrix elements is known to be greater for thicker injection barriers, where transport is limited by tunneling through the injection barrier ~\cite{callebaut_JAP_2005}. For this reason, results are also plotted for different injection barrier widths in Figs.~\ref{JE_classical}(b) and (c). Figure~\ref{JE_classical}(a) shows the results for the same device as in Fig.~\ref{JE_comparison} (4-nm-thick injection barrier), while (b) and (c) show the results for a thicker ($5~$nm) and thinner (3~nm) injection barrier, respectively (but otherwise identical devices). In Fig.~\ref{JE_classical}(a), we see that even for a midinfrared QCL, where transport has previously been reported to be mostly incoherent,~\cite{iotti2001} the difference is noticeable and the current density is overestimated by about 10--60\% by semiclassical methods. The relative difference around threshold ($\sim$48~kV/cm) is about 29\%. This is a much bigger difference than the previously reported value of only a few percent for a different midinfrared QCL design ~\cite{iotti2001}. The difference could be a result of the different device design, or the assumption of a factorized (in-plane and cross-plane) density matrix employed in Ref.~\cite{iotti2001}, while we do not make that assumption. Figure~\ref{JE_classical}(b) shows that, for a thicker injection barrier (5~nm), a semiclassical model fails completely, with a relative difference ranging between 22 and 90\% and about 30\% difference around threshold. The semiclassical results also show peaks and negative differential resistance (NDR) regions that are absent (or much less pronounced) in the DM results. Figure~\ref{JE_classical}(c) shows that, for the thinner barrier (3~nm), the difference is smaller (between 8 and 32\%) and about 9\% around threshold. We can see that the considered device design with a 4-nm-thick injection barrier is on the border of admitting a semiclassical description.

\subsubsection{Optical Gain and Threshold Current Density}
\label{sec:gain_ir}
\begin{figure}[h!]
\centering
\includegraphics[width=\FigureWidth]{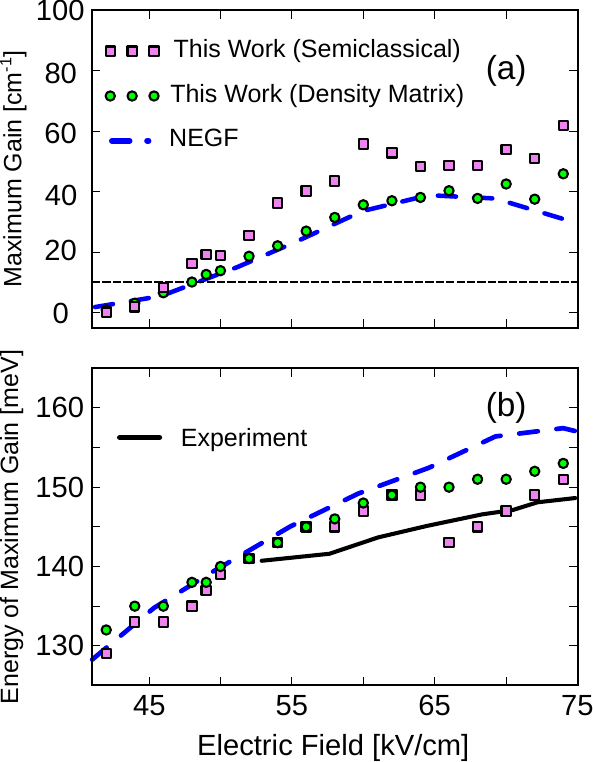}
\caption{
(a) Peak gain vs electric field. (b) Energy corresponding to peak gain as a function of electric field. Results are given for calculations using the full density-matrix (green circles), a semiclassical model (violet squares), and a comparison with NEGF (dashed blue line). Experimental results from Ref.~\cite{bismuto_APL_2010} are denoted by a thick solid black line. The NEGF results are from Ref.~\cite{lindskog_APL_2014}.
}
\label{gain_comparison}
\end{figure}

Figure~\ref{gain_comparison}(a) shows the peak gain as a function of electric field calculated using our density-matrix and semiclassical models. Also shown is a comparison with the results based on NEGF~\cite{lindskog_APL_2014}. The estimated threshold gain~\cite{bismuto_APL_2010,lindskog_APL_2014} $G_\mathrm{th}\simeq 10$~cm$^{-1}$ from experiment E1 is denoted by a horizontal dashed line. From Fig.~\ref{gain_comparison}(a), we see that our DM results predict a threshold field of $E_\mathrm{th}=48.0$~kV/cm. By matching this field to Fig.~\ref{JE_comparison} a threshold current density of $J_\mathrm{th}=1.55$~kA/cm$^2$ is found. The results are in good agreement with experiment ($E_\mathrm{th}=48.0$~kV/cm and $J_\mathrm{th}=1.50$~kA/cm$^2$). Our results are also in fair agreement with the NEGF results ($E_\mathrm{th}=47.6$~kV/cm and $J_\mathrm{th}=1.20$~kA/cm$^2$). In Fig.~\ref{gain_comparison}(a), we see that the semiclassical model gives the same qualitative behavior as the density-matrix results. However, the gain is overestimated for all field values above threshold. The agreement between our DM results and the NEGF results is excellent, except for very high field values ($>$ 70~kV/cm), where our DM model gives slightly higher gain.

Figure~\ref{gain_comparison}(b) shows the energy corresponding to the maximum gain in Fig~\ref{gain_comparison}(a). We see that both the DM and NEGF models give similar results, where the energetic position of the gain maximum is higher than the experimentally observed photon energy for all field values and the difference is greater at higher fields. However, both models agree with experiment slightly above threshold (52~kV/cm). The overestimation of the peak gain energy might look surprising considering the excellent agreement of the energy spacing between the upper and lower lasing level ($\Delta_{8,7}$) shown in Fig.~\ref{ediff}. However, other transitions such as $8\rightarrow 6$ also contribute to gain (more pronounced at higher fields) and the energy spacing $\Delta_{8,6}$ is overestimated by 3--13~meV. Another noticeable feature is the significant disagreement between the DM and semiclassical results at fields higher than 65~kV/cm. We attribute the difference to underestimation of the gain broadening using the semiclassical model, which leads to two distinct gain peaks, corresponding to $\Delta_{8,7}$ and $\Delta_{8,6}$. However, when using the full density matrix, the gain broadening is strong enough to combine the two peaks.

\subsubsection{In-Plane Dynamics}
\label{sec:inplane}
\begin{figure}[h!]
\centering
\includegraphics[width=\FigureWidth]{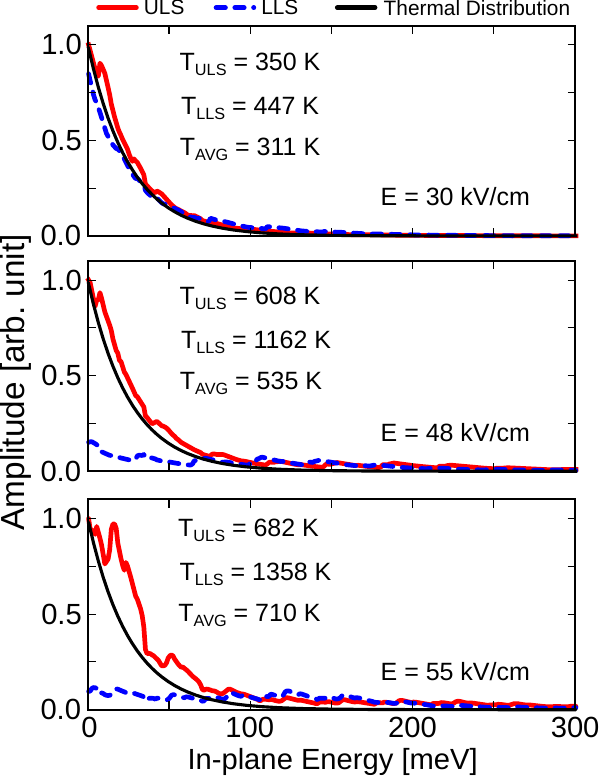}
\caption{
In-plane energy distribution for the ULS (solid red) and LLS (dashed blue) electrons. For comparison, a Maxwellian thermal distribution with a temperature of 300~K is also shown (thin black). Results are shown for electric fields of 30~kV/cm (top), 48~kV/cm (middle), and 55~kV/cm (bottom). For each value of the electric field, the electron temperatures of the ULS and LLS calculated using Eq.\eqref{eq:inplane} are shown, as well as the weighted average electron temperature,  $T_\mathrm{avg}$, obtained using Eq.~\eqref{eq:inplane_av}.
}
\label{inplane_comparison}
\end{figure}

Figure~\ref{inplane_comparison} shows the in-plane distribution $f_{N,0}^{E_k}$ for $N$ corresponding to the upper lasing state (ULS) and lower lasing state (LLS). Results are shown far below threshold (30~kV/cm), at threshold (48~kV/cm), and above threshold (55~kV/cm). As a point of reference, a Maxwellian thermal distribution with a temperature of 300~K is also shown. For each value of electric field, the electron temperatures calculated using Eq.~\eqref{eq:inplane} are shown, as well as the weighted average for all subbands using~\eqref{eq:inplane_av}. For low fields (30~kV/cm), the in-plane distribution is close to a Maxwellian distribution and the weighted average electron temperature is only 11~K higher than the lattice. For higher fields, significant electron heating takes place and in-plane distributions strongly deviate from a Maxwellian. There is a striking difference in shape and temperature between the ULS and LLS distributions at high fields. At threshold ($48$~kV/cm), the ULS is slightly ($73$~K) ``hotter'' than average, with a distribution that bears some resemblance to a thermal distribution, with most electrons having energy below $50$~meV. However, the ULS in-plane distribution has a much thicker tail than the thermal distribution, resulting in a high electron temperature. The LLS is considerably hotter than average electron temperature, with a distribution that is flat up to around $150$~meV (approximate lasing energy) and then slowly decays with a thick tail. These results are intuitive because the LLS has a low occupation and most of the occupation comes from electrons that recently scattered from the ULS with a large excess in-plane energy of $\simeq 150$~meV (corresponding to about $1740$~K), making the LLS very hot. On the other hand, electrons in the ULS come from long-lifetime states in the injector, where electrons have had time to emit many LO-phonons and get rid of excess energy.

We note that the inclusion of the optical-field feedback (not included in this work) is known to reduce the temperature of the LLS, because electrons can then transition from the ULS to the LLS via stimulated emission without acquiring any excess in-plane energy~\cite{matyas_JAP_2011}. Inclusion of electron-electron scattering (also not included in this work) would also reduce the difference between the in-plane distributions of different subbands~\cite{iotti_APL_2001}, even for the relatively low sheet doping density of $n_\mathrm{2D}=7.8\times10^{10}$~cm$^{-2}$ in the considered device. However, we do not expect electron--electron interaction to significantly change the average electron temperature, because electron--electron scattering conserves the total energy of the electron system.

The electron temperatures shown in Fig.~\ref{inplane_comparison} are considerably higher than previously reported for the same device using NEGF~\cite{lindskog_APL_2014}. For example, at a field of $55$~kV/cm, we obtain $T_\mathrm{ULS}=682$~K, compared with NEGF results of $T_\mathrm{ULS}=398$~K~\cite{lindskog_APL_2014}. The agreement is better for the LLS, where we obtain $T_\mathrm{LLS}=1358$~K, while NEGF gives $T_\mathrm{LLS}=1305$~K (private communications with the authors of Ref.~\cite{lindskog_APL_2014}). Despite the disagreement, our results are similar to experimental and theoretical results for different QCL designs with similar wavelengths. For example, at threshold, electronic temperature of $800$~K (at room temperature) has been experimentally estimated~\cite{spagnolo_APL_2004} for the GaAs-based QCL proposed in Ref.~\cite{page2001}, which had a similar lasing wavelength $\lambda \sim 9$~$\upmu$m. Even higher excess electron temperatures have been estimated by Monte Carlo calculations for a $9.4$-$\upmu$m QCL,~\cite{iotti_APL_2001} where the authors obtained electron temperature of about $700$~K at threshold for a lattice temperature of $77$~K.

\begin{figure}[h!]
\centering
\includegraphics[width=\FigureWidth]{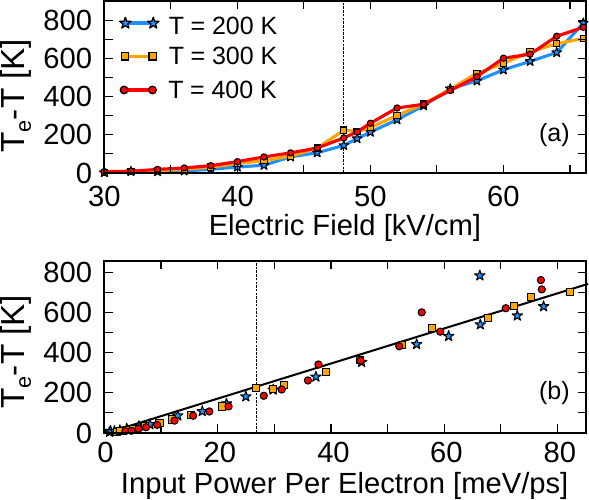}
\caption{
(a) Average excess electron temperature $T_e-T$ as a function of field, where $T_e$ is calculated using Eq.~\eqref{eq:inplane}. Results are shown for $T=200$~K (blue stars), $T=300$~K (orange squares) and $T=400$~K (red circles). The dashed vertical line represent threshold field of 48~kV/cm for the room-temperature results. (b) Average excess electron temperature $T_e-T$ as a function of electrical input power per electron, defined by Eq.~\eqref{eq:P_def}. The black line shows a best fit with the energy-balance equation~\eqref{eq:EBE}. The vertical line has the same meaning as in panel (a).
}
\label{heating_analysis}
\end{figure}

Figure~\ref{heating_analysis}(a) shows the average excess electron temperature $T_e-T$ as a function of electric field for different values of lattice temperature $T$. The average electron temperature is calculated using Eq.~\eqref{eq:inplane_av}. In Fig.~\ref{heating_analysis}(a), we see that considerable electron heating takes place for fields higher than $\sim 40$~kV/cm. For example, at a lattice temperature of $300$~K, we observe an excess electron temperature of $225$~K ($T_e=525$~K) around threshold ($48.0$~kV/cm) and a maximum value of $705$~K ($T_e=1005$~K) for $E=66$~kV/cm. 

Every time an electron traverses a single stage of length $L_P$, it gains energy $\mathcal E_0$, equal to the potential drop over a single period. It is therefore instructive to consider the excess electron temperature as a function of the normalized input power density
\begin{align}
  \label{eq:P_def}
  p_\mathrm{input} = \frac{v}{L_P} \mathcal E_0= \frac{J E}{n_\mathrm{3D}}
\end{align}
where $v$ is the drift velocity, $E$ the electric field, $J$ the current density and $n_\mathrm{3D}$ the average electron density. The normalized electrical power input gives the power transferred to each electron and is more convenient to compare between different devices than the (unnormalized) electrical power input $P_\mathrm{input}=JE$, which depends linearly on the electron density (for low doping, as long as scattering rates and the Hartree potential are weakly affected). Figure~\ref{heating_analysis}(b) shows a scatter plot of the excess electron temperature as a function of the normalized input power for several values of the lattice temperature. The reason for a scatter plot is that the electron temperature is not a single-valued function of $p_\mathrm{input}$ because $J$ is not an invertible function of in-plane energy ($J$ is not a monotonically increasing function of electric field, as there are regions of negative differential resistance, so it cannot be inverted). Figure~\ref{heating_analysis}(b) shows that electron heating is well described with a simple energy-balance equation (EBE)
\begin{align}
 \label{eq:EBE}
  T_e-T = \tau_E \frac{p_\mathrm{input}}{k_B}
\end{align}
with an energy dissipation time $\tau_E$. A least-squares fit to Eq.\eqref{eq:EBE} for the $300$~K results gives $\tau_E=0.79$~ps, which is of the same order of magnitude as typical POP scattering times~\cite{lundstrom2000}.

\subsection{Strain-balanced design}
\label{sec:sb_design}

In order to verify the accuracy of our model for QCLs based on strain-balanced heterostructures, we simulated the 4.6-$\upmu$m QCL proposed in Ref.~\cite{bai_APL_2008}, with additional information in Ref.~\cite{evans_APL_2007}. The device is based on alternating layers of In$_{0.669}$Ga$_{0.331}$As (well material) and In$_{0.362}$Al$_{0.638}$As (barrier material), with an approximate conduction-band discontinuity of $820$~meV, enabling strong confinement of electrons at the considered wavelength of 4.6~$\upmu$m ($\sim$270~meV). All material parameters relating to band structure (e.g., Kane energy) as well as other constants such as phonon energy and interface-roughness parameters are provided in Appendix~\ref{app:matpar}.

Figure~\ref{band_structure_bai} shows the conduction-band diagram (layer structure given in figure caption) and the probability density of the relevant eigenfunctions. The figure shows the 12 considered states belonging to a single period in bold and states belonging to neighboring periods are represented as thin curves. The lasing transition is between states 11 and 9, and state 12 is an unwanted leakage path. State number 1 is the energetically lowest state in the injector (often denoted as the ground state).
\begin{figure}[h]
\centering
\includegraphics[width=\FigureWidth]{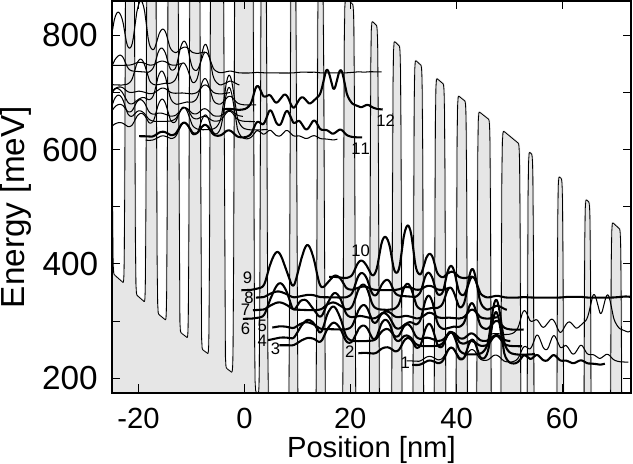}
\caption{
Conduction-band edge (thin curve bounding gray area) and probability densities for the 12 eigenstates used in calculations (bold curves) for a field strength of 78~kV/cm. States belonging to neighboring periods are denoted by thin gray curves. The states are numbered in increasing order of energy, starting with the ground state in the injector. Using this convention, the lasing transition is from state 11 to state 9, and state 12 is an unwanted leakage path. The length of one period is 50.4~nm, with a layer structure (in nanometers), starting with the injector barrier (centered at the origin) \textbf{3.8}/1.2~/\textbf{1.3}/4.3/\textbf{1.3}/3.8/\textbf{1.4}/3.6/\textbf{2.2}/2.8/\textbf{1.7}/2.5/\textbf{1.8}/\underline{2.2} /\underline{\textbf{1.9}}/\underline{2.1}/\underline{\textbf{2.1}}/\underline{2.0}/\underline{\textbf{2.1}}/1.8/\textbf{2.7}/1.8, with barriers denoted bold. Underlined layers are doped to $1.1\times10^{17}$~cm$^{-3}$, resulting in an average charge density of $n_\mathrm{3D}=2.70\times10^{16}$~cm$^{-3}$ over a single stage.
}
\label{band_structure_bai}
\end{figure}

\begin{figure}[h]
\centering
\includegraphics[width=\FigureWidth]{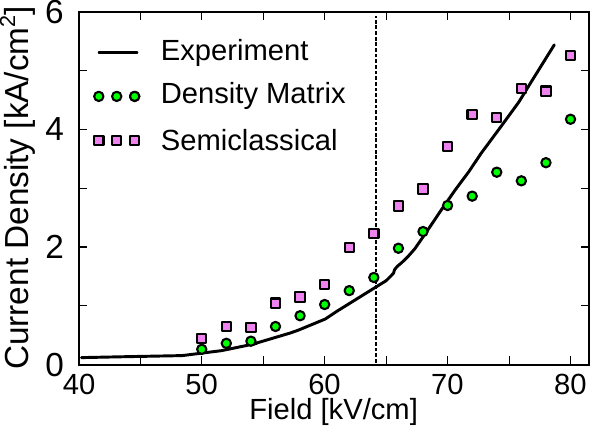}
\caption{
Calculated current density vs electric field obtained using our density-matrix (green circles) and semiclassical (violet squares) models. Also shown are experimental results (black curve) from Ref.~\cite{bai_APL_2008}. Experimentally determined threshold field of 64~kV/cm is denoted by dashed vertical line. Both theoretical results are given for a lattice temperature of 300~K and experimental results are given for a heatsink temperature of 300~K. Experimental results are provided for continuous-wave operation, so some lattice heating is expected to take place for high current density and field.
}
\label{JE_comparison_bai}
\end{figure}

Figure~\ref{JE_comparison_bai} shows results for current density vs electric field calculated using the density-matrix model and a comparison with experiment as well as semiclassical results. We obtain excellent agreement between our density-matrix model and experiment up field strengths of 66~kV/cm, which is slightly above the experimentally determined threshold of 64~kV/cm. Above threshold, current density is underestimated due to the omission of the laser field in the theoretical results. Another possible reason for the underestimation of current density at high fields is lattice heating (experimental results are in continuous-wave operation), which is not included in this work. As in the case of the 8.5~$\upmu$m QCL considered in section~\ref{sec:lm_design}, the semiclassical model overestimates current density for all field values. For example, around threshold (64~kV/cm), the density-matrix model predicts a current density of 1.48~kA/cm$^2$, while the semiclassical model gives 2.23~kA/cm$^2$, which is a relative difference of 40\%. The biggest absolute difference between the density-matrix and semiclassical results is 1.57~kA/cm$^2$, for a field strength of 76~kV/cm, while the biggest relative error is 56\%, at a field strength of 52~kV/cm.

\begin{figure}[h]
\centering
\includegraphics[width=\FigureWidth]{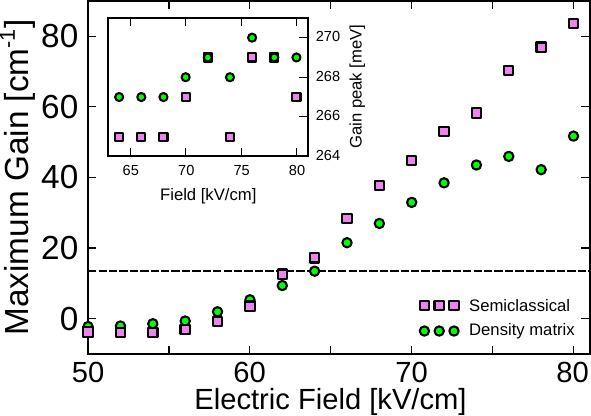}
\caption{
Peak gain vs electric field. Results are given for calculations both with off-diagonal density-matrix elements (green circles) and without them (violet squares). Dashed horizontal line shows the threshold gain $G_\mathrm{th}=13.5$~cm$^{-1}$, calculated using the experimentally determined threshold field of $64$~kV/cm.~\cite{bai_APL_2008} The inset shows the energy position corresponding to peak gain as a function of electric field. All results are for a lattice temperature of 300~K.
}
\label{gain_comparison_bai}
\end{figure}

Figure~\ref{gain_comparison_bai} shows the peak gain as a function of electric field calculated using our density-matrix and semiclassical models. No estimate for the threshold gain $G_\mathrm{th}$ was given in Ref.~\cite{bai_APL_2008} so the threshold gain (vertical dashed line) is simply the peak gain at the experimentally determined threshold field of 64~kV/cm, which gives a threshold gain of 13.5~cm$^{-1}$. From Fig.~\ref{gain_comparison_bai}, we see that the density-matrix and semiclassical models give similar results up to threshold, while for higher fields, the gain is overestimated by the semiclassical model. This is the same trend as for the 8.5-$\upmu$m-QCL considered in section \ref{sec:lm_design}.

The inset of Fig.~\ref{gain_comparison_bai} shows the energy corresponding to the maximum gain for electric field strengths above threshold. From the inset we see that the photon energy corresponding to maximum gain depends weakly on field strength for both DM and semiclassical models, where energies range from 267 to 270~meV (4.60 to 4.65~$\upmu$m) using the DM model and 265 to 269~meV (4.62 to 4.69~$\upmu$m) using the semiclassical model, in very good agreement with experiment (4.6~$\upmu$m).

\section{Conclusion}
\label{sec:conclusion}
We derived a Markovian master equation for the single-electron density matrix that conserves the positivity of the density matrix and is applicable to electron transport in both midinfrared and terahertz quantum cascade lasers. The master equation contains both the coherent (unitary) evolution of the density matrix as well as all relevant elastic and in-elastic scattering mechanism. Nonparabolicity in the band structure was accounted for by using a three-band $\dpr{k}{p}$ model, which includes the conduction band, light-hole band, and the spin-orbit split-off band. The model is an improvement of previous work employing density matrices because it does not rely on a phenomenological treatment of dephasing across thick injection barriers~\cite{terazzi_NJP_2010} or a factorization of the density matrix into cross-plane and in-plane distribution.~\cite{iotti_SST_2004} We demonstrated the validity of the model by simulating QCLs based in lattice-matched as well as strain-balanced heterostructures, with lasing wavelengths of 8.5~$\upmu$m and 4.6~$\upmu$m respectively. We compared our results to experiment as well as theoretical results based on NEGF and a semiclassical model.

For the 8.5-$\upmu$m-QCL, we calculated the current density vs electric field strength and obtained excellent agreement with experiments for electric fields below threshold, while current density was underestimated above threshold due to our omission of the laser electromagnetic field. We obtained a good agreement with NEGF for low field values ($<$ 42~kV/cm) and fair agreement for higher fields. At room temperature, we obtained a threshold electric field of $E_\mathrm{th}=48.0$~kV/cm and threshold current density of $J_\mathrm{th}=1.55$~kA/cm$^2$, in good agreement with experimental results of $E_\mathrm{th}=48.0$~kV/cm and $J_\mathrm{th}=1.50$~kA/cm$^2$ and in fair agreement with NEGF results of $E_\mathrm{th}=47.6$~kV/cm and $J_\mathrm{th}=1.2$~kA/cm$^2$. Comparison with a semiclassical model shows that off-diagonal matrix elements play a significant role. The semiclassical model (off-diagonal matrix elements are ignored) was found to overestimate current density by $29\%$ around threshold and up to $60\%$ for other electric field values. Our results for maximum gain vs field were in good agreement with NEGF except for very high ($>$ 70~kV/cm) field values. The semiclassical model was found to overestimate the peak gain above threshold, while giving fair agreement with DM and NEGF results below threshold. We predict significant electron heating, with in-plane distribution deviating strongly from a thermal distribution, especially the lower lasing state. We obtained an average electron temperature of $535$~K around threshold, which is $235$~K above the lattice temperature of $300$~K. From a simple energy-balance equation, we obtained an energy-dissipation time of $0.79$~ps for the considered device, which is similar to the scattering time due to polar-optical phonons.

For the 4.6-$\upmu$m-QCL, we calculated current density vs electric field strength at room temperature and obtained excellent agreement with experiment for fields up to 66~kV/cm, which is slightly above the experimentally determined threshold of $64$~kV/cm. Above threshold, the calculated current density was below experimental results due to our omission of the laser electric field as well effects of lattice heating in continuous-wave operation. As in the case of the 8.5-$\upmu$m-QCL We found that a semiclassical model overestimates current density for all field values. The semiclassical model was found to overestimate current density by $40\%$ around threshold and up to $58\%$ for other electric field values, demonstrating the need to include off-diagonal elements of the density matrix.

\begin{acknowledgments}
This work was supported by the DOE-BES award DE-SC0008712 and by AFOSR award FA9550-18-1-0340. Preliminary work was partially supported by the Splinter Professorship. Calculations were performed using the resources of the UW-Madison Center For High Throughput Computing (CHTC) in the Department of Computer Sciences.
\end{acknowledgments}


\appendix

\section{Electronic States}
\label{sec:electronic}

Within the envelope function approximation, the electron eigenfunctions near a band extremum $\ve k_0$ ($\Gamma$ valley in this work) can be written as a combination of envelope functions from different bands~\cite{bastard_1988_wave,Chuang1995}
\begin{align}
  \label{eq:env_def}
  \Psi_{n\ve k}(\ve r,z)=  \sum_b \phi_{n,\ve k}^{(b)}(\ve r,z)u_{\ve k=\ve k_0}^{(b)}(\ve r,z),
\end{align}
where the $b$ sum is over the considered bands, $u_{\ve k_0}^{(b)}(\ve k,z)$ is a rapidly varying Bloch function that has the period of the lattice, and $\phi_{n,\ve k}^{(b)}(\ve r,z)$ are slowly varying envelope functions belonging to band $b$. We make the approximation that the envelope functions factor into cross-plane and in-plane direction according to
\begin{align}
  \label{eq:separation}
  \phi_{n,\ve k}^{(b)}(\ve r,z)= N_\mathrm{norm} \psi_n^{(b)}(z)e^{i\dpr{k}{r}},
\end{align}
where $\psi_n^{(b)}(z)=\phi_{n,\ve k=0}^{(b)}(\ve r,z)$ is the envelope function for $\ve k=0$ and $N_\mathrm{norm}$ is a normalization constant. This separation of in- and cross-plane motion is a standard approximation when describing QCLs and other superlattices.~\cite{jirauschek_APR_2014} Without it, a separate eigenvalue problem for each $\ve k$ would need to be calculated, which is very computationally expensive.

In order to model QCLs based on III-V semiconductors, accurate calculation the $\Gamma$-valley eigenstates is crucial. The simplest approach is the Ben Daniel--Duke model, in which only the conduction band is considered ~\cite{bastard_1988_wave}. The Schr{\"o}dinger equation for the conduction-band envelope function $\psi_n^{(c)}(z)$ in Eq.~\eqref{eq:separation} reduces to the 1D eigenvalue problem~\cite{jirauschek_APR_2014}
\begin{subequations}
\begin{align}
  \sth{ -\frac{\hbar^2}{2}\frac{d}{dz}\frac{1}{m^*(z)}\frac{d}{dz}+V^{(c)}(z) }\psi_n^{(c)}(z)
  =E_n\psi_n^{(c)}(z) \\   \label{eq:sch_eff_mass}
  E_{n,\ve k}=E_n + E_k,
\end{align}
\end{subequations}
where $V^{(c)}(z)$ is the conduction-band profile, and $m^*(z)$ is the spatially dependent cross-plane effective mass of the conduction band. Typically, a parabolic dispersion relation $E_k=\hbar^2k^2/(2\mpar)$ is assumed. A single-band approximation such as Eq.~\eqref{eq:sch_eff_mass} can be justified when electrons have energy close to the conduction-band edge (compared with the bandgap $E_\mathrm{gap}$) and quantum well widths are larger than $\sim$10~nm ~\cite{nelson_PRB_1987,bastard_1988_wave}. This is the case for THz QCLs, where single-band approximations can give satisfactory results~\cite{dupont_JAP_2012,jonasson_JCEL_2016}. However, in mid-IR QCLs, the well thicknesses can be below 2~nm~\cite{bismuto_APL_2010} and the energy of the lasing transition can exceed 150~meV, which is a significant fraction of typical bandgaps of the semiconductor materials, especially for low-bandgap materials containing InAs. In this case, including nonparabolicity in the electron Hamiltonian is important.

Nonparabolicity has previously been included in semiclassical ~\cite{matyas_JAP_2011,jirauschek_APR_2014} density-matrix~\cite{terazzi2012,lindskog_APL_2014}, and NEGF-based~\cite{kolek_APL_2012,lindskog_APL_2014,bugajski_PSS_2014} approaches using single-band models, where only the conduction band is modeled, but the influence of the valence bands is included via an energy-dependent effective mass. One of the most important effects of nonparabolicity is the energetic lowering of the high-energy states, such as the upper lasing level. This effect results from higher effective mass at high energies, reducing the kinetic energy term in the Schr{\"o}dinger equation ~\cite{ekenberg_PRB_1989}. However, an energy-dependent effective mass has an undesirable trait in density-matrix and semiclassical models, the resulting eigenfunctions are not orthogonal~\cite{jirauschek_APR_2014}. In this work, we will calculate the bandstructure using a $3$-band $\dpr{k}{p}$ model, which includes the conduction (c), light hole (lh) and spin-orbit split-off (so) bands ~\cite{sirtori_PRB_1994}. The three-band $\dpr{k}{p}$ model has previously been used in semiclassical work on QCLs~\cite{gao_JAP_2007,shi_JAP_2014}; present work is a quantum-mechanical QCL simulation relying this model.

The three-band Hamiltonian can be derived from an $8$-band Hamiltonian for bulk zinc-blende crystals~\cite{bahder_PRB_1990}, which, at in-plane wave vector $\ve k=0$, reduces to $3$ bands due to spin degeneracy and the heavy-hole band factoring out. The $3\times 3$ Hamiltonian for a strained bulk crystal can be written as a sum of three terms
\begin{align}
  \label{eq:kp_split}
  H_B=H_\mathrm{B,edge}+H_\mathrm{B,kin}+H_\mathrm{B,strain}.
\end{align}
The first term in Eq.~\eqref{eq:kp_split} is diagonal and contains the band edges
\begin{align}
 H_\mathrm{B,edge}=
\begin{pmatrix}
 E_c &
 0 &
 0 \\
  0 &
  E_v &
  0 \\
  0 &
  0 &
   E_v-\Delta_\mathrm{so}
\end{pmatrix},
\end{align}
where $E_c$ ($E_v$) is the conduction (valence) band edge and $\Delta_\mathrm{so}$ is the spin-orbit split-off energy.

The second term in Eq.~\eqref{eq:kp_split} contains the kinetic contributions as well as coupling between the different bands which involve the cross-plane wavenumber $k_z$
\begin{align}
 \nonumber
 &H_\mathrm{B,kin}= \\
 &
\begin{pmatrix}
 (1+2 F) \frac{\hbar^2k_z^2}{2m_0} &
 \sqrt{\frac{\hbar^2E_P}{3m_0}}  k_z &
 \sqrt{\frac{\hbar^2E_P}{6m_0}}  k_z \\
  \sqrt{\frac{\hbar^2E_P}{3m_0}}  k_z &
  - (\gamma_1+2\gamma_2)\frac{\hbar^2k_z^2}{2m_0} &
  -\sqrt{2}\gamma_2 \frac{\hbar^2k_z^2}{m_0} \\
  \sqrt{\frac{\hbar^2E_P}{6m_0}}  k_z &
  -\sqrt{2}\gamma_2 \frac{\hbar^2k_z^2}{m_0} &
  -\gamma_1 \frac{\hbar^2k_z^2}{2m_0}
\end{pmatrix},
\end{align}
with $E_P$ the Kane-energy,~\cite{kane_1966} $m_0$ the free-electron mass, $F$ is a correction to the conduction band effective mass due to higher energy $\Gamma$ valleys ~\cite{kane_1966},  $\gamma_1$ and $\gamma_2$ are the modified Luttinger parameters~\cite{pidgeon_PR_1966} given by~\cite{bahder_PRB_1990}
\begin{align}
\begin{aligned}
  \label{eq:lutt1}
  \gamma_1&=\gamma_1^{L}-\frac{E_P}{3E_\mathrm{gap}+\Delta_\mathrm{so}} \\
  \gamma_2&=\gamma_2^{L}-\frac12\frac{E_P}{3E_\mathrm{gap}+\Delta_\mathrm{so}},
\end{aligned}
\end{align}
where $\gamma_1^L$ and $\gamma_2^{L}$ are the standard Luttinger parameters ~\cite{luttinger_PR_1957} and $E_\mathrm{gap}$ is the $\Gamma$-valley bandgap. For a brief overview of the different material parameters, we refer the reader to Ref.~\cite{vurgaftman_JAP_2001}.

The third term in Eq.~\eqref{eq:kp_split} contains the effects of strain for a layer grown on a (001)-oriented (z-direction) substrate ~\cite{Chuang1995},
\begin{align}
 \label{eq:strainpart}
 &H_\mathrm{B,strain}=
 &
\begin{pmatrix}
 A_\eps &
 0 &
 0 \\
  0 &
  - P_\eps+Q_\eps &
  \sqrt{2} Q_\eps \\
  0 &
  \sqrt{2}Q_\eps &
  -P_\eps
\end{pmatrix},
\end{align}
with
\begin{subequations}
\begin{align}
  \label{eq:strain_const}
  A_\eps&= 2 a_c \left( 1- \frac{c_{12}}{c_{11}}\right)\frac{a_0-a}{a} \\
  P_\eps&= 2 a_v \left( 1- \frac{c_{12}}{c_{11}}\right)\frac{a_0-a}{a} \\
  Q_\eps&= - b \left( 1+2\frac{c_{12}}{c_{11}}\right)\frac{a_0-a}{a},
\end{align}
\end{subequations}
where $a_c$, $a_v$ and $b$ are the Bir-Pikus deformation potentials~\cite{bir_1974} with the sign convention of all being negative, $a_0$ and $a$ are the lattice constant of the substrate and the grown material (in the absence of strain), respectively and $c_{11}$ and $c_{12}$ are the elastic stiffness constants of the strained material. Numerical values for all material parameters used in this work are given in Appendix~\ref{app:matpar}.

\begin{figure}[h]
\centering
\includegraphics[width=\FigureWidth]{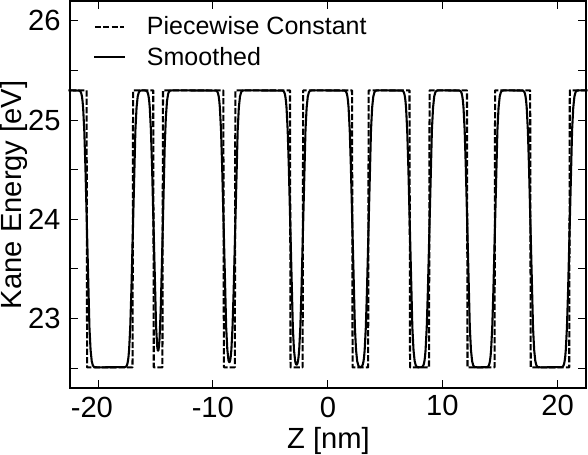}
\caption
{
Kane energy as a function of position for a single period of the InGaAs/InAlAs-based mid-IR QCL described in section~\ref{sec:lm_design}. The dashed curve shows the piecewise constant version and the solid curve is the smoothed version given by Eq.~\eqref{eq:material_smooth}, with $\sigma=0.29$~nm. This is double the value of $\sigma$ we typically use and is chosen for visualization purposes. The region of higher Kane energy corresponds to InGaAs.
}
\label{material_smoothing}
\end{figure}

Equation~\eqref{eq:kp_split} is valid for a bulk III-V semiconductors, where $k_z$ refers to crystal momentum in the [001] direction. However, for a heterostructure under bias, $k_z$ is no longer a good quantum number and must be replaced by a differential operator $k_z\rightarrow -i \pa{}{z}$. In addition, all the material properties become functions of position. Care must be taken when applying Eq.~\eqref{eq:kp_split} to heterostructures because the proper operator symmetrization must be performed to preserve the hermiticity of the Hamiltonian. In this work, we follow the standard operator symmetrization procedure in $\dpr{k}{p}$ theory for heterostructures~\cite{willatzen_2009_kp}
\begin{align}
\begin{aligned}
  \label{eq:symmetrization}
  f(z) k_z  &\rightarrow \frac12 \pth{f(z) k_z + k_z f(z)},  \\
  g(z) k_z^2&\rightarrow k_z g(z) k_z,
\end{aligned}
\end{align}
where $f(z)$ and $g(z)$ are arbitrary functions of position (e.g., Kane energy, or Luttinger parameter). The electron Hamiltonian for a heterostructure can then be written as
\begin{align}
  \label{eq:blockham}
  H_\mathrm{het}=
\begin{pmatrix}
 H_c &
 H_{c,lh} &
 H_{c,so} \\
  H_{c,lh}^\dagger &
  H_{lh} &
  H_{lh,so} \\
  H_{c,so}^\dagger &
  H_{lh,so}^\dagger &
  H_{so}
\end{pmatrix},
\end{align}
with
\begin{align}
\begin{aligned}
  \label{eq:blockdef}
  H_c   &= E_c(z) -\frac{\hbar^2}{2m_0} \paz (1+2 F(z)) \paz + A_\eps(z) ,  \\
  H_{lh} &= E_v(z) + \frac{\hbar^2}{2m_0} \paz (\gamma_1(z)+2\gamma_2(z)) \paz \\
        &   - P_\eps(z) + Q_\eps(z)  , \\
  H_{so} &= E_v(z)-\Delta_\mathrm{so}(z)+ \frac{\hbar^2}{2m_0} \paz \gamma_1(z) \paz
         - P_\eps(z) , \\
  H_{c,lh} & = -\frac{i}{2} \sqrt{\frac{\hbar^2E_P(z)}{3m_0}}  \paz
              - \frac{i}{2} \paz \sqrt{\frac{\hbar^2E_P(z)}{3m_0}} , \\
  H_{c,so} & = -\frac{i}{2} \sqrt{\frac{\hbar^2E_P(z)}{6m_0}}  \paz
              -\frac{i}{2} \paz \sqrt{\frac{\hbar^2E_P(z)}{6m_0}} , \\
  H_{lh,so}& = \sqrt{2} \frac{\hbar^2}{m_0} \paz \gamma_2(z) \paz
             + \sqrt{2}Q_\eps(z) .
\end{aligned}
\end{align}
From Eq.~\eqref{eq:blockdef} we see that the electron Hamiltonian now depends on spatial derivatives of material functions such as the Kane energy $E_P(z)$ and modified Luttinger parameters $\gamma_1(z)$, and results will be sensitive to how an interface between two different materials is treated. One choice is to treat the material functions as piecewise-constant. However, with that choice, spatial derivatives become delta functions (and derivatives of delta functions) that are difficult to treat numerically. In this work we will assume that material parameters are smooth functions that have slow variations on length scales shorter than a single monolayer $L_\mathrm{ML}$ (in heterostructures grown on InP, $L_\mathrm{ML}\simeq 0.29$~nm at room temperature). The material parameters that we use are obtained using
%
\begin{align}
  \label{eq:material_smooth}
  f(z)=\frac{1}{\sqrt{\pi} \sigma } \int f_{\mathrm{pc}}(z') e^{-(z-z')^2/\sigma^2} dz' ,
\end{align}
where we use $\sigma=\tfrac12 L_\mathrm{ML}$ and $f_{\mathrm{pc}}(z)$ is a piecewise constant material function that is discontinuous at material interfaces. As an example, Fig.~\ref{material_smoothing} shows the Kane energy as well as the piecewise-constant variant for a single period of the QCL studied in section~\ref{sec:lm_design}, with $\sigma=0.29$~nm.

The eigenfunctions of the Hamiltonian $H_\mathrm{het}$ given in Eqs.~\eqref{eq:blockham}-\eqref{eq:blockdef} satisfy
\begin{align}
  \label{eq:eigendef}
  H_\mathrm{het}\overline{\psi_n}(z)=E_n\overline{\psi_n}(z),
\end{align}
where $\overline{\psi_n}$ are three-dimensional vectors of envelope functions $\overline{\psi_n}(z)=\pth{\psi^c_n(z),\psi^\mathrm{lh}_n(z),\psi^\mathrm{so}_n(z)}^T$.  We will use Dirac notation $\ket{n}$, with $\braket{z}{n}  =\bar{\psi_n}(z)$ and $\ket{n,\ve k}$ with $\braket{\ve r, z}{n,\ve k}=N_\mathrm{norm} \bar{\psi_n}(z) e^{i\dpr{k}{r}}$ and normalization $\braket{n,\ve k}{n',\ve k'}=\delta_{n,n'}\delta(\ve k-\ve k')$. The electron probability density is obtained by summing over the different bands
\begin{align}
  \label{eq:densdef}
  |\overline{\psi}_n(z)|^2=|\braket{z}{n}|^2=\sum_b |\psi_n^{b}(z)|^2
\end{align}
and matrix elements of an operator B such as $\innp{n}{B}{n'}$ are understood as
\begin{align}
  \label{eq:innrpod}
  \innp{n}{B}{n'} = \sum_b \int dz \sth{\psi_n^{(b)}(z)}^*B^{(b)}\psi_{n'}^{(b)}(z) \ ,
\end{align}
where the sum is over the considered bands, with
\begin{align}
  \label{eq:opdef}
  B =
\begin{pmatrix}
 B^\mathrm{(c)} & 0 & 0 \\
 0 & B^\mathrm{(lh)} & 0 \\
 0 & 0 & B^\mathrm{(so)}
\end{pmatrix} .
\end{align}
Note that $B^{(b)}$ is an operation working on component b of the envelope function $\overline{\psi}_n(z)$. In this work all operators are assumed to be identical for all three bands ($B^{(c)}=B^{(lh)}=B^{(so)}$) except for the interaction potential due to interface roughness. The interaction potential due to interface roughness is not the same for all bands because the band discontinuities at material interfaces are not the same for different bands [see appendix~\ref{sec:roughness}].

\begin{figure}[h]
\centering
\includegraphics[width=2.5in]{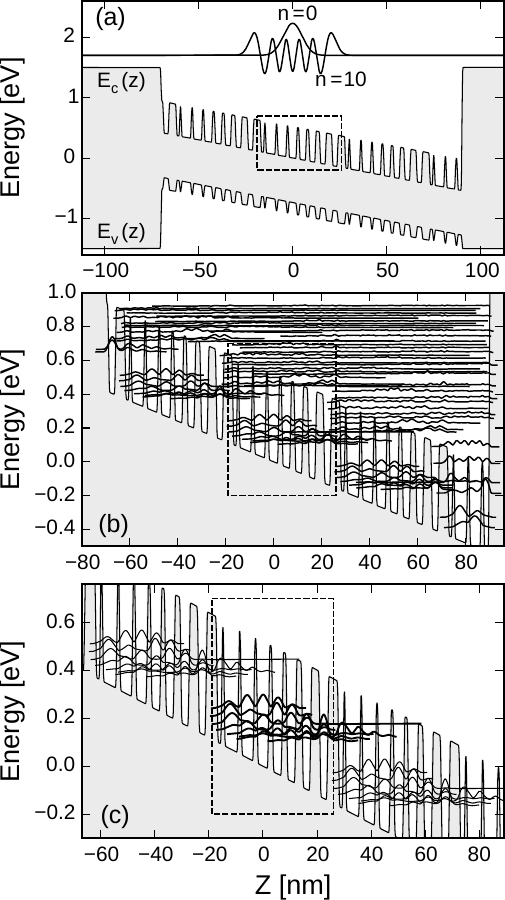}
\caption{
(a) Conduction band $E_c(z)$ and valence band $E_v(z)$ used for the calculation of eigenfunctions. Dashed rectangle shows a single period of the conduction band. Potential barriers are imposed to the far left and far right for both conduction and valence bands. Also show, are two of the Hermite functions ($n=0$ and $n=10$) used as basis functions. (b) After discarding the high and low-energy states, about 80 states are left, most of which have an energy high above the conduction band or are greatly affected by the boundary conditions on the right. (c) After choosing 8 states with the lowest $\tilde E_n$ and discarding copies from adjacent stages (see main text), the states in the central stage (bold curves) are left. States belonging to adjacent periods (thin curves) are calculated by translating states from the central stage in position and energy.
}
\label{eig_calc}
\end{figure}

The eigenfunctions $\ket{n}$ are calculated in a similar manner as in previous semiclassical work ~\cite{gao_JAP_2007,shi_JAP_2014,jirauschek_APR_2014}. A typical computational domain is shown in Fig.~\ref{eig_calc}(a), where the conduction band of a single period of the QCL from Ref.~\cite{bismuto_APL_2010} is marked by a dashed rectangle. The computational domain contains 3--4 periods, padded with tall potential barriers, placed far away from the central region in order for the boundary conditions not to effect the eigenfunctions near the center of the computational domain. We solve the eigenvalue problem in~\eqref{eq:eigendef} using a basis of Hermite functions (eigenfunctions of the harmonic oscillator), with two example basis functions shown in Fig.~\ref{eig_calc}(a). We used Hermite functions because they are easy to compute and required less computational resources than finite-difference methods. We typically use a basis of $\sim 400$ Hermite functions for each band, resulting in a matrix with dimension $1200\times 1200$. The diagonalization procedure results in $1200$ eigenstates, most of which are far above the conduction band or far below the valence bands. For transport calculations, we only include the states close to the conduction-band edge. This bound-state approach is known to produce spurious solutions, which are states with high amplitude far above the conduction band edge. In order to systematically single out the relevant states, we first  ``throw away'' the states that have energies far ($\sim 1$~eV) below or far above the conduction-band edge. This step typically reduces the number of states to $\sim100$. Figure~\ref{eig_calc}(b) shows the conduction-band edge along with the remaining eigenstates. Next we remove the states that are obtained by translation of a state from the previous/next period (e.g., remove duplicates). We do accomplish this by calculating each eigenstate's "center of mass",
\begin{align}
  \label{eq:centermass}
  z_n =\int z |\psi_n(z)|^2 dz,
\end{align}
and only keep states within a range $z\in[z_c,z_c+L_P]$, where $L_P$ is the period length. We will refer to this range as the center period. The choice of $z_c$ is arbitrary and we typically use $z_c=-L_P/4$. Lastly, we follow the procedure in Ref.~\cite{jirauschek_APR_2014} and, for the remaining states, calculate the energy with respect to the conduction-band edge using
\begin{align}
  \label{eq:energy_cb}
  \tilde E_n=E_n-\int E_c(z)|\psi_n(z)|^2 dz .
\end{align}
States with $\tilde E_n<0$ are valence-band states and are discarded. We keep $N_s$ states, with the lowest positive values of $\tilde E_n$. The value of $N_s$ is chosen such that all eigenstates below the conduction band top are included. After this elimination step, $N_s$ is the number of states used in transport calculations. The remaining states for $N_s=8$ are shown in bold in Fig.~\ref{eig_calc}(c). States belonging to other periods (needed to calculate coupling between different periods) are finally obtained by translation in space and energy of the $N_s$ states belonging to the center period. These states are identified by thin curves in Fig.~\ref{eig_calc}(c).

\section{Dissipator for Electron--Phonon Interaction}
\label{sec:phonon}
 Expanding the dissipator  (commutators in Eq.~\eqref{eq:master_schrod}) gives $4$ terms, which can be split into two Hermitian conjugate pairs
\begin{align}
\label{eq:diss_01}
\begin{split}
  &\mathcal D[\rho_e(t)]=\int_0^{\infty} ds \mathrm{Tr}_\mathrm{ph} \{ \\
  &+H_i \rho_e(t)\otimes \rho_\mathrm{ph} e^{-i(H_0+H_\mathrm{ph})s}H_i e^{i(H_0+H_\mathrm{ph})s}  \\
  &-H_ie^{-i(H_0+H_\mathrm{ph})s}H_ie^{i(H_0+H_\mathrm{ph})s}\rho_e(t)\otimes \rho_\mathrm{ph} \} \\
  &  + \mathrm{h.c..}
\end{split}
\end{align}
In order to proceed, we assume a Fr{\"o}lich-type Hamiltonian for the description of an electron interacting with a phonon bath in the single electron approximation~\cite{frolich_PRSA_1937}
\begin{align}
  \label{eq:inter_phonon}
  H_i= \frac{1}{\sqrt{V}} \sum_{\mathbf Q}
  \mathcal M(\mathbf Q)(b_\mathbf{Q}e^{i\mathbf Q\cdot \mathbf R}
             -b^\dagger_\mathbf{Q}e^{-i\mathbf Q\cdot \mathbf R}).
\end{align}
Here, $b_\mathbf{Q}^\dagger$ ($b_\mathbf{Q}$) is the creation (annihilation) operator for a phonon with wave vector $\mathbf Q$ and the associated matrix element $\mathcal M(\mathbf Q)$ and $V$ is the quantization volume. Note that the matrix elements are anti-Hermitian $\mathcal M(\mathbf Q)^*=-\mathcal M(\mathbf Q)$, making the interaction Hamiltonian Hermitian.

Inserting Eq.~\eqref{eq:inter_phonon} into Eq.~\eqref{eq:diss_01} gives
\begin{align}
  \nonumber
  \mathcal D &=\frac{1}{V}\int_0^{\infty} ds
  \sum_{\mathbf Q,\mathbf Q'} \mathcal M(\mathbf Q)^* \mathcal{M}(\mathbf Q') \mathrm{Tr}_\mathrm{ph} \Big\{
  \\ \nonumber
  -&\left( b_\mathbf{Q}e^{i\dpr{Q}{R}}-b^\dagger_\mathbf{Q} e^{+i\dpr{Q}{R}} \right)
  \rho_e(t)\otimes \rho_\mathrm{ph}
  e^{-i(H_0+H_\mathrm{ph})s} \times \\ \nonumber
  &\left( b_\mathbf{Q'}e^{i\dpr{Q'}{R}}-b^\dagger_\mathbf{Q'} e^{-i\dpr{Q'}{R}} \right)
  e^{i(H_0+H_\mathrm{ph})s}  \\ \nonumber
  +&\left( b_\mathbf{Q}e^{i\dpr{Q}{R}}-b^\dagger_\mathbf{Q} e^{-i\dpr{Q}{R}} \right)
  e^{-i(H_0+H_\mathrm{ph})s}\times \\ \nonumber
  &\left( b_\mathbf{Q'}e^{i\dpr{Q'}{R}}-b^\dagger_\mathbf{Q'} e^{-i\dpr{Q'}{R}} \right)
  e^{i(H_0+H_\mathrm{ph})s}
  \rho_e(t)\otimes \rho_\mathrm{ph} \Big\} \\ \label{eq:diss_02}
  +& \mathrm{h.c..}
\end{align}
In order to proceed, we use the fact that the thermal equilibrium-phonon density matrix is diagonal, so only terms with $\mathbf Q=\mathbf Q'$ survive the trace over the phonon degree of freedom. We also use that only terms containing both creation and annihilation operators are nonzero after performing the trace. Using the two simplifications, we get
\begin{align}
  \nonumber
  &\mathcal D=\frac{1}{V}\int_0^{\infty} ds \sum_{\mathbf Q}|\mathcal M(\mathbf Q)|^2
  \mathrm{Tr}_\mathrm{ph} \Big\{ \\ \nonumber
   &+b_\mathbf{Q}e^{i\dpr{Q}{R}} \rho_e(t)\otimes \rho_\mathrm{ph} e^{-iH_0s}e^{-iH_\mathrm{ph}s}
   b_\mathbf{Q}^\dagger e^{iH_\mathrm{ph}s} e^{-i\dpr{Q}{R}}e^{iH_0s} \\ \nonumber
   &+b_\mathbf{Q}^\dagger e^{-i\dpr{Q}{R}} \rho_e(t)\otimes\rho_\mathrm{ph} e^{-iH_0s}e^{-iH_\mathrm{ph}s}
   b_\mathbf{Q}e^{iH_\mathrm{ph}s} e^{i\dpr{Q}{R}}e^{iH_0s} \\ \nonumber
   &-b_\mathbf{Q}e^{i\dpr{Q}{R}}e^{-iH_0s}e^{-iH_\mathrm{ph}s}b_\mathbf{Q}^\dagger e^{iH_\mathrm{ph}s}
   e^{-i\dpr{Q}{R}}e^{iH_0 s} \rho_e(t)\otimes \rho_\mathrm{ph} \\ \nonumber
   &-b_\mathbf{Q}^\dagger e^{-i\dpr{Q}{R}}e^{-iH_0s}e^{-iH_\mathrm{ph}s}b_\mathbf{Q}e^{iH_\mathrm{ph}s}
   e^{i\dpr{Q}{R}}e^{iH_0s}\rho_e(t)\otimes \rho_\mathrm{ph} \Big\} \\ &+ \mathrm{h.c..}
\end{align}
Performing the trace, and going from a sum over $\mathbf Q$ to an integral, we get
\begin{align}
  \nonumber
  &\mathcal D = \frac{1}{(2\pi)^3} \sum_{\pm} \int_0^{\infty}ds \int d^3Q  |\mathcal M(\mathbf Q)|^2\\
  \nonumber
  & \times \big\{ \pth{\tfrac12 \mp \tfrac12 +N_\mathbf{Q}}e^{\mp iE_\mathbf{Q} s}e^{i\dpr{Q}{R}}\rho_e(t)
  e^{-iH_0s}e^{-i\dpr{Q}{R}}e^{iH_0s} \\ \nonumber
 &-\pth{ \tfrac12 \mp \tfrac12 + N_\mathbf{Q}}
  e^{\pm iE_\mathbf{Q} s}e^{-i\dpr{Q}{R}}e^{-iH_0s}e^{i\dpr{Q}{R}}e^{iH_0s}
  \rho_e(t)  \big\} \\ & +\mathrm{h.c.,} \label{eq:aftertrace}
\end{align}
where $E_\mathbf Q$ refers to the energy of a phonon with wave vector $\mathbf Q$ and we have defined the expectation value of the phonon number operator $\mathrm{Tr}_\mathrm{ph} \{\rho_\mathrm{ph} b_\mathbf{Q}^\dagger b_\mathbf{Q}\}=N_\mathbf{Q}$. Note that the upper (lower) sign refers to absorption (emission) terms. In the following, we will assume phonons are dispersionless, so $E_\mathbf Q=E_0$ and $N_\mathbf{Q}=N_0=(e^{E_0/k_BT}-1)^{-1}$, $T$ is the lattice temperature, and $k_B$ is the Boltzmann constant. In order to keep expressions more compact, we define the \emph{scattering weight}
\begin{align}
  \label{eq:w_def}
  \mathcal W^\pm(\mathbf Q)=\pth{ \tfrac12 \mp \tfrac12 + N_0 } |\mathcal M(\mathbf Q)|^2.
\end{align}
We can then write the dissipator due to a single dispersionless phonon type as (after reintroduction of $\hbar$),
\begin{align}
  \nonumber
  &\mathcal D = \frac{1}{\hbar^2(2\pi)^3} \sum_{\pm} \int_0^{\infty}ds \int d^3Q \big\{ \\ \nonumber
  +&\mathcal W^{\pm}(\mathbf Q)e^{\mp iE_0 s/\hbar}
  e^{i\dpr{Q}{R}}\rho_e(t)e^{-iH_0s/\hbar}e^{-i\dpr{Q}{R}}e^{iH_0s/\hbar} \\ \nonumber
 -&\mathcal W^{\pm}(\mathbf Q)e^{\pm iE_0 s/\hbar}e^{-i\dpr{Q}{R}}e^{-iH_0s/\hbar}e^{i\dpr{Q}{R}}e^{iH_0s/\hbar} \rho_e(t)  \big\} \\ \label{eq:aftertrace}
 +& \mathrm{h.c.,}
\end{align}
where the total dissipator will be obtained later by summing over the different relevant phonon branches. The dissipator can be written as $\mathcal D=\mathcal D^{\mathrm{in}}+\mathcal D^{\mathrm{out}}$, corresponding to the positive and negative terms in~\eqref{eq:aftertrace} respectively. We will derive the form of the matrix elements of the in-scattering term. The out-scattering term is less complicated and can be obtained by inspection of the in-scattering term and only the final results will be given.

By using the completeness relation four times, we can write
\begin{align}
  \nonumber
  \mathcal D^{\mathrm{in}}& =\frac{1}{\hbar^2(2\pi)^3}
  \int d^3Q\int_0^{\infty}ds \sum_{n_{1234},\pm}\int d^2k_{1234} \\
  \nonumber
  & \times \innp{n_1,\mathbf k_1}{e^{i\dpr{Q}{R}}}{n_2,\mathbf k_2}e^{\mp iE_0 s/\hbar}
  \mathcal W^{\pm}(\mathbf Q) \\
  \nonumber
  & \times \innp{n_3,\mathbf k_3}{e^{-iH_0s/\hbar}e^{-i\dpr{Q}{R}}e^{iH_0s/\hbar}}{n_4,\mathbf k_4} \\
  \label{eq:completeness}
  & \times \innp{n_2,\mathbf k_2}{\rho_e}{n_3,\mathbf k_3}
  \ket{n_1,\mathbf k_1}\bra{n_4,\mathbf k_4} + \mathrm{h.c.,}
\end{align}
where $\sum{n_{1234}}$ and $\int d^2 k_{1234}$ refer to summation over $n_1$ through $n_4$ and integration of $\mathbf k_1$ through $\mathbf k_4$ respectively. Integration of $\mathbf k_2$ through $\mathbf k_4$ and shifting the integration variable $\mathbf Q\rightarrow -\mathbf Q+\mathbf k_1$ gives
\begin{align}
  \nonumber
  \mathcal D^{\mathrm{in}} =&\frac{1}{\hbar^2(2\pi)^3}
  \int d^3Q\int_0^{\infty}ds \sum_{n_{1234},\pm}\int d^2k_{1} \\ \nonumber
  &\times \mathcal W^\pm(\mathbf Q-\mathbf k_1) \phpr{n_1}{n_2} \phpr{n_3}{n_4}^*
  \rho_{n_2,n_3}^{E_q} \\   \label{eq:afterk24}
  &\times e^{-i\frac{s}{\hbar}(\Delta_{n_3,n_4}-E_{k_1}\pm E_0+E_q ) }
  \ket{n_1,\mathbf k_1}\bra{n_4,\mathbf k_1} + \mathrm{h.c.,}
\end{align}
where $E_k$ is the in-plane dispersion relation, $q=\sqrt{Q_x^2+Q_y^2}$ is the magnitude of the in-plane component of $\ve Q$, $\Delta_{nm}=E_n-E_m$ is the energy spacing of the cross-plane eigenfunctions and
\begin{align}
  \label{eq:phonprod}
  \phpr{n}{m}= \sum_b \int_{-\infty}^{\infty} [ \psi^{(b)}_n(z) ]^* e^{iQ_zz}\psi^{(b)}_m(z) dz\ ,
\end{align}
where the sum is over the considered bands (conduction, light-hole and split-off bands in this work). In order to perform the $s$ integration, we use
\begin{align}
  \label{eq:principal}
  \int_0^\infty e^{-i\Delta s/\hbar}=\pi\hbar \delta(\Delta)-i\hbar \mathcal P\frac{1}{\Delta},
\end{align}
where $\mathcal P$ denotes the Cauchy principal value, which causes a small shift to energies called the \emph{Lamb shift}.~\cite{breuer_2002} By omitting the Lamb shift term and dropping the subscript on the $\mathbf k_1$ integration variable, we get
\begin{align}
  \nonumber
  \mathcal D^{\mathrm{in}} &=\frac{\pi}{\hbar(2\pi)^3}
  \sum_{n_{1234},\pm}\int d^2k \int d^3Q \mathcal W^\pm(\mathbf Q-\mathbf k) \\
  &\phpr{n_1}{n_2} \phpr{n_3}{n_4}^* \rho_{n_2,n_3}^{E_k+\Delta_{n_4n_3}\mp E_0} \\ \nonumber
  & \delta[E_q-(E_k+\Delta_{n_4n_3}\mp E_0)]
  \ket{n_1,\mathbf k}\bra{n_4,\mathbf k} + \mathrm{h.c..}
\end{align}
Multiplying by $\bra{N}$ from the left and $\ket{N+M}$ from the right and renaming the sum variables $n_2\rightarrow n$ and $n_3\rightarrow m$ gives the matrix elements of the in-scattering contribution of the dissipator, giving
\begin{align}
  \nonumber
  \left( \pa{f_{N,M}^{E_k}}{t} \right)^\mathrm{in} &= \frac{\pi}{\hbar(2\pi)^3}
  \sum_{n,m,\pm}\int d^3Q \mathcal W^\pm(\mathbf Q-\mathbf k) \\ \nonumber
  &\phpr{N}{n} \phpr{m}{N+M}^*
   \times \rho_{n,m}^{E_k+\Delta_{N+M,m}\mp E_0} \\ \label{eq:diss_mat}
  & \delta[E_q-(E_k+\Delta_{N+M,m}\mp E_0)]
  + \mathrm{h.c..}
\end{align}
The sums involving $n$ and $m$ are over all integers. However, it is convenient due to numerical reasons to shift the sum variables in such a way that terms with small $|n|$ and $|m|$ dominate, and terms with large $|n|$ and $|m|$ approach zero. This can be done by making the switch $n\rightarrow N+n$ and $m\rightarrow N+M+m$, giving
\begin{align}
  \nonumber
  &\left( \pa{f_{N,M}^{E_k}}{t} \right)^\mathrm{in} = \frac{\pi}{\hbar(2\pi)^3}
  \sum_{n,m,\pm}\int d^3Q \mathcal W^\pm(\mathbf Q-\mathbf k) \times \\ \nonumber
  &\phpr{N}{N+n} \phpr{N+M+m}{N+M}^* \times \\ \nonumber
  &\rho_{N+n,N+M+m}^{E_k+\Delta_{N+M,N+M+m}\mp E_0}
   \delta[E_q-(E_k+\Delta_{N+M,N+M+m}\mp E_0)] \\ \label{eq:diss_mat_01}
  &+ \mathrm{h.c..}
\end{align}
In order to write the RHS in Eq.~\eqref{eq:diss_mat_01} in terms of $f_{N,M}^{E_k}$, we use
\begin{align}
  \label{eq:switch}
  \rho^{E_k}_{N+n,N+M+m}=\rho^{E_k}_{N+n,N+n+(M+m-n)}=f_{N+n,M+m-n}^{E_k}
\end{align}
and get
\begin{align}
  \nonumber
  & \left( \pa{f_{N,M}^{E_k}}{t} \right)^\mathrm{in} = \frac{\pi}{\hbar(2\pi)^3}
  \sum_{n,m,\pm}\int d^3Q \mathcal W^\pm(\mathbf Q-\mathbf k) \\ \nonumber
  & \phpr{N}{N+n} \phpr{N+M+m}{N+M}^* \times \\ \nonumber
  & f_{N+n,M+m-n}^{E_k+\Delta_{N+M,N+M+m}\mp E_0} \delta[E_q-(E_k+\Delta_{N+M,N+M+m}\mp E_0)] \\
  \label{eq:diss_mat_02}
  &+ \mathrm{h.c..}
\end{align}
or
\begin{align}
  \nonumber
  \left( \pa{f_{N,M}^{E_k}}{t} \right)^\mathrm{in} &= \sum_{n,m,\pm}\Gamma_{NMnmE_k}^{\pm,\mathrm{in}}
   f_{N+n,M+m-n}^{E_k+\Delta_{N+M,N+M+m}\mp E_0} \\ \label{eq:gamma_in_first}
   &+ \mathrm{h.c.},
\end{align}
where we have defined the in-scattering rates
\begin{align}
  \nonumber
  \Gamma_{NMnmE_k}^{\pm,\mathrm{in}} & =
  \frac{\pi}{\hbar(2\pi)^3}
  \int d^3Q \mathcal W^\pm(\mathbf Q-\mathbf k)    \\ \nonumber
  & \phpr{N}{N+n} \phpr{N+M+m}{N+M}^* \times \\ \label{eq:gamma_in_def}
  & \delta[E_q-(E_k+\Delta_{N+M,N+M+m}\mp E_0)].
\end{align}
Note that the RHS in Eq.~\eqref{eq:gamma_in_def} looks like it depends on the direction of $\mathbf k$. However, the scattering weight only depends on the magnitude $|\mathbf Q-\mathbf k|$, so the coordinate system for the $\mathbf Q$ integration can always been chosen relative to $\mathbf k$, which means the RHS only depends on the magnitude $|\mathbf k|$.

In order to proceed, we need to specify the in-plane dispersion relation $E_k$, as well as the scattering mechanism, which determines the form of the scattering weight $\mathcal W^\pm(\mathbf Q-\mathbf k)$. In this work, we will assume a parabolic dispersion relation $E_k=\hbar^2k^2/2m_\parallel^*$, with a single in-plane effective mass $m_\parallel^*$, defined as the average effective mass of the $N_s$ considered eigenstates
\begin{align}
  \label{eq:mpar}
  \frac{1}{\mpar}=\frac{1}{N_s} \sum_{n=1}^{N_s}\sum_b \int \frac{|\psi_n^{(b)}(z)|^2}{m_b^*(z)}dz ,
\end{align}
where $m^*_b(z)$ is the position-dependent effective mass of band $b$, calculated using the procedure given in Appendix~\ref{app:matpar}. The approximation of a parabolic in-plane dispersion relation can fail in short-wavelength QCLs, where electrons can have in-plane energies comparable to the material bandgap ~\cite{matyas_JAP_2011}. It is possible to go beyond a parabolic dispersion relation by including a term proportional to $k^4$ (in-plane nonparabolicity). However, the inclusion of a $k^4$ term would significantly complicate the integration over $q$ in Eq.~\eqref{eq:gamma_in_def} and is beyond the scope of this work.

Since we do not have an analytical expression for the eigenfunctions, the products $\phpr{n}{m}$ have to be calculated numerically. For a more compact notation, we define
\begin{align}
  \label{eq:Idef_ph}
    I_{n,m,k,\ell}(Q_z)= \phpr{n}{m} \phpr{k}{\ell}^* \ .
\end{align}
Switching to cylindrical coordinates $d^3Q\rightarrow q dq d\theta dQ_z$, making the change of variables $E_q=\hbar^2q^2/2m_\parallel^*$ and integrating over $E_q$ gives
\begin{align}
  \nonumber
  &\Gamma_{NMnmE_k}^{\pm,\mathrm{in}} = \frac{\mpar}{2\pi\hbar^3}u[E_k+\Delta_{N+M,N+M+m}\mp E_s] \\
  \label{eq:after_Eq}
  & \times \int_{0}^{\infty} dQ_z
  I_{N,N+n,N+M+m,N+M}(Q_z) \\ \nonumber
  & \times \left[\frac{1}{2\pi} \int_0^{2\pi} d\theta
  \mathcal W^\pm(\mathbf Q-\mathbf k) \right]_{E_q=E_k+\Delta_{N+M,N+M+m}\mp E_0} ,
\end{align}
where the term in the square brackets is the scattering weight averaged over the angle between $\mathbf Q$ and $\mathbf k$, with the in-plane part of $\mathbf Q$ constrained by energy conservation, and u[x] is the Heaviside function, which is zero when no positive value of $E_q$ can satisfy energy conservation. Note that we have also taken advantage of the fact that the real part of the integrand is even in $Q_z$, allowing us to limit the range of integration to positive $Q_z$. The resulting expressions from the angle-averaged scattering weight can be quite cumbersome, so we define the angle-averaged scattering weight
\begin{align}
  \label{eq:Wshort}
  \widetilde{\mathcal W}^\pm(E_k,Q_z,E_q)=\frac{1}{2\pi} \int_0^{2\pi} d\theta
  \mathcal W^\pm(\mathbf Q-\mathbf k),
\end{align}
simplifying the in-scattering rate to
\begin{align}
  \nonumber
  \Gamma_{NMnmE_k}^{\pm,\mathrm{in}} &= \frac{\mpar}{2\pi\hbar^3}u[E_k+\Delta_{N+M,N+M+m}\mp E_0] \\
  \nonumber
  & \times \int_{0}^{\infty} dQ_z
  I_{N,N+n,N+M+m,N+M}(Q_z) \\   \label{eq:after_Eq}
  & \times \widetilde{\mathcal W}^\pm(E_k,Q_z,E_k+\Delta_{N+M,N+M+m}\mp E_0) .
\end{align}
In order to proceed we need to specify the scattering weight, which depends on the kind of phonon interaction being considered. Calculations of scattering rates for longitudinal acoustic phonons and polar optical phonons is performed in appendix~\ref{app:rates}.

\section{Dissipator for Elastic Scattering Mechanisms}
\label{sec:elastic}

In addition to electron--phonon scattering, there are numerous elastic scattering mechanisms that are relevant to QCLs. The most important ones are interface roughness, ionized impurities, and alloy scattering ~\cite{lee_PRB_2002,wacker_JSTQE_2013,jirauschek_APR_2014}. The derivation of the corresponding scattering rates is slightly different from the case for phonons and will be demonstrated for the case of interface roughness, which is a very important elastic scattering mechanism for both THz and mid-IR QCLs~\cite{jirauschek_JAP_2009,kubis_PSSC_2008,chiu_APL_2012} and has been found to decrease the upper lasing state lifetime by a factor of $2$ at room temperature in mid-IR QCLs~\cite{chiu_APL_2012}. For other elastic scattering mechanisms, we only briefly discuss the interaction potential and give final expressions for rates in Appendix~\ref{app:rates}.

\subsection{Interface Roughness}
\label{sec:roughness}
For an electron in band b, deviation from a perfect interface between two different semiconductor material can be modeled using an interaction Hamiltonian on the form~\cite{ferry2013}
\begin{align}
  \label{eq:Hrough}
  H_\mathrm{IR}^{(b)}= \sum_i \Delta_i(\ve r) \left. \pa{V^{(b)}}{z}\right|_{z=z_i}
   \simeq \sum_i \Delta V_i^{(b)} \delta(z-z_i) \Delta_i(\ve r) \ ,
\end{align}
where $i$ labels an interface located at position $z_i$, $V^{(b)}(z)$ the band edge corresponding to band b and $\Delta V_i^{(b)}$ is the band discontinuity at interface $i$. The function $\Delta_i(\ve r)$ represents the deviation from a perfect interface at different in-plane positions $\ve r$. For the conduction band of an InGaAs/InAlAs heterostructure, the sign of $\Delta V_i^{(b)}$ is positive when going from well material to barrier material (from the left) and negative when going from barrier to well material. For the light-hole and split-off bands, the signs are opposite (with respect to the conduction band).

The introduction of the interaction Hamiltonian in Eq.~\eqref{eq:Hrough} breaks translational invariance in the $xy$-plane, and the resulting density matrix will no longer be diagonal in the in-plane wave vector. In order to get around this limitation, we follow the treatment in Refs.~\cite{roblin_JAP_1996,lee_PRB_2002,jirauschek_APR_2014} and only consider statistical properties of the interface roughness, which is contained in the correlation function
\begin{align}
  \label{eq:IR_corr}
  \left<  \Delta_i(\ve r) \Delta_j(\ve r')\right>=\delta_{ij} C(|\ve r - \ve r'|)
\end{align}
where $\left< ... \right>$ denotes an average over the macroscopic interface area, $C(|\ve r|)$ is the spatial autocorrelation function and the Kronecker delta function means that we neglect correlations between different interfaces. In this work, we employ a Gaussian correlation function $C(|\ve r|)=\Delta^2e^{-|\ve r|^2/\Lambda^2}$, characterized by standard deviation $\Delta$ and correlation length $\Lambda$. However, calculations will be taken as far as possible without specifying a correlation function. When calculating rates, we will mainly use its Fourier transform,
\begin{align}
  \label{eq:ffour}
  C(|\ve q|)= \int d^2 C(|\ve r|)e^{-i\dpr{r}{q}}.
\end{align}

The derivation for the interface-roughness scattering rates proceeds the same way as for electron--phonon scattering until Eq.~\eqref{eq:diss_01}. For the case of interface roughness, there is no phonon degree of freedom to trace over and we get
\begin{align}
  \nonumber
  \mathcal D_\mathrm{IR}&=\int_0^{\infty} ds \{
   +H_\mathrm{IR}\rho_e(t) e^{-iH_0 s}H_\mathrm{IR} e^{iH_0 s}  \\
    \label{eq:diss_elastic}
  &-H_\mathrm{IR} e^{-iH_0 s} H_\mathrm{IR}e^{iH_0s}\rho_e(t) \}
  + \mathrm{h.c..}
\end{align}
After using the completeness relation 4 times and performing the $s$ integration, the in-scattering term in Eq.~\eqref{eq:diss_elastic} becomes
\begin{align}
\begin{split}
  \mathcal D_\mathrm{IR}^\mathrm{in}&=\frac{\pi}{\hbar}\sum_{n_{1234}} \int d^2 k_{1234}
  \delta[ E(n_4,\ve k_4)-E(n_3,\ve k_3)] \\
  &\innp{n_2,\ve k_2}{\rho_e}{n_3,\ve k_3} \times  \innp{n_1,\ve k_1}{H_\mathrm{IR}}{n_2,\ve k_2} \\
  &\innp{n_3,\ve k_3}{H_\mathrm{IR}}{n_4,\ve k_4}
  \ket{n_1,\ve k_1} \bra{n_4,\ve k_4} + \mathrm{h.c.}.
\end{split}
\end{align}
Integrating over $\ve k_3$ gives
\begin{align}
  \nonumber
  \mathcal D_\mathrm{IR}^\mathrm{in}&=\frac{\pi}{\hbar}\sum_{n_{1234}} \int d^2 k_{124}
  \delta[ E(n_4,\ve k_4)-E(n_3,\ve k_2)]
  \rho_{n_2,n_3}^{E_{k_2}} \\ \label{eq:beforeint}
  &\times  \innp{n_1,\ve k_1}{H_\mathrm{IR}}{n_2,\ve k_2}
  \innp{n_3,\ve k_2}{H_\mathrm{IR}}{n_4,\ve k_4} \\ \nonumber
  &\times \ket{n_1,\ve k_1} \bra{n_4,\ve k_4} + \mathrm{h.c.}
\end{align}
Now let us focus on the part of the integrand containing the interaction Hamiltonian
\begin{align}
  \nonumber
  &\innp{n_1,\ve k_1}{H_\mathrm{IR}}{n_2,\ve k_2} \innp{n_3,\ve k_2}{H_\mathrm{IR}}{n_4,\ve k_4}= \\
    \label{eq:integrand_IR}
  & \sum_{i,j} \irpr{n_1}{n_2}{z_i}\irpr{n_3}{n_4}{z_j}^* \times \\
  & \int d^2 r d^2 r' \braket{\ve r}{\ve k_1}^*\braket{\ve r}{\ve k_2} \nonumber
    \braket{\ve r'}{\ve k_2}^*\braket{\ve r'}{\ve k_4} \Delta_i(\ve r)\Delta_j(\ve r'),
\end{align}
where we have defined
\begin{align}
  \label{eq:irprod}
  \irpr{n}{m}{z_i}= \sum_b \Delta V_i^{(b)} \psi_n^*(z_i)\psi_m(z_i),
\end{align}
where the sum is over the conduction (c), light hole (lh), and spin-orbit split-off bands (so). We use a convention where the band discontinuity $\Delta V_i^{(b)}$ is positive for the conduction band. In order to proceed, we replace the product $\Delta_i(\ve r)\Delta_j(\ve r')$ with  the Fourier decomposition of its spatial average using Eqs.~\eqref{eq:IR_corr}-\eqref{eq:ffour} and get
\begin{align}
  \nonumber
  &\innp{n_1,\ve k_1}{H_\mathrm{IR}}{n_2,\ve k_2} \innp{n_3,\ve k_2}{H_\mathrm{IR}}{n_4,\ve k_4}= \\
    \label{eq:integrand_IR2}
   & \sum_{i} \irpr{n_1}{n_2}{z_i}\irpr{n_3}{n_4}{z_i}^*
   \frac{1}{(2\pi)^2} \\ \nonumber
  &\int d^2q \delta[\ve k_2-(\ve k_1-\ve q)]\delta[\ve k_4 -(\ve k_2+\ve q)] C(|\ve q|) \ .
\end{align}
Inserting Eq.~\eqref{eq:integrand_IR2} into Eq.~\eqref{eq:beforeint} and integrating over $\ve k_4$ and $\ve k_2$, dropping the subscript index of $\ve k_1$ and shifting the integration variable $\ve q\rightarrow -\ve q+ \ve k$ gives
\begin{align}
  \nonumber
  \mathcal D_\mathrm{IR}^\mathrm{in}&=\frac{\pi}{\hbar(2\pi)^2}
  \sum_{n_{1234}} \int d^2k \int d^2q
  \delta[E_q-\Delta_{n_4,n_3}-E_{k}] \\   \label{eq:after_avg}
  &\sum_i \irpr{n_1}{n_2}{z_i}\irpr{n_3}{n_4}{z_i}^*
  C(|\ve q-\ve k|) \\ \nonumber
  &\rho_{n_2,n_3}^{E_q} \ket{n_1,\ve k}\bra{n_4,\ve k_1}
  + \mathrm{h.c..}
\end{align}
Switching to polar coordinates $d^2q\rightarrow q dqd\theta$ and performing the change of variables $E_q=\hbar^2q^2/(2\mpar)$ gives
\begin{align}
  \nonumber
  \mathcal D_\mathrm{IR}^\mathrm{in}&=\frac{\mpar}{2\hbar^3} \sum_{n_{1234}} \int d^2k
  \sum_i \irpr{n_1}{n_2}{z_i}\irpr{n_3}{n_4}{z_i}^* \\ \nonumber
  &\times \frac{1}{2\pi}\int_0^{2\pi} \int_0^{\infty} dE_q
  \delta[E_q-\Delta_{n_4,n_3}-E_{k}] \\   \label{eq:after_avg2}
  &\times C(|\ve q-\ve k|) \rho_{n_2,n_3}^{E_q} \ket{n_1,\ve k}\bra{n_4,\ve k_1}
  + \mathrm{h.c..}
\end{align}
Performing the $E_q$ integration, renaming the dummy variables $n_3\rightarrow m$ and $n_2\rightarrow n$ and multiplying from the left by $\bra{N}$ and from the right with $\ket{N+M}$ gives
\begin{align}
\label{eq:EOM_IR}
\begin{split}
  &\pth{\pa{f_{N,M}^{E_k}}{t}}^\mathrm{in}_\mathrm{IR}
  =\frac{\mpar}{2\hbar^3}
  \sum_{n,m} u[\Delta_{N+M,m}+E_k] \\
  &\times \sum_i \irpr{N}{n}{z_i}\irpr{m}{N+M}{z_i}^* \\
  &\times \tilde C(E_k,\Delta_{N+M,m}+E_k) \rho_{nm}^{\Delta_{N+M,m}+E_k} + \mathrm{h.c..}
\end{split}
\end{align}
where we have defined the angle-averaged correlation function
\begin{align}
  \label{eq:angleav_corr}
  &\tilde C(E_k,E_q)=\\ \nonumber
  & \frac{1}{2\pi}\int_0^{2\pi}d \theta
  C\pth{ \sqrt{\frac{2\mpar}{\hbar^2}} \sqrt{E_k+E_q-2\sqrt{E_kE_q}\cos(\theta)}  } .
\end{align}
Shifting the sum variables $n\rightarrow N+n$, $m\rightarrow N+M+m$ and using Eq.~\eqref{eq:switch}, we can write
\begin{align}
  \nonumber
  &\pth{\pa{f_{N,M}^{E_k}}{t}}^\mathrm{in}_\mathrm{IR}
  =\frac{\mpar}{2\hbar^3} \sum_{n,m} u[\Delta_{N+M,N+M+m}+E_k] \\ \nonumber
  &\times\sum_i \irpr{N}{N+n}{z_i}\irpr{N+M+m}{N+M}{z_i}^* \\ \nonumber
  &\times \tilde C(E_k,\Delta_{N+M,N+M+m}+E_k)
  f_{N+n,M+m-n}^{\Delta_{N+M,N+M+m}+E_k} + \mathrm{h.c.} \\ \label{eq:EOM_IR}
  &=2 \sum_{n,m} \Gamma^{\mathrm{in},\mathrm{IR},\pm}_{NMnmE_k} f_{N+n,M+m-n}^{\Delta_{N+M,N+M+m}+E_k} + \mathrm{h.c.},
\end{align}
where we have defined $\Gamma^{\mathrm{in},\mathrm{IR},\pm}_{NMnmE_k}$, the in-scattering rate due to interface roughness. The purpose of the factor of two in the definition of the rate is so we can write
\begin{align}
  \pth{\pa{f_{N,M}^{E_k}}{t}}^\mathrm{in}_\mathrm{IR}
  &=\sum_{n,m,\pm} \Gamma^{\mathrm{in},\mathrm{IR},\pm}_{NMnmE_k}
  f_{N+n,M+m-n}^{\Delta_{N+M,N+M+m}+E_k} + \mathrm{h.c.}.
\end{align}
In other words, we have split the interface roughness scattering into two identical absorption and emission terms to be consistent with the notation for phonon scattering in Section~\ref{sec:phonon}. In order to calculate the rates, we need to choose a correlation function. The rates are calculated in Appendix~\ref{app:rates} for a Gaussian correlation function, which is often used to model QCLs ~\cite{lindskog_APL_2014,jirauschek_APR_2014}. The reason we choose a Gaussian correlation function is that it results in a simple analytical expression for the angle-averaged correlation function $\tilde C$, given by Eq.~\eqref{eq:angleav_corr}. Another popular choice is an exponential correlation function~\cite{lee_PRB_2002,wacker_JSTQE_2013,jirauschek_APR_2014}. However, angle-averaged expressions for the correlation function are more complicated, involving elliptical integrals~\cite{lee_PRB_2002,jirauschek_APR_2014}.

\subsection{Ionized Impurities}
The effects of the long-range Coulomb interaction from ionized impurities is already included to lowest order with a mean-field treatment (Poisson's equation). The goal of this section is to include higher order effects by including scattering of electrons with ionized impurities. We assume a 3D doping density of the form
\begin{align}
  \label{eq:deltadope}
  N(z)=N_\mathrm{2D}\delta(z-z_\ell) \ ,
\end{align}
which is a delta-doped sheet at cross-plane position $z_\ell$. The advantage of Eq.~\eqref{eq:deltadope} is that an arbitrary doping density can be approximated by a linear combination of closely spaced delta-doped sheets. Typically, in QCLs, only one~\cite{dupont_JAP_2012} or a few~\cite{evans_APL_2007,bismuto_APL_2010} layers are doped and the doping is kept away from the active region to minimize nonradiative transitions from the upper to lower lasing level. Note that, in this context, a layer refers to a barrier or a well, not atomic layers. The corresponding scattering rates due to the ionized-impurity scattering distribution in Eq.~\eqref{eq:deltadope} is calculated in Appendix~\ref{subsec:ii}.

\subsection{Random Alloy Scattering}
In ternary semiconductor alloys $A_xB_{1-x}C$, the energy of band $b$ can be calculated using the virtual crystal approximation,~\cite{ferry2013} where the alloy band energy is obtained by averaging over the band profiles of the two materials $AC$ and $BC$
\begin{align}
\label{eq:virtual}
\begin{split}
  V_{b,\mathrm{VCA}}(\ve R) =
  \sum_n \big[ & x V_{b,\mathrm{AC}}(\ve R-\ve R_n) \\
               &+ (1-x)V_{b,\mathrm{BC}}(\ve R-\ve R_n) \big],
\end{split}
\end{align}
where the sum is over all unit cells $n$ and $V_{b,\mathrm{AC}}(\ve R-\ve R_n)$ and $V_{b,\mathrm{BC}}(\ve R-\ve R_n)$ are the corresponding bands of the AC and BC material respectively. The difference between an alloy and a binary compound is that, in alloys, periodicity of the crystal is broken by the random distribution of $AC$ and $BC$ unit cells. This deviation from a perfect lattice can be modeled as a scattering potential, which can be considered much weaker than Eq.~\eqref{eq:virtual}. The scattering potential is given by the difference between the real band energy and the band energy obtained using the virtual crystal approximation. We will follow the treatment of alloy scattering in Refs.~\cite{ando_JPSJ_1982,wacker_JSTQE_2013,jirauschek_APR_2014} and write the scattering potential (which we assume is the same for all bands) as
\begin{align}
  \label{eq:alloydef}
  V_\mathrm{alloy}(\ve R)= \sum_n C_n^x\delta(\ve R-\ve R_n),
\end{align}
where the sum is over all unit cells and $C_n^x$ is a random variable, analogous to the interface-roughness correlation function in Eq.~\eqref{eq:IR_corr}, with
\begin{align}
\begin{aligned}
  \label{eq:corr_all}
  \left< C_n^x \right>&=0 \\
  \left< C_n^x C_m^x \right> &= \delta_{nm} x_n(1-x_n) (V_n \Omega)^2,
\end{aligned}
\end{align}
where $\Omega=a^3/4$ is equal to one-fourth of the volume of a unit cell and $V_n$ the alloy scattering potential in unit cell $n$, which can be approximated as the conduction band offset of the two binary materials, corresponding to $x=0$ and $x=1$. The presence of the Kronecker delta function $\delta_{nm}$ means that the location of A and B atoms are assumed to be completely uncorrelated. The derivation of corresponding scattering rates proceeds in the same way as for interface roughness scattering and results are given in Appendix~\ref{app:rates}. The final expressions for the rates are considerably simpler than the ones due to interface roughness due to the local nature of the interaction potential in Eq.~\eqref{eq:alloydef}.


\section{Is the Dissipator Really of the Lindblad Form?}
\label{sec:lindblad}

For each of the scattering mechanisms considered in this work, it can be shown that the corresponding dissipator is of the Lindblad form \cite{Lindblad1976}. In fact, each dissipator was derived according to a procedure that guarantees a Lindbladian dissipator \cite{breuer_2002,KarimiDissertation2017,Karimi_PRB_2016,jonasson_JCEL_2016}. However, one can readily convince oneself of this fact. In what follows, we prove the Lindblad form for the electron--phonon interaction dissipator as an example.

According to Eq.~(\ref{eq:aftertrace}) of the manuscript, we can write the dissipator due to a single dispersionless phonon type as
\begin{align}
  \nonumber
  &\mathcal D = \frac{1}{\hbar^2(2\pi)^3} \sum_{\pm} \int_0^{\infty}ds \int d^3Q \big\{ \\ \nonumber
  +&\mathcal W^{\pm}(\mathbf Q)e^{\mp iE_0 s/\hbar}
  e^{i\dpr{Q}{R}}\rho_e(t)e^{-iH_0s/\hbar}e^{-i\dpr{Q}{R}}e^{iH_0s/\hbar} \\ \nonumber
 -&\mathcal W^{\pm}(\mathbf Q)e^{\pm iE_0 s/\hbar}e^{-i\dpr{Q}{R}}e^{-iH_0s/\hbar}e^{i\dpr{Q}{R}}e^{iH_0s/\hbar} \rho_e(t)  \big\} \\
 +& \mathrm{h.c.,}
\end{align}
The dissipator can be written as a summation of in- and out-scattering terms ($\mathcal D=\mathcal D^{\mathrm{in}}+\mathcal D^{\mathrm{out}}$), corresponding to the positive and negative terms in Eq.~(\ref{eq:aftertrace}), respectively. By using the completeness relation four times, we can write
\begin{align}
  \nonumber
  \mathcal D^{\mathrm{in}}& =\frac{1}{\hbar^2(2\pi)^3}
  \int d^3Q\int_0^{\infty}ds \sum_{n_{1234},\pm}\int d^2k_{1234} \\
  \nonumber
  & \times \innp{n_1,\mathbf k_1}{e^{i\dpr{Q}{R}}}{n_2,\mathbf k_2}e^{\mp iE_0 s/\hbar}
  \mathcal W^{\pm}(\mathbf Q) \\
  \nonumber
  & \times \innp{n_2,\mathbf k_2}{\rho_e}{n_3,\mathbf k_3}\\
  & \times \innp{n_3,\mathbf k_3}{e^{-iH_0s/\hbar}e^{-i\dpr{Q}{R}}e^{iH_0s/\hbar}}{n_4,\mathbf k_4}
  \label{Eq2}
  \ket{n_1,\mathbf k_1}\bra{n_4,\mathbf k_4} + \mathrm{h.c.},
\end{align}
where $\sum{n_{1234}}$ and $\int d^2 k_{1234}$ refer to summation over $n_1$ through $n_4$ and integration of $\mathbf k_1$ through $\mathbf k_4$, respectively.
Since, $\rho$ and $\mathcal{D}$ are diagonal in $\mathbf{k}$, this can be further simplified to

\begin{align}
  \nonumber
  \mathcal D^{\mathrm{in}}& =\frac{1}{\hbar^2(2\pi)^3}
  \int d^3Q\int_0^{\infty}ds \sum_{n_{1234},\pm}\int d^2k_{1,2} \\
  \nonumber
  & \times \innp{n_1,\mathbf k_1}{e^{i\dpr{Q}{R}}}{n_2,\mathbf k_2}e^{-i(\Delta_{n3,n4}\pm E_0 +E_{\mathbf k_2} - E_{\mathbf k_1}) s/\hbar}
  \mathcal W^{\pm}(\mathbf Q) \\
  \nonumber
  & \times \innp{n_2,\mathbf k_2}{\rho_e}{n_3,\mathbf k_2}\\
  & \times \innp{n_3,\mathbf k_2}{e^{-i\dpr{Q}{R}}}{n_4,\mathbf k_1}
  \label{Eq3}
  \ket{n_1,\mathbf k_1}\bra{n_4,\mathbf k_1} + \mathrm{h.c.}.
\end{align}

\noindent By defining
\begin{align}
\mathcal{A}_Q = \innp{n_1,\mathbf k_1}{e^{+i\dpr{Q}{R}}}{n_2,\mathbf k_2}  \ket{n_1,\mathbf k_1}\bra{n_2,\mathbf k_2}\, ,\label{Eq4}\\
\Gamma_{n_{1} n_{2}}^{\mathbf{k}_{1}\mathbf{k}_2}= \frac{\mathcal{W}^{\pm}(\mathbf Q)}{\hbar^2(2\pi)^3}\int_0^{\infty}ds  e^{-i(\Delta_{n1,n2}\pm E_0 +E_{\mathbf k_2} - E_{\mathbf k_1}) s/\hbar}\, ,\label{Eq5}
\end{align}

\noindent Equation (\ref{Eq3}) can be written compactly as

\begin{align}
\mathcal{D}^{\mathrm{in}} = 2\sum_{n_{1,2,3,4}}\int d^3Q\int d^2\mathbf{k}_{1,2} \;\Gamma \mathcal{A}\rho\mathcal{A}^{\dagger}.
\end{align}
Note that $\Gamma$, as defined in Eq.~(\ref{Eq5}), is Hermitian (transposition of the top two or bottom two indices results in the complex conjugate of the original). Similarly, the out-scattering terms can be written as
\begin{align}
  \nonumber
  \mathcal D^{\mathrm{out}}& =-\frac{1}{\hbar^2(2\pi)^3}
  \int d^3Q\int_0^{\infty}ds \sum_{n_{1234},\pm}\int d^2k_{1234} \\
  \nonumber
  & \times \innp{n_1,\mathbf k_1}{e^{-i\dpr{Q}{R}}}{n_2,\mathbf k_2}e^{ i(\Delta_{n4,n3}\pm E_0+ E_{\mathbf{k}_1} - E_{\mathbf{k}_2}) s/\hbar}
  \mathcal W^{\pm}(\mathbf Q) \\
  \nonumber
  & \times \innp{n_2,\mathbf k_2}{e^{+i\dpr{Q}{R}}}{n_3,\mathbf k_1}\\
  & \times \innp{n_3,\mathbf k_2}{\rho_e}{n_4,\mathbf k_2}\
  \label{Eq7}
  \ket{n_1,\mathbf k_1}\bra{n_4,\mathbf k_1} + \mathrm{h.c.}\, .
\end{align}

\noindent Using Eqs.~(\ref{Eq4}) and (\ref{Eq5}), $\mathcal{D}^{out}$ is written as
\begin{align}
\mathcal{D}^{\mathrm{out}} = -\sum_{n_{1,2,3,4}}\int d^3Q\int d^2\mathbf{k}_{1,2} \;\Gamma\left(\mathcal{A}^{\dagger}\mathcal{A}\rho + \rho\mathcal{A}^{\dagger}\mathcal{A}\right)], .
\end{align}
\noindent Summing in- and out-scattering terms would give us the total dissipator for electron-phonon interaction of the form

\begin{align}
\mathcal{D} = \sum_{n_{1,2,3,4}}\int d^3Q\int d^2\mathbf{k}_{1,2} \;\Gamma\left(2\mathcal{A}\rho\mathcal{A}^{\dagger} - \left\{\mathcal{A}^{\dagger}\mathcal{A},\rho\right\}\right),
\end{align}
which is of the Lindblad form.


\section{Calculation of Scattering Rates}
\label{app:rates}
\subsection{Longitudinal Acoustic Phonons}
We will start with longitudinal acoustic (LA) phonons, which have the following scattering weight for absorption~\cite{ferry2013}
\begin{align}
  \label{eq:WLA}
  \mathcal W^+_\mathrm{LA}(\mathbf Q)=N_Q \frac{\hbar^2 D_\mathrm{ac}^2Q^2}{2\rho E_Q}
  = \left( e^{E_Q/kT}-1 \right)^{-1} \frac{D_\mathrm{ac}^2E_Q}{2\rho v_s^2},
\end{align}
where we have used the LA phonon dispersion relation $E_Q=\hbar v_s Q$, where $v_s$ is the speed of sound in the medium, $\rho$ the mass density and $D_\mathrm{ac}$ is the acoustic deformation potential. We cannot directly use the expression in Eq.~\eqref{eq:WLA} because the phonon energy is not constant, which was an approximation we used to derive Eq.~\eqref{eq:after_Eq}. It is possible to discretize the relevant range of $Q$ and treat LA phonons with different magnitude $Q$ as independent scattering mechanism. However, we would need to calculate and store the corresponding rates for each value of $Q$, which is prohibitive in terms of CPU time as well as memory use. Another option is to consider the limit $E_Q\ll kT$ and treat the LA phonons as an elastic scattering mechanism. However, because the LA phonons can exchange an arbitrarily small energy with the lattice, they play the important role of low energy thermalization at energies much smaller than the polar optical phonon. A second reason to include a nonzero energy for the LA phonons is that they tend to smooth the in-plane energy distribution of the density matrix, greatly facilitating numerical convergence. For this reason we follow the treatment in Ref.~\cite{lee_PRB_2002} and approximate acoustic phonons as having a single energy $E_\mathrm{LA}\ll kT$. Typically we will assume the LA phonon energy is equal to the energy spacing in the simulation ($0.5$-$2$~meV). Using this approximation, we get
\begin{align}
  \label{eq:WLA_final}
  \mathcal W^\pm_\mathrm{LA}(\mathbf Q)=\left( \tfrac12\mp \tfrac12 + N_\mathrm{LA} \right )
  \frac{D_\mathrm{ac}^2E_\mathrm{LA}}{2\rho v_s^2},
\end{align}
with $N_\mathrm{LA}=( \operatorname{exp}(E_\mathrm{LA}/kT)-1 )^{-1}$. The expression in Eq.~\eqref{eq:WLA_final} is isotropic (does not depend on $\mathbf Q$), so the angular integration in Eq.~\eqref{eq:after_Eq} is trivial and we get
\begin{align}
\label{eq:LA_in}
\begin{split}
  \Gamma_{NMnmE_k}^{\pm,\mathrm{in},\mathrm{LA}} &=
  \left(\tfrac12\mp \tfrac12 +N_\mathrm{LA} \right)
  \frac{D_\mathrm{ac}^2E_\mathrm{LA}\mpar}{4\pi \hbar^3\rho v_s^2} \\
  &u[E_k+\Delta_{N+M,N+M+m}\mp E_\mathrm{LA}] \\
  &\int_{0}^{\infty} dQ_z I_{N,N+n,N+M+m,N+M}(Q_z) .
\end{split}
\end{align}
The corresponding out scattering rates are
\begin{align}
\label{eq:LA_out}
\begin{split}
  &\Gamma_{NMnmE_k}^{\pm,\mathrm{out},\mathrm{LA}} =
  \left(\tfrac12\mp \tfrac12 +N_\mathrm{LA} \right)
  \frac{D_\mathrm{ac}^2E_\mathrm{LA}\mpar}{4\pi \hbar^3\rho v_s^2} \\
  & \times u[E_k+\Delta_{N+n,N+M+m}\pm E_\mathrm{LA}] \\
  & \times \int_{0}^{\infty} dQ_z
  I_{N+M,N+M+m,N+M+m,N+n}(Q_z) .
\end{split}
\end{align}

\subsection{Polar Longitudinal-Optical Phonons}
The scattering weight for polar longitudinal-optical (LO) phonons is~\cite{ferry2013}
\begin{align}
  \label{eq:W_pop}
  \mathcal W^\pm_\mathrm{LO}(\mathbf Q)= \pth{\tfrac12\mp \tfrac12 + N_\mathrm{LO}} \beta_\mathrm{LO}
  \frac{Q^2}{(Q^2+Q_D^2)^2} \ ,
\end{align}
with
\begin{align}
  \label{eq:beta_po}
  \beta_\mathrm{LO}= \frac{e^2 E_\mathrm{LO}}{2\eps_0}
  \pth{\frac{1}{\eps_s^\infty}-\frac{1}{\eps_s^0}},
\end{align}
where $\eps_0$ is the permittivity of vacuum, $\eps_s^\infty$ and $\eps_s^0$ are the high-frequency and low-frequency relative bulk permittivities respectively, $E_\mathrm{LO}$ is the polar-optical phonon energy and $Q_D$ is the Debye screening wave vector defined by $Q_D^2=n_{3D}e^2/(\eps_s^\infty k_BT)$, with $n_{3D}$ the average 3D electron density in the device. We note that QCLs are typicially low-doped, so the effect of screening is minimal. However, the screening facilitates numerical calculations by avoiding the singularity when $|\mathbf Q|=0$ in Eq.~\eqref{eq:W_pop}. Integration over the angle between $\mathbf Q$ and $\mathbf k$ gives
\begin{align}
\label{eq:polar_theta}
\begin{split}
  & \frac{\widetilde{\mathcal W}^\pm_\mathrm{LO}(E_k,Q_z,E_q) }
         {\pth{\tfrac12 \mp \tfrac12+N_\mathrm{LO} } \beta_\mathrm{LO}}
   = \frac{1}{2\pi}\int_0^{2\pi} \frac{|\ve Q-\ve k|^2}{( |\ve Q-\ve k|^2+Q_D^2)^2} \\
  &= \frac{1}{2\pi}\int_0^{2\pi} d\theta
  \frac{Q_z^2+q^2+k^2-2qk\cos(\theta)}{( q_z^2+q^2+k^2-2kq\cos(\theta) +Q_D^2 )^2} \\
  &= \frac{1}{Q_z^2+q^2+k^2} \\
  &\times \left[  1+\frac{Q_D^2}{Q_z^2+q^2+k^2} - \frac{4k^2q^2}{(Q_z^2+q^2+k^2)^2 } \right] \\
  & \times \left[ \left(1+\frac{Q_D^2}{Q_z^2+q^2+k^2}\right)^2 -\frac{4k^2q^2}{(Q_z^2+q^2+k^2)^2}
  \right]^{-\frac32},
\end{split}
\end{align}
which can be written as
\begin{align}
\label{eq:angle_pop}
\begin{split}
  &\widetilde{\mathcal W}^\pm_\mathrm{LO}(E_k,Q_z,E_q)
  = \frac{\hbar^2}{2m^*}\frac{\pth{\tfrac12 \mp \tfrac12+N_\mathrm{LO} }
  \beta_\mathrm{LO}}{E_z+E_k+E_q} \\
  &\times \left[  1+\frac{E_D}{E_z+E_q+E_k} - \frac{4E_kE_q}{(E_z+E_q+E_k)^2 } \right] \\
  &\times \left[ \left(1+\frac{E_D}{E_z+E_q+E_k}\right)^2 -\frac{4E_kE_q}{(E_z+E_q+E_k)^2}
  \right]^{-\frac32} \ ,
\end{split}
\end{align}
with $E_z=\hbar^2Q_z^2/(2\mpar)$. Note that this definition of $E_z$ is only for convenience, it has nothing to do with energy quantization in the $z$-direction. The in- and out-scattering rates can now be written as
\begin{align}
  \nonumber
  \Gamma_{NMnmE_k}^{\pm,\mathrm{in},\mathrm{LO}} &=
  \frac{\mpar}{2\pi\hbar^3}u[E_k+\Delta_{N+M,N+M+m}\mp E_\mathrm{LO}] \\
  \nonumber
  & \times \int_{0}^{\infty} dQ_z
  I_{N,N+n,N+M+m,N+M}(Q_z) \\   \label{eq:gamma_pop_in}
  & \times \widetilde{\mathcal W}^\pm_\mathrm{LO}(E_k,Q_z,E_k+\Delta_{N+M,N+M+m}\mp E_\mathrm{LO})
  \\ \nonumber
  \Gamma_{NMnmE_k}^{\pm,\mathrm{out},\mathrm{LO}}
  &= \frac{\mpar}{2\pi\hbar^3} u[E_k+\Delta_{N+n,N+M+m}\pm E_\mathrm{LO}] \\
  \nonumber
  & \times \int_{0}^{\infty} dQ_z
  I_{N+M,N+M+m,N+M+m,N+n}(Q_z) \\   \label{eq:gamma_pop_out}
  & \times \widetilde{\mathcal W}^\pm_\mathrm{LO}(E_k,Q_z,E_k+\Delta_{N+n,N+M+m}\pm E_\mathrm{LO}),
\end{align}
with the angle-averaged scattering weights $\widetilde{\mathcal W}^\pm_\mathrm{LO}$ given by Eq.~\eqref{eq:angle_pop}.

\subsection{Interface Roughness}
In order to calculate the scattering rates due to interface roughness in Eq.~\eqref{eq:EOM_IR}, we need to calculate the angle-averaged correlation function $\tilde C(E_k,E_q)$. In this work, we employ a Gaussian correlation function $C(|\ve r|)=\Delta^2e^{-|\ve r|^2/\Lambda^2}$, with a corresponding Fourier transform $C(|\ve q|)=\pi\Delta^2\Lambda^2e^{-|\ve q|^2\Lambda^2/4}$. The angle-averaged correlation function can be calculated using
\begin{align}
  \nonumber
  &\tilde C(E_k,E_q)=\frac{1}{2\pi}\int_0^{2\pi}C(|\ve q-\ve k|) d\theta \\ \nonumber
  &=\frac{1}{2\pi}\int_0^{2\pi} d\theta
  C\pth{ \sqrt{\frac{\mpar}{2\hbar^2}} \sqrt{E_k+E_q-2\sqrt{E_kE_q}\cos(\theta)} } \\ \nonumber
  &=\Delta^2\Lambda^2\int_0^{\pi} d\theta
   e^{ -\frac{\mpar\Lambda^2}{2\hbar^2}(E_k+E_q-2\sqrt{E_kE_q}\cos(\theta))  } \\ \label{eq:corr_av_g}
  &=\pi\Delta^2\Lambda^2 e^{ -\frac{\mpar\Lambda^2}{2\hbar^2}(E_k+E_q)}
    I_0(\sqrt{E_kE_q}\mpar\Lambda^2/\hbar^2),
\end{align}
where we have used Eq.~(3.339) from Ref.~\cite{gradshteyn2007}
\begin{align}
  \label{eq:bessel_id}
  \int_0^{\pi} e^{-x\cos(\theta)} d\theta= \pi I_0(x),
\end{align}
with $I_0(x)$, the modified Bessel function of the first kind of order zero. The in-scattering rates can now be written as
\begin{align}
  \nonumber
  \Gamma^{\mathrm{in},\mathrm{IR},\pm}_{NMnmE_k} &=
  \frac{\mpar}{4\hbar^3} u[\Delta_{N+M,N+M+m}+E_k] \\ \nonumber
  &\times\sum_i \irpr{N}{N+n}{z_i}\irpr{N+M+m}{N+M}{z_i}^* \\ \nonumber
  &\times \tilde C(E_k,\Delta_{N+M,N+M+m}+E_k),
\end{align}
with the corresponding out-scattering rates
\begin{align}
  \nonumber
  &\Gamma^{\mathrm{out},\mathrm{IR},\pm}_{NMnmE_k} =
  \frac{\mpar}{4\hbar^3} u[\Delta_{N+n,N+M+m}+E_k] \\ \nonumber
  &\times\sum_i \irpr{N+M}{N+M+m}{z_i}\irpr{N+M+m}{N+n}{z_i}^* \\ \label{eq:IR_rates}
  &\times \tilde C(E_k,\Delta_{N+n,N+M+m}+E_k) ,
\end{align}
where we have split the rates into identical absorption ($+$) and emission ($-$) terms.

\subsection{Ionized Impurities}
\label{subsec:ii}

The potential due to a single ionized impurity situated at $\ve R_i$, can be written as
\begin{align}
  \label{eq:Hion_single}
  V(\ve R)=-\frac{e^2}{4\pi \eps} \frac{e^{-\beta|\ve R-\ve R_i|}}{|\ve R-\ve R_i|},
\end{align}
where $\beta$ is the inverse screening length. Now suppose there are many impurities, situated on a sheet $\ve R_i=(\ve r_i,z_\ell)$. The resulting interaction Hamiltonian can be written as
\begin{align}
  \nonumber
  H_\mathrm{II}&=-\frac{e^2}{4\pi\eps} \sum_{i}
  \frac{e^{-\beta\sqrt{ (\ve r-\ve r_i)^2+(z-z_\ell)^2 }}}{\sqrt{(\ve r-\ve r_i)^2+(z-z_\ell)^2}}
  \\ \label{eq:Hion}
  &=\frac{-e^2}{8\pi^2\eps}\sum_i \int d^2q e^{i\dpr{q}{(r-r_i)}}
    \frac{e^{-\sqrt{\beta^2+q^2}|z-z_\ell|}}{\sqrt{\beta^2+q^2}} ,
\end{align}
where the second line in Eq.~\eqref{eq:Hion} is the 2D Fourier decomposition of the first line. We now proceed as we did with interface roughness scattering up until Eq.~\eqref{eq:beforeint} and get the in-scattering part of the dissipator due to ionized-impurities
\begin{align}
\label{eq:beforeint2}
\begin{split}
  &\mathcal D_\mathrm{II}^\mathrm{in}=\frac{\pi}{\hbar}\sum_{n_{1234}} \int d^2 k_{124}
  \delta[ E(n_4,\ve k_4)-E(n_3,\ve k_2)]
  \rho_{n_2,n_3}^{E_{k_2}} \\
  &\times  \innp{n_1,\ve k_1}{H_\mathrm{II}}{n_2,\ve k_2}
  \innp{n_3,\ve k_2}{H_\mathrm{II}}{n_4,\ve k_4}
  \ket{n_1,\ve k_1} \bra{n_4,\ve k_4} \\
  &+ \mathrm{h.c..}
\end{split}
\end{align}
Inserting Eq.~\ref{eq:Hion} into Eq.~\eqref{eq:beforeint2}, we can simplify the integrand using
\begin{align}
  \nonumber
  & \innp{n_1,\ve k_1}{H_\mathrm{II}}{n_2,\ve k_2}
  \innp{n_3,\ve k_2}{H_\mathrm{II}}{n_4,\ve k_4} \\ \nonumber
  &= \frac{e^4}{64\pi^4\eps^2}
   \sum_{r_i,r_j}\int d^2q d^2q' \frac{1}{\sqrt{\beta^2+q^2}}\frac{1}{\sqrt{\beta^2+q'^2}} \\
  \nonumber
  &\times e^{-i\dpr{q}{r}_i}e^{-i\dpr{q'}{r}_j} \iipr{n_1}{n_2}{q}{\ell}\iipr{n_2}{n_3}{q'}{\ell} \\
  \label{eq:integrand_II}
  &\times \braket{\ve k_1}{\ve k_2+\ve q}\braket{\ve k_2}{\ve k_4-\ve q'} \ .
\end{align}
with
\begin{align}
  \label{eq:II_prod}
  \iipr{n}{m}{q}{\ell}= \sum_b \int dz [\psi_n^{(b)}(z)]^*\psi_m^{(b)}(z)
  \frac{e^{-\sqrt{\beta^2+q^2}|z-z_\ell| }}{\sqrt{\beta^2+q^2}},
\end{align}
where the sum is over the considered bands. Equation~\eqref{eq:integrand_II} depends the detailed distribution of impurities, just as Eq.~\eqref{eq:integrand_IR} depends on the detailed in-plane interface roughness $\Delta_i(\ve r)$. By averaging over all possible distribution of impurities $\ve r_i$ and $\ve r_j$, we can see that terms with $\ve r_i=\ve r_j$ and $\ve q=\ve q'$ dominate. Limiting the sum to $\ve r_i=\ve r_j$ and going from sum over $\ve r_i$ to integral gives
\begin{align}
  \label{eq:sumtoint_ion}
\begin{split}
  \sum_{i}e^{-i\ve r_i\cdot(\ve q-\ve q')} &\simeq \int d^2 r_i N_\mathrm{2D}e^{-i\ve r_i\cdot(\ve q-\ve q')} \\
   &=(2\pi)^2 N_\mathrm{2D}\delta[\ve q-\ve q'].
\end{split}
\end{align}
Inserting Eq.~\eqref{eq:sumtoint_ion} into Eq.~\eqref{eq:integrand_II} gives
\begin{align}
  \nonumber
  & \innp{n_1,\ve k_1}{H_\mathrm{II}}{n_2,\ve k_2}
  \innp{n_3,\ve k_2}{H_\mathrm{II}}{n_4,\ve k_4} \\ \nonumber
  &= \frac{e^4}{16\pi^2\eps^2} N_\mathrm{2D} \int d^2q \frac{1}{\beta^2+q^2}
     \iipr{n_1}{n_2}{q}{\ell}\iipr{n_2}{n_3}{q}{\ell} \\ \label{eq:ii_last}
  &\times \delta[\ve k_2-(\ve k_1-\ve q)]\delta[\ve k_4-(\ve k_2+\ve q)].
\end{align}
Inserting Eq.~\eqref{eq:ii_last} into Eq.~\eqref{eq:beforeint2} and integrating over $\ve k_4$ and $\ve k_2$, dropping the subscript on $\ve k_1$ and shifting the integration variable $\ve q\rightarrow -\ve q+\ve k$ gives
\begin{align}
  \nonumber
  \mathcal D_\mathrm{II}^\mathrm{in}= -\frac{e^4 N_\mathrm{2D}}{16\pi \eps^2\hbar}
  &\sum_{n_{1234}} \int d^2 k \int d^2q
  \delta(E_q-\Delta_{n_4,n_3}-E_k) \\   \label{eq:ii_skipalot}
  &\times \iipr{n_1}{n_2}{|\ve q-\ve k|}{\ell} \iipr{n_3}{n_4}{|\ve q-\ve k|}{\ell} \rho_{n_2,n_3}^{E_q} .
\end{align}
Calculating the scattering rates from Eq.~\eqref{eq:ii_last} proceeds the same way as for interface roughness in Eqs.~\eqref{eq:after_avg}-\eqref{eq:EOM_IR}. The total rate due to ionized impurities is obtained by summing over all delta-doped sheets located at positions $z=z_\ell$ with sheet densities $N_{\mathrm{2D},\ell}$. The result for the in-scattering rates is
\begin{align}
\begin{split}
  &\Gamma_{NMnmE_k}^{\pm,\mathrm{in},\mathrm{II}}=
  \frac{e^2\mpar}{16\eps^2\hbar^3}
  u[E_k+\Delta_{N+M,N+M+m}] \sum_{\ell} \frac{N_{\mathrm{2D},\ell}}{2\pi} \\
  &\times \Big[ \int_0^{2\pi} d\theta
   \iipr{N}{N+n}{|\ve q-\ve k|}{\ell} \\
  &\times \iiprc{N+M+m}{N+M}{|\ve q-\ve k|}{\ell} \Big]_{E_q=\Delta_{N+M,N+M+m}+E_k}\\
\end{split}
\end{align}
and the corresponding out-scattering rates are given by
\begin{align}
\begin{split}
  &\Gamma_{NMnmE_k}^{\pm,\mathrm{out},\mathrm{II}} =
  \frac{e^2\mpar}{16\eps^2\hbar^3}
  u[E_k+\Delta_{N+n,N+M+m}] \sum_{\ell} \frac{N_{\mathrm{2D},\ell}}{2\pi} \\
  &\times \Big[  \int_0^{2\pi} d \theta
   \iipr{N+M}{N+M+m}{|\ve q-\ve k|}{\ell} \\
  &\times \iiprc{N+M+m}{N+n}{|\ve q-\ve k|}{\ell} \Big]_{E_q=\Delta_{N+n,N+M+m}+E_k} .
\end{split}
\end{align}
The products involving the wavefunctions $\iipr{n}{m}{|\ve q-\ve k|}{\ell}$ are more complicated than in the case for both the phonon and interface roughness interaction. This is because the interaction potential for ionized impurities in Eq.~\eqref{eq:Hion} does not factor into the cross plane and in-plane directions and the products involving eigenfunctions $\psi_n(z)$ are coupled with the in-plane motion. To overcome this issue, we numerically calculate and store the quantity
\begin{align}
  \label{eq:ii_temp}
  F_{nm\ell}(b)=\frac{1}{b}\int \psi^*_n(z)\psi_m(z)e^{-b|z-z_\ell|} dz
\end{align}
using an evenly spaced mesh $[b_\mathrm{min},b_\mathrm{max}]$ for the relevant range of $b$. Typical values are $b_\mathrm{min}=\beta$ and $b_\mathrm{max}=10$~nm$^{-1}$. We then perform the $\theta$ integration numerically with the stored values of $F_{nm\ell}(b)$ and convert between $b$ and $\theta$ using
\begin{align}
  \label{eq:bdef}
  b(\theta)=\sqrt{ \frac{2\mpar}{\hbar^2} }\sqrt{E_D+E_k+E_q-2\sqrt{E_kE_q}\cos(\theta)} .
\end{align}
with $E_D=\hbar^2 \beta^2/(2\mpar)$ the Debye energy.

\subsection{Alloy Scattering}
\label{subsec:allloy}
Inserting the alloy interaction potential given by Eq.~\eqref{eq:integrand_IR} into Eq.~\eqref{eq:beforeint} gives the in-scattering part of the dissipator due to random alloy scattering
\begin{align}
\label{eq:beforeint_alloy}
\begin{split}
  \mathcal D_\mathrm{alloy}^\mathrm{in}&=\frac{\pi}{\hbar}\sum_{n_{1234}} \int d^2 k_{124}
  \delta[ E(n_4,\ve k_4)-E(n_3,\ve k_2)]
  \rho_{n_2,n_3}^{E_{k_2}} \\
  &\times  \innp{n_1,\ve k_1}{V_\mathrm{alloy}}{n_2,\ve k_2}
  \innp{n_3,\ve k_2}{V_\mathrm{alloy}}{n_4,\ve k_4} \\
  & \times \ket{n_1,\ve k_1} \bra{n_4,\ve k_4} + \mathrm{h.c..}
\end{split}
\end{align}
The matrix elements involving the interaction potential can be simplified using
\begin{align}
\begin{split}
  &\innp{n_1,\ve k_1}{V_\mathrm{alloy}}{n_2,\ve k_2}
  \innp{n_3,\ve k_2}{V_\mathrm{alloy}}{n_4,\ve k_4} \\
  &= \sum_{i,j} \aprod{n_1}{n_2}{z_i} \aprod{n_3}{n_4}{z_j} C_i^x C_j^x \\
  &\times \innp{\ve k_1}{\delta(\ve r-\ve r_i)}{\ve k_2}
  \innp{\ve k_2}{\delta(\ve r-\ve r_j)}{\ve k_4} \\
  &= \Omega_0 \sum_i \aprod{n_1}{n_2}{z_i} \aprod{n_3}{n_4}{z_i} x_i(1-x_i) V_i^2 \\
  &\times \innp{\ve k_1}{\delta(\ve r-\ve r_i)}{\ve k_2}
  \innp{\ve k_2}{\delta(\ve r-\ve r_i)}{\ve k_4} \ ,
\end{split}
\end{align}
where we have replaced the product $C_i^xC_j^x$ with the correlation function in Eq.~\eqref{eq:corr_all} and defined the alloy matrix element
\begin{align}
  \label{eq:alloy_mat}
  \aprod{n}{m}{z_i}= \sum_b [\psi^{(b)}_n(z_i)]^* \psi^{(b)}_m(z_i),
\end{align}
where the sum is over the considered bands. Going from sum to integral using $\sum_i\rightarrow \frac{1}{\Omega_0}\int d^3 R_i$ and following the same procedure as for interface roughness scattering in Eqs.~\eqref{eq:beforeint}-\eqref{eq:EOM_IR} gives the scattering rates
\begin{align}
\begin{split}
  &\Gamma_{NMnmE_k}^{\pm,\mathrm{in},\mathrm{alloy}}=
  \frac{\mpar\Omega_0}{4\hbar^3}
  u[E_k+\Delta_{N+M,N+M+m}] \\ \nonumber
  & \times \int dz \aprod{N}{N+n}{z} \aprod{N+M+m}{N+M}{z}\\ \nonumber
  & \times x(z)(1-x(z))V(z)^2 \\ \nonumber
\end{split}
\end{align}
and
\begin{align}
\begin{split}
  &\Gamma_{NMnmE_k}^{\pm,\mathrm{out},\mathrm{alloy}}=
  \frac{\mpar\Omega_0}{4\hbar^3}
  u[E_k+\Delta_{N+n,N+M+m}] \\ \nonumber
  & \times \int dz \aprod{N+M}{N+M+m}{z} \\
  & \aprod{N+M+m}{N+n}{z}  \times x(z)(1-x(z))V(z)^2,
\end{split}
\end{align}
where $x(z)$ is the alloy fraction at position $z$ and $V(z)$ is the alloy strength parameter at position $z$, which we assume is the same for all bands.

\section{Material parameters}
\label{app:matpar}

Table~\ref{tab:parameters} lists all material specific parameters relating to the III-V binaries used in this work, as well as other parameters such as polar-optical-phonon energy and density. The temperature-dependent lattice constant is calculated using
\begin{align}
  \label{eq:alc}
  a(T)=a(T=300\ \mathrm{K}) + \alpha_L(T-300\ \mathrm{ K}),
\end{align}
with $a(T=300\mathrm{ K})$ the lattice constant at 300~K and the thermal expansion coefficient $\alpha_L$ given in table~\ref{tab:parameters}. In order to calculate the temperature-dependent $\Gamma$-valley bandgap, we use the empirical Varshni functional form~\cite{varshni_physica_1967}
\begin{align}
  \label{eq:bandgap_temp}
  E_\mathrm{gap}(T)=E_\mathrm{gap}(T=0)-\frac{\alpha^\Gamma T^2}{T+\beta^\Gamma},
\end{align}
where $E_\mathrm{gap}(T=0)$ is the zero-temperature $\Gamma$-valley bandgap and $\alpha^\Gamma$ and $\beta^\Gamma$ are the Varshni parameters given in table~\ref{tab:parameters}. All parameters relating to bandstructure in table~\ref{tab:parameters} are taken from Ref.~\cite{vurgaftman_JAP_2001} with the exception of Kane energy for GaAs and InAs, which were chosen to give room-temperature effective electron masses of $0.063 m_0$ and $0.023 m_0$, respectively ($m_0$ is the free electron mass). Note that the conduction-band effective mass can be written in terms of the parameters in table~\ref{tab:parameters}
\begin{align}
  \label{eq:kp_elmass}
  \frac{m_0}{m^*_c} = (1+2F) +
  \frac{E_P(E_\mathrm{gap}+2\Delta_\mathrm{so}/3)}{E_\mathrm{gap}(E_\mathrm{gap}+\Delta_\mathrm{so})} .
\end{align}

\begin{table*}[h!]
\caption{Various material parameters for GaAs, AlAs and InAs.}
\label{tab:parameters}
\begin{tabular}{lcccc}
 Parameter    & Symbol             &  GaAs & AlAs & InAs \\
 \hline
 Lattice constant at 300~K (\AA) &$a(T=300\text{ K})$  & $5.65325$ & $5.6611$ & $6.0583$ \\
 Thermal expansion coefficient ($10^{-5}$\AA/K)  &$\alpha_\mathrm{L}$ & $3.88$ & $2.90$ & $2.74$ \\
 Bandgap ($\Gamma$ valley) at 0~K (eV) &$E_\mathrm{gap}(T=0)$ & $1.519$   & $3.099$  & $0.417$ \\
 Varshni parameter (meV/K) &$\alpha^\Gamma$   & $0.5405$  & $0.885$  & $0.276$ \\
 Varshni parameter (K)     &$\beta^\Gamma$         & $204$     & $530$    & $93$ \\
 Spin-orbit splitting (eV) &$\Delta_\mathrm{so}$    & $0.341$   & $0.28$   & $0.39$ \\
 Luttinger parameter       &$\gamma_1^L$              & $6.98$    & $3.76$   & $20.0$ \\
 Luttinger parameter       &$\gamma_2^L$              & $2.06$    & $0.82$   & $8.50$ \\
 Kane energy (eV)          &$E_P$       &  $28.6$   &  $21.1$   & $20.7$ \\
 F parameter               &$F$         &  $-1.94$  & $-0.48$  & $-2.9$  \\
 Valence-band offset (eV)  &VBO        & $-0.80$   & $-1.33$  & $-0.59$ \\
 Conduction band deformation potential (eV) &$a_c$  & $-7.17$   & $-5.64$  & $-5.08$ \\
 Valence band deformation potential (eV) &$a_v$  & $-1.16$   & $-2.47$  & $-1.00$ \\
 Shear deformation potential (eV) &$b$  & $-2.00$   & $-2.30$  & $-1.80$ \\
 Elastic constant (GPa)    &$c_{11}$   & $122.1$   & $125.0$  & $83.3$ \\
 Elastic constant (GPa)    &$c_{12}$   & $56.6$    & $53.4$   & $45.3$ \\
 Polar-optical phonon energy (meV) &$E_\mathrm{POP}$ & $35$      & $50$      & $30$   \\
 Mass density (g/cm$^3$)   &$\rho$         & $5.32$   & $3.76$    & $5.68$  \\
 Low-frequency relative dielectric constant &$\eps_s^{0}$ & $12.90$   & $10.06$    & $15.15$  \\
 High-frequency relative dielectric constant &$\eps_s^{\infty}$  & $10.89$   & $8.16$    & $12.30$  \\
\end{tabular}
\end{table*}

Material parameters such as Kane energy for  ternary alloys ($A_{1-x}B_xC$) as a function of alloy composition $x$ are obtained using a second order interpolation formula on the form~\cite{vechten_PRB_1970}
\begin{align}
\label{eq:alloy_int}
\begin{aligned}
  E_P(A_{1-x}B_xC) &= (1-x)E_P(A) + x E_P(B)  \\
  & -x(1-x)C_b,
\end{aligned}
\end{align}
where $E_P(A)$ and $E_P(B)$ are the Kane energies of binary AC and BC respectively and the bowing parameter $C_b$ accounts for deviation from linear interpolation. Generalization to other material parameters is straightforward. Exceptions to Eq.~\eqref{eq:alloy_int} are interpolations for the Kane energy and F parameters of the In$_{1-x}$Ga$_x$As alloy, which contain a third order term due to the sensitivity of energy spacing to the effective mass of the well material. This third order term is described in more detail later in this appendix.

Nonzero bowing parameters for the In$_{1-x}$Ga$_x$As and In$_{1-x}$Al$_x$As alloys are given in table~\ref{tab:bowing}. All parameters are obtained from Ref.~\cite{vurgaftman_JAP_2001}, except for the Kane energy and F parameter in In$_{1-x}$Ga$_x$As, which in this work were chosen to give a room-temperature electron effective mass of $0.041 m_0$ for In$_{0.53}$Ga$_{0.47}$As (lattice-matched to InP). Note that when interpolating Luttinger parameters, the light holes ($m_\mathrm{lh}^*$) and heavy hole ($m_\mathrm{hh}^*$) masses should be interpolated using the bowing parameters in table~\ref{tab:bowing} and corresponding Luttinger parameters calculated afterwards using
\begin{align}
\label{eq:luttholes}
\begin{aligned}
 \frac{m_0}{m_\mathrm{lh}^*}&=\gamma_1^L+2\gamma_2^L \\
 \frac{m_0}{m_\mathrm{hh}^*}&=\gamma_1^L-2\gamma_2^L . \\
\end{aligned}
\end{align}
\begin{table}[h!]
\caption{Nonzero bowing parameters for the In$_{1-x}$Ga$_x$As and In$_{1-x}$Al$_x$As alloys.}
\label{tab:bowing}
\begin{tabular}{lcc}
 Parameter  &  In$_{1-x}$Ga$_x$As & In$_{1-x}$Al$_x$As \\
 \hline
 $E_\mathrm{gap}^\Gamma$ (eV) & $0.477$ & $0.70$ \\
 $\Delta_\mathrm{so}$ (eV) & $0.15$ & $0.15$ \\
 $m_\mathrm{hh}^*$ & $-0.145$ & $0.00$ \\
 $m_\mathrm{lh}^*$ & $0.0202$ & $0.00$ \\
 $E_P$ (eV) & $3.20$  & $-4.81$  \\
 $F$        & $-1.00$  & $-4.44$  \\
 VBO (eV) & $-0.38$ & $-0.64$ \\
 $a_c$ (eV) & $2.61$ & $-1.40$ \\
\end{tabular}
\end{table}

As shown in Sec.~\ref{sec:lm_design}, the material parameters presented in tables~\ref{tab:parameters} and \ref{tab:bowing} give results in very good agreement with experiment as well as theory based on NEGF for lattice-matched designs. However, for the strain-balanced design in section~\ref{sec:sb_design}, the peak gain wavelength was overestimated by $0.2$~$\upmu$m, with an associated shift to lower electric fields in the current density vs electric field curve. We attribute this disagreement to underestimation of the electron mass in the well material (lower mass means smaller energy spacing) and addressed the issue by adding a cubic term
\begin{align}
  \label{eq:cubic_int}
  -x(1-x)(x-0.47)D
\end{align}
to the alloy interpolation formula [Eq.~\eqref{eq:alloy_int}] for Kane energy and F parameter in In$_{1-x}$Ga$_x$As. This functional form was chosen to not change the lattice-matched results for $x=0.47$, or the binary results ($x=0$ and $x=1$). Values of $D=32$~eV for Kane energy and $D=15$ for the F parameter were found to reproduce experimental results with an energy spacing of $270$~meV (corresponding to $\lambda=4.6$~$\upmu$m) between the upper and lower lasing states. These values of $D$ might seem excessively high. However, $D$ is multiplied by 3 numbers that have magnitudes smaller than 1.

We will end this Appendix with a discussion of the parameters related to interface-roughness scattering. We used the same interface-roughness correlation length $\Lambda=9$~nm as in Refs.~\cite{lindskog_APL_2014,bugajski_PSS_2014}. We estimate the RMS value for interface roughness ($\Delta$) by using the fact that interface-roughness scattering is the main broadening mechanism of the gain spectrum.~\cite{faist_book_2013} We followed the procedure in Ref.~\cite{bugajski_PSS_2014} and fix the correlation length at $9$~nm and vary $\Delta$ such that the gain spectra produces a peak with a FWHM of $25$~meV, matching experimental results for electroluminescence spectra at a field strength of $53$~kV/cm.~\cite{bismuto_APL_2010} We note that the resulting value of $\Delta=0.08$~nm is smaller than the value of $\Delta=0.1$~nm used in Ref.~\cite{lindskog_APL_2014} to simulate the same device. We attribute the difference to the different electron Hamiltonian used in this work (three-band $\dpr{k}{p}$), resulting in higher amplitude of eigenfunctions at material interfaces.


\begin{thebibliography}{86}%
\makeatletter
\providecommand \@ifxundefined [1]{%
 \@ifx{#1\undefined}
}%
\providecommand \@ifnum [1]{%
 \ifnum #1\expandafter \@firstoftwo
 \else \expandafter \@secondoftwo
 \fi
}%
\providecommand \@ifx [1]{%
 \ifx #1\expandafter \@firstoftwo
 \else \expandafter \@secondoftwo
 \fi
}%
\providecommand \natexlab [1]{#1}%
\providecommand \enquote  [1]{``#1''}%
\providecommand \bibnamefont  [1]{#1}%
\providecommand \bibfnamefont [1]{#1}%
\providecommand \citenamefont [1]{#1}%
\providecommand \href@noop [0]{\@secondoftwo}%
\providecommand \href [0]{\begingroup \@sanitize@url \@href}%
\providecommand \@href[1]{\@@startlink{#1}\@@href}%
\providecommand \@@href[1]{\endgroup#1\@@endlink}%
\providecommand \@sanitize@url [0]{\catcode `\\12\catcode `\$12\catcode
  `\&12\catcode `\#12\catcode `\^12\catcode `\_12\catcode `\%12\relax}%
\providecommand \@@startlink[1]{}%
\providecommand \@@endlink[0]{}%
\providecommand \url  [0]{\begingroup\@sanitize@url \@url }%
\providecommand \@url [1]{\endgroup\@href {#1}{\urlprefix }}%
\providecommand \urlprefix  [0]{URL }%
\providecommand \Eprint [0]{\href }%
\providecommand \doibase [0]{https://doi.org/}%
\providecommand \selectlanguage [0]{\@gobble}%
\providecommand \bibinfo  [0]{\@secondoftwo}%
\providecommand \bibfield  [0]{\@secondoftwo}%
\providecommand \translation [1]{[#1]}%
\providecommand \BibitemOpen [0]{}%
\providecommand \bibitemStop [0]{}%
\providecommand \bibitemNoStop [0]{.\EOS\space}%
\providecommand \EOS [0]{\spacefactor3000\relax}%
\providecommand \BibitemShut  [1]{\csname bibitem#1\endcsname}%
\let\auto@bib@innerbib\@empty
\bibitem [{\citenamefont {Faist}\ \emph {et~al.}(1994)\citenamefont {Faist},
  \citenamefont {Capasso}, \citenamefont {Sivco}, \citenamefont {Sirtori},
  \citenamefont {Hutchinson},\ and\ \citenamefont {Cho}}]{faist_book_2013}%
  \BibitemOpen
  \bibfield  {author} {\bibinfo {author} {\bibfnamefont {J.}~\bibnamefont
  {Faist}}, \bibinfo {author} {\bibfnamefont {F.}~\bibnamefont {Capasso}},
  \bibinfo {author} {\bibfnamefont {D.~L.}\ \bibnamefont {Sivco}}, \bibinfo
  {author} {\bibfnamefont {C.}~\bibnamefont {Sirtori}}, \bibinfo {author}
  {\bibfnamefont {A.~L.}\ \bibnamefont {Hutchinson}},\ and\ \bibinfo {author}
  {\bibfnamefont {A.~Y.}\ \bibnamefont {Cho}},\ }\bibfield  {title} {\bibinfo
  {title} {Quantum cascade laser},\ }\href@noop {} {\bibfield  {journal}
  {\bibinfo  {journal} {Science}\ }\textbf {\bibinfo {volume} {264}},\ \bibinfo
  {pages} {553} (\bibinfo {year} {1994})}\BibitemShut {NoStop}%
\bibitem [{\citenamefont {Vitiello}\ \emph {et~al.}(2015)\citenamefont
  {Vitiello}, \citenamefont {Scalari}, \citenamefont {Williams},\ and\
  \citenamefont {De~Natale}}]{vitiello_OE_2015}%
  \BibitemOpen
  \bibfield  {author} {\bibinfo {author} {\bibfnamefont {M.~S.}\ \bibnamefont
  {Vitiello}}, \bibinfo {author} {\bibfnamefont {G.}~\bibnamefont {Scalari}},
  \bibinfo {author} {\bibfnamefont {B.}~\bibnamefont {Williams}},\ and\
  \bibinfo {author} {\bibfnamefont {P.}~\bibnamefont {De~Natale}},\ }\bibfield
  {title} {\bibinfo {title} {Quantum cascade lasers: 20 years of challenges},\
  }\href@noop {} {\bibfield  {journal} {\bibinfo  {journal} {Opt. Express}\
  }\textbf {\bibinfo {volume} {23}},\ \bibinfo {pages} {5167} (\bibinfo {year}
  {2015})}\BibitemShut {NoStop}%
\bibitem [{\citenamefont {Capasso}(2010)}]{capasso_OE_2010}%
  \BibitemOpen
  \bibfield  {author} {\bibinfo {author} {\bibfnamefont {F.}~\bibnamefont
  {Capasso}},\ }\bibfield  {title} {\bibinfo {title} {High-performance
  midinfrared quantum cascade lasers},\ }\href@noop {} {\bibfield  {journal}
  {\bibinfo  {journal} {Optical Engineering}\ }\textbf {\bibinfo {volume}
  {49}},\ \bibinfo {pages} {111102} (\bibinfo {year} {2010})}\BibitemShut
  {NoStop}%
\bibitem [{\citenamefont {Yao}\ \emph {et~al.}(2012)\citenamefont {Yao},
  \citenamefont {Hoffman},\ and\ \citenamefont {Gmachl}}]{yao_NATP_2012}%
  \BibitemOpen
  \bibfield  {author} {\bibinfo {author} {\bibfnamefont {Y.}~\bibnamefont
  {Yao}}, \bibinfo {author} {\bibfnamefont {A.~J.}\ \bibnamefont {Hoffman}},\
  and\ \bibinfo {author} {\bibfnamefont {C.~F.}\ \bibnamefont {Gmachl}},\
  }\bibfield  {title} {\bibinfo {title} {Mid-infrared quantum cascade lasers},\
  }\href@noop {} {\bibfield  {journal} {\bibinfo  {journal} {Nat. Photon}\
  }\textbf {\bibinfo {volume} {6}},\ \bibinfo {pages} {432} (\bibinfo {year}
  {2012})}\BibitemShut {NoStop}%
\bibitem [{\citenamefont {Razeghi}\ \emph {et~al.}(2013)\citenamefont
  {Razeghi}, \citenamefont {Bandyopadhyay}, \citenamefont {Bai}, \citenamefont
  {Lu},\ and\ \citenamefont {Slivken}}]{razeghi_OME_2013}%
  \BibitemOpen
  \bibfield  {author} {\bibinfo {author} {\bibfnamefont {M.}~\bibnamefont
  {Razeghi}}, \bibinfo {author} {\bibfnamefont {N.}~\bibnamefont
  {Bandyopadhyay}}, \bibinfo {author} {\bibfnamefont {Y.}~\bibnamefont {Bai}},
  \bibinfo {author} {\bibfnamefont {Q.}~\bibnamefont {Lu}},\ and\ \bibinfo
  {author} {\bibfnamefont {S.}~\bibnamefont {Slivken}},\ }\bibfield  {title}
  {\bibinfo {title} {Recent advances in mid infrared (3-5$\mu$m) quantum
  cascade lasers},\ }\href@noop {} {\bibfield  {journal} {\bibinfo  {journal}
  {Opt. Mater. Express}\ }\textbf {\bibinfo {volume} {3}},\ \bibinfo {pages}
  {1872} (\bibinfo {year} {2013})}\BibitemShut {NoStop}%
\bibitem [{\citenamefont {Curl}\ \emph {et~al.}(2010)\citenamefont {Curl},
  \citenamefont {Capasso}, \citenamefont {Gmachl}, \citenamefont {Kosterev},
  \citenamefont {McManus}, \citenamefont {Lewicki}, \citenamefont {Pusharsky},
  \citenamefont {Wysocki},\ and\ \citenamefont {Tittel}}]{curl_CPL_2010}%
  \BibitemOpen
  \bibfield  {author} {\bibinfo {author} {\bibfnamefont {R.~F.}\ \bibnamefont
  {Curl}}, \bibinfo {author} {\bibfnamefont {F.}~\bibnamefont {Capasso}},
  \bibinfo {author} {\bibfnamefont {C.}~\bibnamefont {Gmachl}}, \bibinfo
  {author} {\bibfnamefont {A.~A.}\ \bibnamefont {Kosterev}}, \bibinfo {author}
  {\bibfnamefont {B.}~\bibnamefont {McManus}}, \bibinfo {author} {\bibfnamefont
  {R.}~\bibnamefont {Lewicki}}, \bibinfo {author} {\bibfnamefont
  {M.}~\bibnamefont {Pusharsky}}, \bibinfo {author} {\bibfnamefont
  {G.}~\bibnamefont {Wysocki}},\ and\ \bibinfo {author} {\bibfnamefont {F.~K.}\
  \bibnamefont {Tittel}},\ }\bibfield  {title} {\bibinfo {title} {Quantum
  cascade lasers in chemical physics},\ }\href@noop {} {\bibfield  {journal}
  {\bibinfo  {journal} {Chem. Phys. Lett.}\ }\textbf {\bibinfo {volume}
  {487}},\ \bibinfo {pages} {1} (\bibinfo {year} {2010})}\BibitemShut {NoStop}%
\bibitem [{\citenamefont {Bartalini}\ \emph {et~al.}(2013)\citenamefont
  {Bartalini}, \citenamefont {Vitiello},\ and\ \citenamefont
  {De~Natale}}]{bartalini_MST_2014}%
  \BibitemOpen
  \bibfield  {author} {\bibinfo {author} {\bibfnamefont {S.}~\bibnamefont
  {Bartalini}}, \bibinfo {author} {\bibfnamefont {M.}~\bibnamefont
  {Vitiello}},\ and\ \bibinfo {author} {\bibfnamefont {P.}~\bibnamefont
  {De~Natale}},\ }\bibfield  {title} {\bibinfo {title} {Quantum cascade lasers:
  a versatile source for precise measurements in the mid/far-infrared range},\
  }\href@noop {} {\bibfield  {journal} {\bibinfo  {journal} {Meas. Sci.
  Technol}\ }\textbf {\bibinfo {volume} {25}},\ \bibinfo {pages} {012001}
  (\bibinfo {year} {2013})}\BibitemShut {NoStop}%
\bibitem [{\citenamefont {Dupont}\ \emph {et~al.}(2010)\citenamefont {Dupont},
  \citenamefont {Fathololoumi},\ and\ \citenamefont {Liu}}]{dupont_PRB_2010}%
  \BibitemOpen
  \bibfield  {author} {\bibinfo {author} {\bibfnamefont {E.}~\bibnamefont
  {Dupont}}, \bibinfo {author} {\bibfnamefont {S.}~\bibnamefont
  {Fathololoumi}},\ and\ \bibinfo {author} {\bibfnamefont {H.}~\bibnamefont
  {Liu}},\ }\bibfield  {title} {\bibinfo {title} {Simplified density-matrix
  model applied to three-well terahertz quantum cascade lasers},\ }\href@noop
  {} {\bibfield  {journal} {\bibinfo  {journal} {Phys. Rev. B.}\ }\textbf
  {\bibinfo {volume} {81}},\ \bibinfo {pages} {205311} (\bibinfo {year}
  {2010})}\BibitemShut {NoStop}%
\bibitem [{\citenamefont {Jirauschek}\ and\ \citenamefont
  {Kubis}(2014)}]{jirauschek_APR_2014}%
  \BibitemOpen
  \bibfield  {author} {\bibinfo {author} {\bibfnamefont {C.}~\bibnamefont
  {Jirauschek}}\ and\ \bibinfo {author} {\bibfnamefont {T.}~\bibnamefont
  {Kubis}},\ }\bibfield  {title} {\bibinfo {title} {Modeling techniques for
  quantum cascade lasers},\ }\href@noop {} {\bibfield  {journal} {\bibinfo
  {journal} {Appl. Phys. Rev.}\ }\textbf {\bibinfo {volume} {1}},\ \bibinfo
  {pages} {011307} (\bibinfo {year} {2014})}\BibitemShut {NoStop}%
\bibitem [{\citenamefont {Indjin}\ \emph
  {et~al.}(2002{\natexlab{a}})\citenamefont {Indjin}, \citenamefont {Harrison},
  \citenamefont {Kelsall},\ and\ \citenamefont {Ikoni{\'c}}}]{indjin_JAP_2002}%
  \BibitemOpen
  \bibfield  {author} {\bibinfo {author} {\bibfnamefont {D.}~\bibnamefont
  {Indjin}}, \bibinfo {author} {\bibfnamefont {P.}~\bibnamefont {Harrison}},
  \bibinfo {author} {\bibfnamefont {R.}~\bibnamefont {Kelsall}},\ and\ \bibinfo
  {author} {\bibfnamefont {Z.}~\bibnamefont {Ikoni{\'c}}},\ }\bibfield  {title}
  {\bibinfo {title} {Self-consistent scattering theory of transport and output
  characteristics of quantum cascade lasers},\ }\href@noop {} {\bibfield
  {journal} {\bibinfo  {journal} {J. Appl. Phys.}\ }\textbf {\bibinfo {volume}
  {91}},\ \bibinfo {pages} {9019} (\bibinfo {year}
  {2002}{\natexlab{a}})}\BibitemShut {NoStop}%
\bibitem [{\citenamefont {Indjin}\ \emph
  {et~al.}(2002{\natexlab{b}})\citenamefont {Indjin}, \citenamefont {Harrison},
  \citenamefont {Kelsall},\ and\ \citenamefont {Ikoni{\'c}}}]{indjin_APL_2002}%
  \BibitemOpen
  \bibfield  {author} {\bibinfo {author} {\bibfnamefont {D.}~\bibnamefont
  {Indjin}}, \bibinfo {author} {\bibfnamefont {P.}~\bibnamefont {Harrison}},
  \bibinfo {author} {\bibfnamefont {R.}~\bibnamefont {Kelsall}},\ and\ \bibinfo
  {author} {\bibfnamefont {Z.}~\bibnamefont {Ikoni{\'c}}},\ }\bibfield  {title}
  {\bibinfo {title} {Influence of leakage current on temperature performance of
  gaas/algaas quantum cascade lasers},\ }\href@noop {} {\bibfield  {journal}
  {\bibinfo  {journal} {Appl. Phys. Lett.}\ }\textbf {\bibinfo {volume} {81}},\
  \bibinfo {pages} {400} (\bibinfo {year} {2002}{\natexlab{b}})}\BibitemShut
  {NoStop}%
\bibitem [{\citenamefont {Mir{\v{c}}eti{\'c}}\ \emph
  {et~al.}(2005)\citenamefont {Mir{\v{c}}eti{\'c}}, \citenamefont {Indjin},
  \citenamefont {Ikoni{\'c}}, \citenamefont {Harrison}, \citenamefont
  {Milanovi{\'c}},\ and\ \citenamefont {Kelsall}}]{mircetic_JAP_2005}%
  \BibitemOpen
  \bibfield  {author} {\bibinfo {author} {\bibfnamefont {A.}~\bibnamefont
  {Mir{\v{c}}eti{\'c}}}, \bibinfo {author} {\bibfnamefont {D.}~\bibnamefont
  {Indjin}}, \bibinfo {author} {\bibfnamefont {Z.}~\bibnamefont {Ikoni{\'c}}},
  \bibinfo {author} {\bibfnamefont {P.}~\bibnamefont {Harrison}}, \bibinfo
  {author} {\bibfnamefont {V.}~\bibnamefont {Milanovi{\'c}}},\ and\ \bibinfo
  {author} {\bibfnamefont {R.~W.}\ \bibnamefont {Kelsall}},\ }\bibfield
  {title} {\bibinfo {title} {Towards automated design of quantum cascade
  lasers},\ }\href@noop {} {\bibfield  {journal} {\bibinfo  {journal} {J. Appl.
  Phys.}\ }\textbf {\bibinfo {volume} {97}},\ \bibinfo {pages} {084506}
  (\bibinfo {year} {2005})}\BibitemShut {NoStop}%
\bibitem [{\citenamefont {Wang}\ \emph {et~al.}(2018)\citenamefont {Wang},
  \citenamefont {Grillot},\ and\ \citenamefont {Wang}}]{wang2018rate}%
  \BibitemOpen
  \bibfield  {author} {\bibinfo {author} {\bibfnamefont {X.-G.}\ \bibnamefont
  {Wang}}, \bibinfo {author} {\bibfnamefont {F.}~\bibnamefont {Grillot}},\ and\
  \bibinfo {author} {\bibfnamefont {C.}~\bibnamefont {Wang}},\ }\bibfield
  {title} {\bibinfo {title} {Rate equation modeling of the frequency noise and
  the intrinsic spectral linewidth in quantum cascade lasers},\ }\href@noop {}
  {\bibfield  {journal} {\bibinfo  {journal} {Opt. Express}\ }\textbf {\bibinfo
  {volume} {26}},\ \bibinfo {pages} {2325} (\bibinfo {year}
  {2018})}\BibitemShut {NoStop}%
\bibitem [{\citenamefont {Iotti}\ and\ \citenamefont
  {Rossi}(2001{\natexlab{a}})}]{iotti2001}%
  \BibitemOpen
  \bibfield  {author} {\bibinfo {author} {\bibfnamefont {R.~C.}\ \bibnamefont
  {Iotti}}\ and\ \bibinfo {author} {\bibfnamefont {F.}~\bibnamefont {Rossi}},\
  }\bibfield  {title} {\bibinfo {title} {Nature of charge transport in
  quantum-cascade lasers},\ }\href@noop {} {\bibfield  {journal} {\bibinfo
  {journal} {Phys. Rev. Lett.}\ }\textbf {\bibinfo {volume} {87}},\ \bibinfo
  {pages} {146603} (\bibinfo {year} {2001}{\natexlab{a}})}\BibitemShut
  {NoStop}%
\bibitem [{\citenamefont {Callebaut}\ \emph {et~al.}(2004)\citenamefont
  {Callebaut}, \citenamefont {Kumar}, \citenamefont {Williams}, \citenamefont
  {Hu},\ and\ \citenamefont {Reno}}]{callebaut_APL_2004}%
  \BibitemOpen
  \bibfield  {author} {\bibinfo {author} {\bibfnamefont {H.}~\bibnamefont
  {Callebaut}}, \bibinfo {author} {\bibfnamefont {S.}~\bibnamefont {Kumar}},
  \bibinfo {author} {\bibfnamefont {B.~S.}\ \bibnamefont {Williams}}, \bibinfo
  {author} {\bibfnamefont {Q.}~\bibnamefont {Hu}},\ and\ \bibinfo {author}
  {\bibfnamefont {J.~L.}\ \bibnamefont {Reno}},\ }\bibfield  {title} {\bibinfo
  {title} {Importance of electron-impurity scattering for electron transport in
  terahertz quantum-cascade lasers},\ }\href@noop {} {\bibfield  {journal}
  {\bibinfo  {journal} {Appl. Phys. Lett.}\ }\textbf {\bibinfo {volume} {84}},\
  \bibinfo {pages} {645} (\bibinfo {year} {2004})}\BibitemShut {NoStop}%
\bibitem [{\citenamefont {Gao}\ \emph {et~al.}(2007)\citenamefont {Gao},
  \citenamefont {Botez},\ and\ \citenamefont {Knezevic}}]{gao_JAP_2007}%
  \BibitemOpen
  \bibfield  {author} {\bibinfo {author} {\bibfnamefont {X.}~\bibnamefont
  {Gao}}, \bibinfo {author} {\bibfnamefont {D.}~\bibnamefont {Botez}},\ and\
  \bibinfo {author} {\bibfnamefont {I.}~\bibnamefont {Knezevic}},\ }\bibfield
  {title} {\bibinfo {title} {X-valley leakage in gaas-based midinfrared quantum
  cascade lasers: A monte carlo study},\ }\href@noop {} {\bibfield  {journal}
  {\bibinfo  {journal} {J. Appl. Phys.}\ }\textbf {\bibinfo {volume} {101}},\
  \bibinfo {pages} {063101} (\bibinfo {year} {2007})}\BibitemShut {NoStop}%
\bibitem [{\citenamefont {Shi}\ and\ \citenamefont
  {Knezevic}(2014)}]{shi_JAP_2014}%
  \BibitemOpen
  \bibfield  {author} {\bibinfo {author} {\bibfnamefont {Y.}~\bibnamefont
  {Shi}}\ and\ \bibinfo {author} {\bibfnamefont {I.}~\bibnamefont {Knezevic}},\
  }\bibfield  {title} {\bibinfo {title} {Nonequilibrium phonon effects in
  midinfrared quantum cascade lasers},\ }\href@noop {} {\bibfield  {journal}
  {\bibinfo  {journal} {J. Appl. Phys.}\ }\textbf {\bibinfo {volume} {116}},\
  \bibinfo {pages} {123105} (\bibinfo {year} {2014})}\BibitemShut {NoStop}%
\bibitem [{\citenamefont {Borowik}\ \emph {et~al.}(2017)\citenamefont
  {Borowik}, \citenamefont {Thobel},\ and\ \citenamefont
  {Adamowicz}}]{borowik2017monte}%
  \BibitemOpen
  \bibfield  {author} {\bibinfo {author} {\bibfnamefont {P.}~\bibnamefont
  {Borowik}}, \bibinfo {author} {\bibfnamefont {J.-L.}\ \bibnamefont
  {Thobel}},\ and\ \bibinfo {author} {\bibfnamefont {L.}~\bibnamefont
  {Adamowicz}},\ }\bibfield  {title} {\bibinfo {title} {Monte carlo modeling
  applied to studies of quantum cascade lasers},\ }\href@noop {} {\bibfield
  {journal} {\bibinfo  {journal} {Optical and Quantum Electronics}\ }\textbf
  {\bibinfo {volume} {49}},\ \bibinfo {pages} {96} (\bibinfo {year}
  {2017})}\BibitemShut {NoStop}%
\bibitem [{\citenamefont {Willenberg}\ \emph {et~al.}(2003)\citenamefont
  {Willenberg}, \citenamefont {D{\"o}hler},\ and\ \citenamefont
  {Faist}}]{willenberg_PRB_2003}%
  \BibitemOpen
  \bibfield  {author} {\bibinfo {author} {\bibfnamefont {H.}~\bibnamefont
  {Willenberg}}, \bibinfo {author} {\bibfnamefont {G.}~\bibnamefont
  {D{\"o}hler}},\ and\ \bibinfo {author} {\bibfnamefont {J.}~\bibnamefont
  {Faist}},\ }\bibfield  {title} {\bibinfo {title} {Intersubband gain in a
  bloch oscillator and quantum cascade laser},\ }\href@noop {} {\bibfield
  {journal} {\bibinfo  {journal} {Phys. Rev. B.}\ }\textbf {\bibinfo {volume}
  {67}},\ \bibinfo {pages} {085315} (\bibinfo {year} {2003})}\BibitemShut
  {NoStop}%
\bibitem [{\citenamefont {Kumar}\ and\ \citenamefont
  {Hu}(2009)}]{kumar_PRB_2009}%
  \BibitemOpen
  \bibfield  {author} {\bibinfo {author} {\bibfnamefont {S.}~\bibnamefont
  {Kumar}}\ and\ \bibinfo {author} {\bibfnamefont {Q.}~\bibnamefont {Hu}},\
  }\bibfield  {title} {\bibinfo {title} {Coherence of resonant-tunneling
  transport in terahertz quantum-cascade lasers},\ }\href@noop {} {\bibfield
  {journal} {\bibinfo  {journal} {Phys. Rev. B.}\ }\textbf {\bibinfo {volume}
  {80}},\ \bibinfo {pages} {245316} (\bibinfo {year} {2009})}\BibitemShut
  {NoStop}%
\bibitem [{\citenamefont {Weber}\ \emph {et~al.}(2009)\citenamefont {Weber},
  \citenamefont {Wacker},\ and\ \citenamefont {Knorr}}]{weber_PRB_2009}%
  \BibitemOpen
  \bibfield  {author} {\bibinfo {author} {\bibfnamefont {C.}~\bibnamefont
  {Weber}}, \bibinfo {author} {\bibfnamefont {A.}~\bibnamefont {Wacker}},\ and\
  \bibinfo {author} {\bibfnamefont {A.}~\bibnamefont {Knorr}},\ }\bibfield
  {title} {\bibinfo {title} {Density-matrix theory of the optical dynamics and
  transport in quantum cascade structures: The role of coherence},\ }\href@noop
  {} {\bibfield  {journal} {\bibinfo  {journal} {Phys. Rev. B.}\ }\textbf
  {\bibinfo {volume} {79}},\ \bibinfo {pages} {165322} (\bibinfo {year}
  {2009})}\BibitemShut {NoStop}%
\bibitem [{\citenamefont {Terazzi}\ and\ \citenamefont
  {Faist}(2010)}]{terazzi_NJP_2010}%
  \BibitemOpen
  \bibfield  {author} {\bibinfo {author} {\bibfnamefont {R.}~\bibnamefont
  {Terazzi}}\ and\ \bibinfo {author} {\bibfnamefont {J.}~\bibnamefont
  {Faist}},\ }\bibfield  {title} {\bibinfo {title} {A density matrix model of
  transport and radiation in quantum cascade lasers},\ }\href@noop {}
  {\bibfield  {journal} {\bibinfo  {journal} {New J. Phys.}\ }\textbf {\bibinfo
  {volume} {12}},\ \bibinfo {pages} {033045} (\bibinfo {year}
  {2010})}\BibitemShut {NoStop}%
\bibitem [{\citenamefont {Burnett}\ \emph {et~al.}(2018)\citenamefont
  {Burnett}, \citenamefont {Pan}, \citenamefont {Chui},\ and\ \citenamefont
  {Williams}}]{burnett2018robust}%
  \BibitemOpen
  \bibfield  {author} {\bibinfo {author} {\bibfnamefont {B.~A.}\ \bibnamefont
  {Burnett}}, \bibinfo {author} {\bibfnamefont {A.}~\bibnamefont {Pan}},
  \bibinfo {author} {\bibfnamefont {C.~O.}\ \bibnamefont {Chui}},\ and\
  \bibinfo {author} {\bibfnamefont {B.~S.}\ \bibnamefont {Williams}},\
  }\bibfield  {title} {\bibinfo {title} {Robust density matrix simulation of
  terahertz quantum cascade lasers},\ }\href@noop {} {\bibfield  {journal}
  {\bibinfo  {journal} {IEEE Transactions on Terahertz Science and Technology}\
  }\textbf {\bibinfo {volume} {8}},\ \bibinfo {pages} {492} (\bibinfo {year}
  {2018})}\BibitemShut {NoStop}%
\bibitem [{\citenamefont {Pan}\ \emph {et~al.}(2017)\citenamefont {Pan},
  \citenamefont {Burnett}, \citenamefont {Chui},\ and\ \citenamefont
  {Williams}}]{pan2017density}%
  \BibitemOpen
  \bibfield  {author} {\bibinfo {author} {\bibfnamefont {A.}~\bibnamefont
  {Pan}}, \bibinfo {author} {\bibfnamefont {B.~A.}\ \bibnamefont {Burnett}},
  \bibinfo {author} {\bibfnamefont {C.~O.}\ \bibnamefont {Chui}},\ and\
  \bibinfo {author} {\bibfnamefont {B.~S.}\ \bibnamefont {Williams}},\
  }\bibfield  {title} {\bibinfo {title} {Density matrix modeling of quantum
  cascade lasers without an artificially localized basis: A generalized
  scattering approach},\ }\href@noop {} {\bibfield  {journal} {\bibinfo
  {journal} {Phys. Rev. B.}\ }\textbf {\bibinfo {volume} {96}},\ \bibinfo
  {pages} {085308} (\bibinfo {year} {2017})}\BibitemShut {NoStop}%
\bibitem [{\citenamefont {Demi{\'c}}\ \emph {et~al.}(2017)\citenamefont
  {Demi{\'c}}, \citenamefont {Grier}, \citenamefont {Ikoni{\'c}}, \citenamefont
  {Valavanis}, \citenamefont {Evans}, \citenamefont {Mohandas}, \citenamefont
  {Li}, \citenamefont {Linfield}, \citenamefont {Davies},\ and\ \citenamefont
  {Indjin}}]{demic2017infinite}%
  \BibitemOpen
  \bibfield  {author} {\bibinfo {author} {\bibfnamefont {A.}~\bibnamefont
  {Demi{\'c}}}, \bibinfo {author} {\bibfnamefont {A.}~\bibnamefont {Grier}},
  \bibinfo {author} {\bibfnamefont {Z.}~\bibnamefont {Ikoni{\'c}}}, \bibinfo
  {author} {\bibfnamefont {A.}~\bibnamefont {Valavanis}}, \bibinfo {author}
  {\bibfnamefont {C.~A.}\ \bibnamefont {Evans}}, \bibinfo {author}
  {\bibfnamefont {R.}~\bibnamefont {Mohandas}}, \bibinfo {author}
  {\bibfnamefont {L.}~\bibnamefont {Li}}, \bibinfo {author} {\bibfnamefont
  {E.~H.}\ \bibnamefont {Linfield}}, \bibinfo {author} {\bibfnamefont {A.~G.}\
  \bibnamefont {Davies}},\ and\ \bibinfo {author} {\bibfnamefont
  {D.}~\bibnamefont {Indjin}},\ }\bibfield  {title} {\bibinfo {title}
  {Infinite-period density-matrix model for terahertz-frequency quantum cascade
  lasers},\ }\href@noop {} {\bibfield  {journal} {\bibinfo  {journal} {IEEE
  Transactions on Terahertz Science and Technology}\ }\textbf {\bibinfo
  {volume} {7}},\ \bibinfo {pages} {368} (\bibinfo {year} {2017})}\BibitemShut
  {NoStop}%
\bibitem [{\citenamefont {Jirauschek}(2017)}]{jirauschek2017density}%
  \BibitemOpen
  \bibfield  {author} {\bibinfo {author} {\bibfnamefont {C.}~\bibnamefont
  {Jirauschek}},\ }\bibfield  {title} {\bibinfo {title} {Density matrix monte
  carlo modeling of quantum cascade lasers},\ }\href@noop {} {\bibfield
  {journal} {\bibinfo  {journal} {J. Appl. Phys.}\ }\textbf {\bibinfo {volume}
  {122}},\ \bibinfo {pages} {133105} (\bibinfo {year} {2017})}\BibitemShut
  {NoStop}%
\bibitem [{\citenamefont {Riesch}\ \emph {et~al.}(2018)\citenamefont {Riesch},
  \citenamefont {Tchipev}, \citenamefont {Senninger}, \citenamefont
  {Bungartz},\ and\ \citenamefont {Jirauschek}}]{riesch2018performance}%
  \BibitemOpen
  \bibfield  {author} {\bibinfo {author} {\bibfnamefont {M.}~\bibnamefont
  {Riesch}}, \bibinfo {author} {\bibfnamefont {N.}~\bibnamefont {Tchipev}},
  \bibinfo {author} {\bibfnamefont {S.}~\bibnamefont {Senninger}}, \bibinfo
  {author} {\bibfnamefont {H.-J.}\ \bibnamefont {Bungartz}},\ and\ \bibinfo
  {author} {\bibfnamefont {C.}~\bibnamefont {Jirauschek}},\ }\bibfield  {title}
  {\bibinfo {title} {Performance evaluation of numerical methods for the
  maxwell--liouville--von neumann equations},\ }\href@noop {} {\bibfield
  {journal} {\bibinfo  {journal} {Optical and Quantum Electronics}\ }\textbf
  {\bibinfo {volume} {50}},\ \bibinfo {pages} {112} (\bibinfo {year}
  {2018})}\BibitemShut {NoStop}%
\bibitem [{\citenamefont {Riesch}\ and\ \citenamefont
  {Jirauschek}(2019)}]{riesch2019analyzing}%
  \BibitemOpen
  \bibfield  {author} {\bibinfo {author} {\bibfnamefont {M.}~\bibnamefont
  {Riesch}}\ and\ \bibinfo {author} {\bibfnamefont {C.}~\bibnamefont
  {Jirauschek}},\ }\bibfield  {title} {\bibinfo {title} {Analyzing the
  positivity preservation of numerical methods for the liouville-von neumann
  equation},\ }\href@noop {} {\bibfield  {journal} {\bibinfo  {journal} {J.
  Comput. Phys.}\ }\textbf {\bibinfo {volume} {390}},\ \bibinfo {pages} {290}
  (\bibinfo {year} {2019})}\BibitemShut {NoStop}%
\bibitem [{\citenamefont {Demic}\ \emph {et~al.}(2019)\citenamefont {Demic},
  \citenamefont {Ikonic}, \citenamefont {Kelsall},\ and\ \citenamefont
  {Indjin}}]{Demic2019_AIPAdvances}%
  \BibitemOpen
  \bibfield  {author} {\bibinfo {author} {\bibfnamefont {A.}~\bibnamefont
  {Demic}}, \bibinfo {author} {\bibfnamefont {Z.}~\bibnamefont {Ikonic}},
  \bibinfo {author} {\bibfnamefont {R.~W.}\ \bibnamefont {Kelsall}},\ and\
  \bibinfo {author} {\bibfnamefont {D.}~\bibnamefont {Indjin}},\ }\bibfield
  {title} {\bibinfo {title} {Density matrix superoperator for periodic quantum
  systems and its application to quantum cascade laser structures},\ }\href
  {https://doi.org/10.1063/1.5095246} {\bibfield  {journal} {\bibinfo
  {journal} {AIP Advances}\ }\textbf {\bibinfo {volume} {9}},\ \bibinfo {pages}
  {095019} (\bibinfo {year} {2019})},\ \Eprint
  {https://arxiv.org/abs/https://doi.org/10.1063/1.5095246}
  {https://doi.org/10.1063/1.5095246} \BibitemShut {NoStop}%
\bibitem [{\citenamefont {Jirauschek}\ \emph {et~al.}(2019)\citenamefont
  {Jirauschek}, \citenamefont {Riesch},\ and\ \citenamefont
  {Tzenov}}]{Jirauschek2019_MaxwellBloch_AdvancedTheoSimulation}%
  \BibitemOpen
  \bibfield  {author} {\bibinfo {author} {\bibfnamefont {C.}~\bibnamefont
  {Jirauschek}}, \bibinfo {author} {\bibfnamefont {M.}~\bibnamefont {Riesch}},\
  and\ \bibinfo {author} {\bibfnamefont {P.}~\bibnamefont {Tzenov}},\
  }\bibfield  {title} {\bibinfo {title} {Optoelectronic device simulations
  based on macroscopic maxwell–bloch equations},\ }\href
  {https://doi.org/10.1002/adts.201900018} {\bibfield  {journal} {\bibinfo
  {journal} {Advanced Theory and Simulations}\ }\textbf {\bibinfo {volume}
  {2}},\ \bibinfo {pages} {1900018} (\bibinfo {year} {2019})},\ \Eprint
  {https://arxiv.org/abs/https://onlinelibrary.wiley.com/doi/pdf/10.1002/adts.201900018}
  {https://onlinelibrary.wiley.com/doi/pdf/10.1002/adts.201900018} \BibitemShut
  {NoStop}%
\bibitem [{\citenamefont {Lee}\ and\ \citenamefont
  {Wacker}(2002)}]{lee_PRB_2002}%
  \BibitemOpen
  \bibfield  {author} {\bibinfo {author} {\bibfnamefont {S.-C.}\ \bibnamefont
  {Lee}}\ and\ \bibinfo {author} {\bibfnamefont {A.}~\bibnamefont {Wacker}},\
  }\bibfield  {title} {\bibinfo {title} {Nonequilibrium green’s function
  theory for transport and gain properties of quantum cascade structures},\
  }\href@noop {} {\bibfield  {journal} {\bibinfo  {journal} {Phys. Rev. B.}\
  }\textbf {\bibinfo {volume} {66}},\ \bibinfo {pages} {245314} (\bibinfo
  {year} {2002})}\BibitemShut {NoStop}%
\bibitem [{\citenamefont {Bugajski}\ \emph {et~al.}(2014)\citenamefont
  {Bugajski}, \citenamefont {Gutowski}, \citenamefont {Karbownik},
  \citenamefont {Kolek}, \citenamefont {Ha{\l}da{\'s}}, \citenamefont
  {Pier{\'s}ci{\'n}ski}, \citenamefont {Pier{\'s}ci{\'n}ska}, \citenamefont
  {Kubacka-Traczyk}, \citenamefont {Sankowska}, \citenamefont {Trajnerowicz}
  \emph {et~al.}}]{bugajski_PSS_2014}%
  \BibitemOpen
  \bibfield  {author} {\bibinfo {author} {\bibfnamefont {M.}~\bibnamefont
  {Bugajski}}, \bibinfo {author} {\bibfnamefont {P.}~\bibnamefont {Gutowski}},
  \bibinfo {author} {\bibfnamefont {P.}~\bibnamefont {Karbownik}}, \bibinfo
  {author} {\bibfnamefont {A.}~\bibnamefont {Kolek}}, \bibinfo {author}
  {\bibfnamefont {G.}~\bibnamefont {Ha{\l}da{\'s}}}, \bibinfo {author}
  {\bibfnamefont {K.}~\bibnamefont {Pier{\'s}ci{\'n}ski}}, \bibinfo {author}
  {\bibfnamefont {D.}~\bibnamefont {Pier{\'s}ci{\'n}ska}}, \bibinfo {author}
  {\bibfnamefont {J.}~\bibnamefont {Kubacka-Traczyk}}, \bibinfo {author}
  {\bibfnamefont {I.}~\bibnamefont {Sankowska}}, \bibinfo {author}
  {\bibfnamefont {A.}~\bibnamefont {Trajnerowicz}}, \emph {et~al.},\ }\bibfield
   {title} {\bibinfo {title} {Mid-ir quantum cascade lasers: Device technology
  and non-equilibrium green's function modeling of electro-optical
  characteristics},\ }\href@noop {} {\bibfield  {journal} {\bibinfo  {journal}
  {Physica Status Solidi B}\ }\textbf {\bibinfo {volume} {251}},\ \bibinfo
  {pages} {1144} (\bibinfo {year} {2014})}\BibitemShut {NoStop}%
\bibitem [{\citenamefont {Wacker}\ \emph {et~al.}(2013)\citenamefont {Wacker},
  \citenamefont {Lindskog},\ and\ \citenamefont {Winge}}]{wacker_JSTQE_2013}%
  \BibitemOpen
  \bibfield  {author} {\bibinfo {author} {\bibfnamefont {A.}~\bibnamefont
  {Wacker}}, \bibinfo {author} {\bibfnamefont {M.}~\bibnamefont {Lindskog}},\
  and\ \bibinfo {author} {\bibfnamefont {D.~O.}\ \bibnamefont {Winge}},\
  }\bibfield  {title} {\bibinfo {title} {Nonequilibrium green’s function
  model for simulation of quantum cascade laser devices under operating
  conditions},\ }\href@noop {} {\bibfield  {journal} {\bibinfo  {journal} {IEEE
  J. Sel. Top. Quantum Electron.}\ }\textbf {\bibinfo {volume} {19}},\ \bibinfo
  {pages} {1} (\bibinfo {year} {2013})}\BibitemShut {NoStop}%
\bibitem [{\citenamefont {Lindskog}\ \emph {et~al.}(2014)\citenamefont
  {Lindskog}, \citenamefont {Wolf}, \citenamefont {Trinite}, \citenamefont
  {Liverini}, \citenamefont {Faist}, \citenamefont {Maisons}, \citenamefont
  {Carras}, \citenamefont {Aidam}, \citenamefont {Ostendorf},\ and\
  \citenamefont {Wacker}}]{lindskog_APL_2014}%
  \BibitemOpen
  \bibfield  {author} {\bibinfo {author} {\bibfnamefont {M.}~\bibnamefont
  {Lindskog}}, \bibinfo {author} {\bibfnamefont {J.}~\bibnamefont {Wolf}},
  \bibinfo {author} {\bibfnamefont {V.}~\bibnamefont {Trinite}}, \bibinfo
  {author} {\bibfnamefont {V.}~\bibnamefont {Liverini}}, \bibinfo {author}
  {\bibfnamefont {J.}~\bibnamefont {Faist}}, \bibinfo {author} {\bibfnamefont
  {G.}~\bibnamefont {Maisons}}, \bibinfo {author} {\bibfnamefont
  {M.}~\bibnamefont {Carras}}, \bibinfo {author} {\bibfnamefont
  {R.}~\bibnamefont {Aidam}}, \bibinfo {author} {\bibfnamefont
  {R.}~\bibnamefont {Ostendorf}},\ and\ \bibinfo {author} {\bibfnamefont
  {A.}~\bibnamefont {Wacker}},\ }\bibfield  {title} {\bibinfo {title}
  {Comparative analysis of quantum cascade laser modeling based on density
  matrices and non-equilibrium green's functions},\ }\href@noop {} {\bibfield
  {journal} {\bibinfo  {journal} {Appl. Phys. Lett.}\ }\textbf {\bibinfo
  {volume} {105}},\ \bibinfo {pages} {103106} (\bibinfo {year}
  {2014})}\BibitemShut {NoStop}%
\bibitem [{\citenamefont {Kolek}\ \emph {et~al.}(2014)\citenamefont {Kolek},
  \citenamefont {Ha{\l}da{\'s}}, \citenamefont {Bugajski}, \citenamefont
  {Pier{\'s}ci{\'n}ski},\ and\ \citenamefont {Gutowski}}]{kolek_JSTQE_2015}%
  \BibitemOpen
  \bibfield  {author} {\bibinfo {author} {\bibfnamefont {A.}~\bibnamefont
  {Kolek}}, \bibinfo {author} {\bibfnamefont {G.}~\bibnamefont
  {Ha{\l}da{\'s}}}, \bibinfo {author} {\bibfnamefont {M.}~\bibnamefont
  {Bugajski}}, \bibinfo {author} {\bibfnamefont {K.}~\bibnamefont
  {Pier{\'s}ci{\'n}ski}},\ and\ \bibinfo {author} {\bibfnamefont
  {P.}~\bibnamefont {Gutowski}},\ }\bibfield  {title} {\bibinfo {title} {Impact
  of injector doping on threshold current of mid-infrared quantum cascade
  laser--non-equilibrium green’s function analysis},\ }\href@noop {}
  {\bibfield  {journal} {\bibinfo  {journal} {IEEE J. Sel. Top. Quantum
  Electron.}\ }\textbf {\bibinfo {volume} {21}},\ \bibinfo {pages} {124}
  (\bibinfo {year} {2014})}\BibitemShut {NoStop}%
\bibitem [{\citenamefont {Ha{\l}da{\'s}}(2019)}]{haldas2019implementation}%
  \BibitemOpen
  \bibfield  {author} {\bibinfo {author} {\bibfnamefont {G.}~\bibnamefont
  {Ha{\l}da{\'s}}},\ }\bibfield  {title} {\bibinfo {title} {Implementation of
  non-uniform mesh in non-equilibrium green’s function simulations of quantum
  cascade lasers},\ }\href@noop {} {\bibfield  {journal} {\bibinfo  {journal}
  {J. Comput. Electron.}\ ,\ \bibinfo {pages} {1}} (\bibinfo {year}
  {2019})}\BibitemShut {NoStop}%
\bibitem [{\citenamefont {Kolek}(2019)}]{kolek2019implementation}%
  \BibitemOpen
  \bibfield  {author} {\bibinfo {author} {\bibfnamefont {A.}~\bibnamefont
  {Kolek}},\ }\bibfield  {title} {\bibinfo {title} {Implementation of
  light--matter interaction in negf simulations of qcl},\ }\href@noop {}
  {\bibfield  {journal} {\bibinfo  {journal} {Optical and Quantum Electronics}\
  }\textbf {\bibinfo {volume} {51}},\ \bibinfo {pages} {171} (\bibinfo {year}
  {2019})}\BibitemShut {NoStop}%
\bibitem [{\citenamefont {Grange}\ \emph {et~al.}(2019)\citenamefont {Grange},
  \citenamefont {Stark}, \citenamefont {Scalari}, \citenamefont {Faist},
  \citenamefont {Persichetti}, \citenamefont {Di~Gaspare}, \citenamefont
  {De~Seta}, \citenamefont {Ortolani}, \citenamefont {Paul}, \citenamefont
  {Capellini} \emph {et~al.}}]{grange2019room}%
  \BibitemOpen
  \bibfield  {author} {\bibinfo {author} {\bibfnamefont {T.}~\bibnamefont
  {Grange}}, \bibinfo {author} {\bibfnamefont {D.}~\bibnamefont {Stark}},
  \bibinfo {author} {\bibfnamefont {G.}~\bibnamefont {Scalari}}, \bibinfo
  {author} {\bibfnamefont {J.}~\bibnamefont {Faist}}, \bibinfo {author}
  {\bibfnamefont {L.}~\bibnamefont {Persichetti}}, \bibinfo {author}
  {\bibfnamefont {L.}~\bibnamefont {Di~Gaspare}}, \bibinfo {author}
  {\bibfnamefont {M.}~\bibnamefont {De~Seta}}, \bibinfo {author} {\bibfnamefont
  {M.}~\bibnamefont {Ortolani}}, \bibinfo {author} {\bibfnamefont {D.~J.}\
  \bibnamefont {Paul}}, \bibinfo {author} {\bibfnamefont {G.}~\bibnamefont
  {Capellini}}, \emph {et~al.},\ }\bibfield  {title} {\bibinfo {title} {Room
  temperature operation of n-type ge/sige terahertz quantum cascade lasers
  predicted by non-equilibrium green's functions},\ }\href@noop {} {\bibfield
  {journal} {\bibinfo  {journal} {Appl. Phys. Lett.}\ }\textbf {\bibinfo
  {volume} {114}},\ \bibinfo {pages} {111102} (\bibinfo {year}
  {2019})}\BibitemShut {NoStop}%
\bibitem [{\citenamefont {Callebaut}\ and\ \citenamefont
  {Hu}(2005)}]{callebaut_JAP_2005}%
  \BibitemOpen
  \bibfield  {author} {\bibinfo {author} {\bibfnamefont {H.}~\bibnamefont
  {Callebaut}}\ and\ \bibinfo {author} {\bibfnamefont {Q.}~\bibnamefont {Hu}},\
  }\bibfield  {title} {\bibinfo {title} {Importance of coherence for electron
  transport in terahertz quantum cascade lasers},\ }\href@noop {} {\bibfield
  {journal} {\bibinfo  {journal} {J. Appl. Phys.}\ }\textbf {\bibinfo {volume}
  {98}},\ \bibinfo {pages} {104505} (\bibinfo {year} {2005})}\BibitemShut
  {NoStop}%
\bibitem [{\citenamefont {Jonasson}\ \emph
  {et~al.}(2016{\natexlab{a}})\citenamefont {Jonasson}, \citenamefont {Mei},
  \citenamefont {Karimi}, \citenamefont {Kirch}, \citenamefont {Botez},
  \citenamefont {Mawst},\ and\ \citenamefont {Knezevic}}]{jonasson_PHT_2016}%
  \BibitemOpen
  \bibfield  {author} {\bibinfo {author} {\bibfnamefont {O.}~\bibnamefont
  {Jonasson}}, \bibinfo {author} {\bibfnamefont {S.}~\bibnamefont {Mei}},
  \bibinfo {author} {\bibfnamefont {F.}~\bibnamefont {Karimi}}, \bibinfo
  {author} {\bibfnamefont {J.}~\bibnamefont {Kirch}}, \bibinfo {author}
  {\bibfnamefont {D.}~\bibnamefont {Botez}}, \bibinfo {author} {\bibfnamefont
  {L.}~\bibnamefont {Mawst}},\ and\ \bibinfo {author} {\bibfnamefont
  {I.}~\bibnamefont {Knezevic}},\ }\bibfield  {title} {\bibinfo {title}
  {Quantum transport simulation of high-power 4.6-$\mu$m quantum cascade
  lasers},\ }in\ \href@noop {} {\emph {\bibinfo {booktitle} {Photonics}}},\
  Vol.~\bibinfo {volume} {3}\ (\bibinfo {organization} {Multidisciplinary
  Digital Publishing Institute},\ \bibinfo {year} {2016})\ p.~\bibinfo {pages}
  {38}\BibitemShut {NoStop}%
\bibitem [{\citenamefont {Iotti}\ and\ \citenamefont
  {Rossi}(2005)}]{iotti_RPP_2005}%
  \BibitemOpen
  \bibfield  {author} {\bibinfo {author} {\bibfnamefont {R.~C.}\ \bibnamefont
  {Iotti}}\ and\ \bibinfo {author} {\bibfnamefont {F.}~\bibnamefont {Rossi}},\
  }\bibfield  {title} {\bibinfo {title} {Microscopic theory of
  semiconductor-based optoelectronic devices},\ }\href@noop {} {\bibfield
  {journal} {\bibinfo  {journal} {Rep. Prog. Phys.}\ }\textbf {\bibinfo
  {volume} {68}},\ \bibinfo {pages} {2533} (\bibinfo {year}
  {2005})}\BibitemShut {NoStop}%
\bibitem [{\citenamefont {Jonasson}\ \emph
  {et~al.}(2016{\natexlab{b}})\citenamefont {Jonasson}, \citenamefont
  {Karimi},\ and\ \citenamefont {Knezevic}}]{jonasson_JCEL_2016}%
  \BibitemOpen
  \bibfield  {author} {\bibinfo {author} {\bibfnamefont {O.}~\bibnamefont
  {Jonasson}}, \bibinfo {author} {\bibfnamefont {F.}~\bibnamefont {Karimi}},\
  and\ \bibinfo {author} {\bibfnamefont {I.}~\bibnamefont {Knezevic}},\
  }\bibfield  {title} {\bibinfo {title} {Partially coherent electron transport
  in terahertz quantum cascade lasers based on a markovian master equation for
  the density matrix},\ }\href@noop {} {\bibfield  {journal} {\bibinfo
  {journal} {J. Comput. Electron.}\ }\textbf {\bibinfo {volume} {15}},\
  \bibinfo {pages} {1192} (\bibinfo {year} {2016}{\natexlab{b}})}\BibitemShut
  {NoStop}%
\bibitem [{\citenamefont {Lindblad}(1976)}]{Lindblad1976}%
  \BibitemOpen
  \bibfield  {author} {\bibinfo {author} {\bibfnamefont {G.}~\bibnamefont
  {Lindblad}},\ }\bibfield  {title} {\bibinfo {title} {On the generators of
  quantum dynamical semigroups},\ }\href {https://doi.org/10.1007/BF01608499}
  {\bibfield  {journal} {\bibinfo  {journal} {Commun. Math. Phys.}\ }\textbf
  {\bibinfo {volume} {48}},\ \bibinfo {pages} {119} (\bibinfo {year} {1976})},\
  \Eprint {https://arxiv.org/abs/https://doi.org/10.1007/BF01608499}
  {https://doi.org/10.1007/BF01608499} \BibitemShut {NoStop}%
\bibitem [{\citenamefont {Breuer}\ \emph {et~al.}(2002)\citenamefont {Breuer},
  \citenamefont {Petruccione} \emph {et~al.}}]{breuer_2002}%
  \BibitemOpen
  \bibfield  {author} {\bibinfo {author} {\bibfnamefont {H.-P.}\ \bibnamefont
  {Breuer}}, \bibinfo {author} {\bibfnamefont {F.}~\bibnamefont {Petruccione}},
  \emph {et~al.},\ }\href@noop {} {\emph {\bibinfo {title} {The theory of open
  quantum systems}}}\ (\bibinfo  {publisher} {Oxford University Press on
  Demand},\ \bibinfo {year} {2002})\BibitemShut {NoStop}%
\bibitem [{\citenamefont {Knezevic}\ and\ \citenamefont
  {Novakovic}(2013)}]{knezevic_JCEL_2013}%
  \BibitemOpen
  \bibfield  {author} {\bibinfo {author} {\bibfnamefont {I.}~\bibnamefont
  {Knezevic}}\ and\ \bibinfo {author} {\bibfnamefont {B.}~\bibnamefont
  {Novakovic}},\ }\bibfield  {title} {\bibinfo {title} {Time-dependent
  transport in open systems based on quantum master equations},\ }\href@noop {}
  {\bibfield  {journal} {\bibinfo  {journal} {J. Comput. Electron.}\ }\textbf
  {\bibinfo {volume} {12}},\ \bibinfo {pages} {363} (\bibinfo {year}
  {2013})}\BibitemShut {NoStop}%
\bibitem [{\citenamefont {Karimi}(2017)}]{KarimiDissertation2017}%
  \BibitemOpen
  \bibfield  {author} {\bibinfo {author} {\bibfnamefont {F.}~\bibnamefont
  {Karimi}},\ }\emph {\bibinfo {title} {Quantum transport theory on optical and
  plasmonic properties of nanomaterials}},\ \href@noop {} {Ph.D. thesis},\
  \bibinfo  {school} {University of Wisconsin -- Madison} (\bibinfo {year}
  {2017})\BibitemShut {NoStop}%
\bibitem [{\citenamefont {Karimi}\ \emph {et~al.}(2016)\citenamefont {Karimi},
  \citenamefont {Davoody},\ and\ \citenamefont {Knezevic}}]{Karimi_PRB_2016}%
  \BibitemOpen
  \bibfield  {author} {\bibinfo {author} {\bibfnamefont {F.}~\bibnamefont
  {Karimi}}, \bibinfo {author} {\bibfnamefont {A.~H.}\ \bibnamefont
  {Davoody}},\ and\ \bibinfo {author} {\bibfnamefont {I.}~\bibnamefont
  {Knezevic}},\ }\bibfield  {title} {\bibinfo {title} {Dielectric function and
  plasmons in graphene: A self-consistent-field calculation within a markovian
  master equation formalism},\ }\href
  {https://doi.org/10.1103/PhysRevB.93.205421} {\bibfield  {journal} {\bibinfo
  {journal} {Phys. Rev. B}\ }\textbf {\bibinfo {volume} {93}},\ \bibinfo
  {pages} {205421} (\bibinfo {year} {2016})}\BibitemShut {NoStop}%
\bibitem [{\citenamefont {Fischetti}(1999)}]{fischetti1999}%
  \BibitemOpen
  \bibfield  {author} {\bibinfo {author} {\bibfnamefont {M.}~\bibnamefont
  {Fischetti}},\ }\bibfield  {title} {\bibinfo {title} {Master-equation
  approach to the study of electronic transport in small semiconductor
  devices},\ }\href@noop {} {\bibfield  {journal} {\bibinfo  {journal} {Phys.
  Rev. B.}\ }\textbf {\bibinfo {volume} {59}},\ \bibinfo {pages} {4901}
  (\bibinfo {year} {1999})}\BibitemShut {NoStop}%
\bibitem [{\citenamefont {Haug}\ and\ \citenamefont {Koch}(2009)}]{haug_2004}%
  \BibitemOpen
  \bibfield  {author} {\bibinfo {author} {\bibfnamefont {H.}~\bibnamefont
  {Haug}}\ and\ \bibinfo {author} {\bibfnamefont {S.~W.}\ \bibnamefont
  {Koch}},\ }\href@noop {} {\emph {\bibinfo {title} {Quantum Theory of the
  Optical and Electronic Properties of Semiconductors: Fivth Edition}}}\
  (\bibinfo  {publisher} {World Scientific Publishing Company},\ \bibinfo
  {year} {2009})\BibitemShut {NoStop}%
\bibitem [{\citenamefont {Calecki}\ \emph {et~al.}(1984)\citenamefont
  {Calecki}, \citenamefont {Palmier},\ and\ \citenamefont
  {Chomette}}]{calecki_JPC_1984}%
  \BibitemOpen
  \bibfield  {author} {\bibinfo {author} {\bibfnamefont {D.}~\bibnamefont
  {Calecki}}, \bibinfo {author} {\bibfnamefont {J.}~\bibnamefont {Palmier}},\
  and\ \bibinfo {author} {\bibfnamefont {A.}~\bibnamefont {Chomette}},\
  }\bibfield  {title} {\bibinfo {title} {Hopping conduction in multiquantum
  well structures},\ }\href@noop {} {\bibfield  {journal} {\bibinfo  {journal}
  {J. Phys. C: Solid state}\ }\textbf {\bibinfo {volume} {17}},\ \bibinfo
  {pages} {5017} (\bibinfo {year} {1984})}\BibitemShut {NoStop}%
\bibitem [{\citenamefont {Gelmont}\ \emph {et~al.}(1996)\citenamefont
  {Gelmont}, \citenamefont {Gorfinkel},\ and\ \citenamefont
  {Luryi}}]{gelmont_APL_1996}%
  \BibitemOpen
  \bibfield  {author} {\bibinfo {author} {\bibfnamefont {B.}~\bibnamefont
  {Gelmont}}, \bibinfo {author} {\bibfnamefont {V.}~\bibnamefont {Gorfinkel}},\
  and\ \bibinfo {author} {\bibfnamefont {S.}~\bibnamefont {Luryi}},\ }\bibfield
   {title} {\bibinfo {title} {Theory of the spectral line shape and gain in
  quantum wells with intersubband transitions},\ }\href@noop {} {\bibfield
  {journal} {\bibinfo  {journal} {Appl. Phys. Lett.}\ }\textbf {\bibinfo
  {volume} {68}},\ \bibinfo {pages} {2171} (\bibinfo {year}
  {1996})}\BibitemShut {NoStop}%
\bibitem [{\citenamefont {Jirauschek}\ and\ \citenamefont
  {Lugli}(2009)}]{jirauschek_JAP_2009}%
  \BibitemOpen
  \bibfield  {author} {\bibinfo {author} {\bibfnamefont {C.}~\bibnamefont
  {Jirauschek}}\ and\ \bibinfo {author} {\bibfnamefont {P.}~\bibnamefont
  {Lugli}},\ }\bibfield  {title} {\bibinfo {title} {Monte-carlo-based spectral
  gain analysis for terahertz quantum cascade lasers},\ }\href@noop {}
  {\bibfield  {journal} {\bibinfo  {journal} {J. Appl. Phys.}\ }\textbf
  {\bibinfo {volume} {105}},\ \bibinfo {pages} {123102} (\bibinfo {year}
  {2009})}\BibitemShut {NoStop}%
\bibitem [{\citenamefont {Saad}(2003)}]{saad_iter_2003}%
  \BibitemOpen
  \bibfield  {author} {\bibinfo {author} {\bibfnamefont {Y.}~\bibnamefont
  {Saad}},\ }\href@noop {} {\emph {\bibinfo {title} {Iterative methods for
  sparse linear systems}}},\ Vol.~\bibinfo {volume} {82}\ (\bibinfo
  {publisher} {siam},\ \bibinfo {year} {2003})\BibitemShut {NoStop}%
\bibitem [{\citenamefont {Bismuto}\ \emph {et~al.}(2010)\citenamefont
  {Bismuto}, \citenamefont {Terazzi}, \citenamefont {Beck},\ and\ \citenamefont
  {Faist}}]{bismuto_APL_2010}%
  \BibitemOpen
  \bibfield  {author} {\bibinfo {author} {\bibfnamefont {A.}~\bibnamefont
  {Bismuto}}, \bibinfo {author} {\bibfnamefont {R.}~\bibnamefont {Terazzi}},
  \bibinfo {author} {\bibfnamefont {M.}~\bibnamefont {Beck}},\ and\ \bibinfo
  {author} {\bibfnamefont {J.}~\bibnamefont {Faist}},\ }\bibfield  {title}
  {\bibinfo {title} {Electrically tunable, high performance quantum cascade
  laser},\ }\href@noop {} {\bibfield  {journal} {\bibinfo  {journal} {Appl.
  Phys. Lett.}\ }\textbf {\bibinfo {volume} {96}},\ \bibinfo {pages} {141105}
  (\bibinfo {year} {2010})}\BibitemShut {NoStop}%
\bibitem [{\citenamefont {M{\'a}ty{\'a}s}\ \emph {et~al.}(2011)\citenamefont
  {M{\'a}ty{\'a}s}, \citenamefont {Lugli},\ and\ \citenamefont
  {Jirauschek}}]{matyas_JAP_2011}%
  \BibitemOpen
  \bibfield  {author} {\bibinfo {author} {\bibfnamefont {A.}~\bibnamefont
  {M{\'a}ty{\'a}s}}, \bibinfo {author} {\bibfnamefont {P.}~\bibnamefont
  {Lugli}},\ and\ \bibinfo {author} {\bibfnamefont {C.}~\bibnamefont
  {Jirauschek}},\ }\bibfield  {title} {\bibinfo {title} {Photon-induced carrier
  transport in high efficiency midinfrared quantum cascade lasers},\
  }\href@noop {} {\bibfield  {journal} {\bibinfo  {journal} {J. Appl. Phys.}\
  }\textbf {\bibinfo {volume} {110}},\ \bibinfo {pages} {013108} (\bibinfo
  {year} {2011})}\BibitemShut {NoStop}%
\bibitem [{\citenamefont {Iotti}\ and\ \citenamefont
  {Rossi}(2001{\natexlab{b}})}]{iotti_APL_2001}%
  \BibitemOpen
  \bibfield  {author} {\bibinfo {author} {\bibfnamefont {R.~C.}\ \bibnamefont
  {Iotti}}\ and\ \bibinfo {author} {\bibfnamefont {F.}~\bibnamefont {Rossi}},\
  }\bibfield  {title} {\bibinfo {title} {Carrier thermalization versus
  phonon-assisted relaxation in quantum-cascade lasers: A monte carlo
  approach},\ }\href@noop {} {\bibfield  {journal} {\bibinfo  {journal} {Appl.
  Phys. Lett.}\ }\textbf {\bibinfo {volume} {78}},\ \bibinfo {pages} {2902}
  (\bibinfo {year} {2001}{\natexlab{b}})}\BibitemShut {NoStop}%
\bibitem [{\citenamefont {Spagnolo}\ \emph {et~al.}(2004)\citenamefont
  {Spagnolo}, \citenamefont {Scamarcio}, \citenamefont {Page},\ and\
  \citenamefont {Sirtori}}]{spagnolo_APL_2004}%
  \BibitemOpen
  \bibfield  {author} {\bibinfo {author} {\bibfnamefont {V.}~\bibnamefont
  {Spagnolo}}, \bibinfo {author} {\bibfnamefont {G.}~\bibnamefont {Scamarcio}},
  \bibinfo {author} {\bibfnamefont {H.}~\bibnamefont {Page}},\ and\ \bibinfo
  {author} {\bibfnamefont {C.}~\bibnamefont {Sirtori}},\ }\bibfield  {title}
  {\bibinfo {title} {Simultaneous measurement of the electronic and lattice
  temperatures in gaas/al 0.45 ga 0.55 as quantum-cascade lasers: influence on
  the optical performance},\ }\href@noop {} {\bibfield  {journal} {\bibinfo
  {journal} {Appl. Phys. Lett.}\ }\textbf {\bibinfo {volume} {84}},\ \bibinfo
  {pages} {3690} (\bibinfo {year} {2004})}\BibitemShut {NoStop}%
\bibitem [{\citenamefont {Page}\ \emph {et~al.}(2001)\citenamefont {Page},
  \citenamefont {Becker}, \citenamefont {Robertson}, \citenamefont {Glastre},
  \citenamefont {Ortiz},\ and\ \citenamefont {Sirtori}}]{page2001}%
  \BibitemOpen
  \bibfield  {author} {\bibinfo {author} {\bibfnamefont {H.}~\bibnamefont
  {Page}}, \bibinfo {author} {\bibfnamefont {C.}~\bibnamefont {Becker}},
  \bibinfo {author} {\bibfnamefont {A.}~\bibnamefont {Robertson}}, \bibinfo
  {author} {\bibfnamefont {G.}~\bibnamefont {Glastre}}, \bibinfo {author}
  {\bibfnamefont {V.}~\bibnamefont {Ortiz}},\ and\ \bibinfo {author}
  {\bibfnamefont {C.}~\bibnamefont {Sirtori}},\ }\bibfield  {title} {\bibinfo
  {title} {300 k operation of a gaas-based quantum-cascade laser at $\lambda$ 9
  $\mu$m},\ }\href@noop {} {\bibfield  {journal} {\bibinfo  {journal} {Appl.
  Phys. Lett.}\ }\textbf {\bibinfo {volume} {78}},\ \bibinfo {pages} {3529}
  (\bibinfo {year} {2001})}\BibitemShut {NoStop}%
\bibitem [{\citenamefont {Lundstrom}(2009)}]{lundstrom2000}%
  \BibitemOpen
  \bibfield  {author} {\bibinfo {author} {\bibfnamefont {M.}~\bibnamefont
  {Lundstrom}},\ }\href@noop {} {\emph {\bibinfo {title} {Fundamentals of
  carrier transport}}}\ (\bibinfo  {publisher} {Cambridge university press},\
  \bibinfo {year} {2009})\BibitemShut {NoStop}%
\bibitem [{\citenamefont {Bai}\ \emph {et~al.}(2008)\citenamefont {Bai},
  \citenamefont {Darvish}, \citenamefont {Slivken}, \citenamefont {Zhang},
  \citenamefont {Evans}, \citenamefont {Nguyen},\ and\ \citenamefont
  {Razeghi}}]{bai_APL_2008}%
  \BibitemOpen
  \bibfield  {author} {\bibinfo {author} {\bibfnamefont {Y.}~\bibnamefont
  {Bai}}, \bibinfo {author} {\bibfnamefont {S.}~\bibnamefont {Darvish}},
  \bibinfo {author} {\bibfnamefont {S.}~\bibnamefont {Slivken}}, \bibinfo
  {author} {\bibfnamefont {W.}~\bibnamefont {Zhang}}, \bibinfo {author}
  {\bibfnamefont {A.}~\bibnamefont {Evans}}, \bibinfo {author} {\bibfnamefont
  {J.}~\bibnamefont {Nguyen}},\ and\ \bibinfo {author} {\bibfnamefont
  {M.}~\bibnamefont {Razeghi}},\ }\bibfield  {title} {\bibinfo {title} {Room
  temperature continuous wave operation of quantum cascade lasers with
  watt-level optical power},\ }\href@noop {} {\bibfield  {journal} {\bibinfo
  {journal} {Appl. Phys. Lett.}\ }\textbf {\bibinfo {volume} {92}},\ \bibinfo
  {pages} {101105} (\bibinfo {year} {2008})}\BibitemShut {NoStop}%
\bibitem [{\citenamefont {Evans}\ \emph {et~al.}(2007)\citenamefont {Evans},
  \citenamefont {Darvish}, \citenamefont {Slivken}, \citenamefont {Nguyen},
  \citenamefont {Bai},\ and\ \citenamefont {Razeghi}}]{evans_APL_2007}%
  \BibitemOpen
  \bibfield  {author} {\bibinfo {author} {\bibfnamefont {A.}~\bibnamefont
  {Evans}}, \bibinfo {author} {\bibfnamefont {S.}~\bibnamefont {Darvish}},
  \bibinfo {author} {\bibfnamefont {S.}~\bibnamefont {Slivken}}, \bibinfo
  {author} {\bibfnamefont {J.}~\bibnamefont {Nguyen}}, \bibinfo {author}
  {\bibfnamefont {Y.}~\bibnamefont {Bai}},\ and\ \bibinfo {author}
  {\bibfnamefont {M.}~\bibnamefont {Razeghi}},\ }\bibfield  {title} {\bibinfo
  {title} {Buried heterostructure quantum cascade lasers with high
  continuous-wave wall plug efficiency},\ }\href@noop {} {\bibfield  {journal}
  {\bibinfo  {journal} {Appl. Phys. Lett.}\ }\textbf {\bibinfo {volume} {91}},\
  \bibinfo {pages} {071101} (\bibinfo {year} {2007})}\BibitemShut {NoStop}%
\bibitem [{\citenamefont {Iotti}\ and\ \citenamefont
  {Rossi}(2004)}]{iotti_SST_2004}%
  \BibitemOpen
  \bibfield  {author} {\bibinfo {author} {\bibfnamefont {R.~C.}\ \bibnamefont
  {Iotti}}\ and\ \bibinfo {author} {\bibfnamefont {F.}~\bibnamefont {Rossi}},\
  }\bibfield  {title} {\bibinfo {title} {Microscopic theory of quantum-cascade
  lasers},\ }\href@noop {} {\bibfield  {journal} {\bibinfo  {journal}
  {Semiconductor Science and Technology}\ }\textbf {\bibinfo {volume} {19}},\
  \bibinfo {pages} {S323} (\bibinfo {year} {2004})}\BibitemShut {NoStop}%
\bibitem [{\citenamefont {Bastard}(1990)}]{bastard_1988_wave}%
  \BibitemOpen
  \bibfield  {author} {\bibinfo {author} {\bibfnamefont {G.}~\bibnamefont
  {Bastard}},\ }\bibfield  {title} {\bibinfo {title} {Wave mechanics applied to
  semiconductor heterostructures},\ }\href@noop {} {\  (\bibinfo {year}
  {1990})}\BibitemShut {NoStop}%
\bibitem [{\citenamefont {Chuang}\ and\ \citenamefont
  {Chuang}(1995)}]{Chuang1995}%
  \BibitemOpen
  \bibfield  {author} {\bibinfo {author} {\bibfnamefont {S.~L.}\ \bibnamefont
  {Chuang}}\ and\ \bibinfo {author} {\bibfnamefont {S.~L.}\ \bibnamefont
  {Chuang}},\ }\bibfield  {title} {\bibinfo {title} {Physics of optoelectronic
  devices},\ }\href@noop {} {\  (\bibinfo {year} {1995})}\BibitemShut {NoStop}%
\bibitem [{\citenamefont {Nelson}\ \emph {et~al.}(1987)\citenamefont {Nelson},
  \citenamefont {Miller},\ and\ \citenamefont {Kleinman}}]{nelson_PRB_1987}%
  \BibitemOpen
  \bibfield  {author} {\bibinfo {author} {\bibfnamefont {D.}~\bibnamefont
  {Nelson}}, \bibinfo {author} {\bibfnamefont {R.}~\bibnamefont {Miller}},\
  and\ \bibinfo {author} {\bibfnamefont {D.}~\bibnamefont {Kleinman}},\
  }\bibfield  {title} {\bibinfo {title} {Band nonparabolicity effects in
  semiconductor quantum wells},\ }\href@noop {} {\bibfield  {journal} {\bibinfo
   {journal} {Phys. Rev. B.}\ }\textbf {\bibinfo {volume} {35}},\ \bibinfo
  {pages} {7770} (\bibinfo {year} {1987})}\BibitemShut {NoStop}%
\bibitem [{\citenamefont {Dupont}\ \emph {et~al.}(2012)\citenamefont {Dupont},
  \citenamefont {Fathololoumi}, \citenamefont {Wasilewski}, \citenamefont
  {Aers}, \citenamefont {Laframboise}, \citenamefont {Lindskog}, \citenamefont
  {Razavipour}, \citenamefont {Wacker}, \citenamefont {Ban},\ and\
  \citenamefont {Liu}}]{dupont_JAP_2012}%
  \BibitemOpen
  \bibfield  {author} {\bibinfo {author} {\bibfnamefont {E.}~\bibnamefont
  {Dupont}}, \bibinfo {author} {\bibfnamefont {S.}~\bibnamefont
  {Fathololoumi}}, \bibinfo {author} {\bibfnamefont {Z.}~\bibnamefont
  {Wasilewski}}, \bibinfo {author} {\bibfnamefont {G.}~\bibnamefont {Aers}},
  \bibinfo {author} {\bibfnamefont {S.}~\bibnamefont {Laframboise}}, \bibinfo
  {author} {\bibfnamefont {M.}~\bibnamefont {Lindskog}}, \bibinfo {author}
  {\bibfnamefont {S.}~\bibnamefont {Razavipour}}, \bibinfo {author}
  {\bibfnamefont {A.}~\bibnamefont {Wacker}}, \bibinfo {author} {\bibfnamefont
  {D.}~\bibnamefont {Ban}},\ and\ \bibinfo {author} {\bibfnamefont
  {H.}~\bibnamefont {Liu}},\ }\bibfield  {title} {\bibinfo {title} {A phonon
  scattering assisted injection and extraction based terahertz quantum cascade
  laser},\ }\href@noop {} {\bibfield  {journal} {\bibinfo  {journal} {J. Appl.
  Phys.}\ }\textbf {\bibinfo {volume} {111}},\ \bibinfo {pages} {073111}
  (\bibinfo {year} {2012})}\BibitemShut {NoStop}%
\bibitem [{\citenamefont {Terazzi}(2012)}]{terazzi2012}%
  \BibitemOpen
  \bibfield  {author} {\bibinfo {author} {\bibfnamefont {R.~L.}\ \bibnamefont
  {Terazzi}},\ }\emph {\bibinfo {title} {Transport in quantum cascade
  lasers}},\ \href@noop {} {Ph.D. thesis},\ \bibinfo  {school} {ETH Zurich}
  (\bibinfo {year} {2012})\BibitemShut {NoStop}%
\bibitem [{\citenamefont {Kolek}\ \emph {et~al.}(2012)\citenamefont {Kolek},
  \citenamefont {Ha{\l}da{\'s}},\ and\ \citenamefont
  {Bugajski}}]{kolek_APL_2012}%
  \BibitemOpen
  \bibfield  {author} {\bibinfo {author} {\bibfnamefont {A.}~\bibnamefont
  {Kolek}}, \bibinfo {author} {\bibfnamefont {G.}~\bibnamefont
  {Ha{\l}da{\'s}}},\ and\ \bibinfo {author} {\bibfnamefont {M.}~\bibnamefont
  {Bugajski}},\ }\bibfield  {title} {\bibinfo {title} {Nonthermal carrier
  distributions in the subbands of 2-phonon resonance mid-infrared quantum
  cascade laser},\ }\href@noop {} {\bibfield  {journal} {\bibinfo  {journal}
  {Appl. Phys. Lett.}\ }\textbf {\bibinfo {volume} {101}},\ \bibinfo {pages}
  {061110} (\bibinfo {year} {2012})}\BibitemShut {NoStop}%
\bibitem [{\citenamefont {Ekenberg}(1989)}]{ekenberg_PRB_1989}%
  \BibitemOpen
  \bibfield  {author} {\bibinfo {author} {\bibfnamefont {U.}~\bibnamefont
  {Ekenberg}},\ }\bibfield  {title} {\bibinfo {title} {Nonparabolicity effects
  in a quantum well: sublevel shift, parallel mass, and landau levels},\
  }\href@noop {} {\bibfield  {journal} {\bibinfo  {journal} {Phys. Rev. B.}\
  }\textbf {\bibinfo {volume} {40}},\ \bibinfo {pages} {7714} (\bibinfo {year}
  {1989})}\BibitemShut {NoStop}%
\bibitem [{\citenamefont {Sirtori}\ \emph {et~al.}(1994)\citenamefont
  {Sirtori}, \citenamefont {Capasso}, \citenamefont {Faist},\ and\
  \citenamefont {Scandolo}}]{sirtori_PRB_1994}%
  \BibitemOpen
  \bibfield  {author} {\bibinfo {author} {\bibfnamefont {C.}~\bibnamefont
  {Sirtori}}, \bibinfo {author} {\bibfnamefont {F.}~\bibnamefont {Capasso}},
  \bibinfo {author} {\bibfnamefont {J.}~\bibnamefont {Faist}},\ and\ \bibinfo
  {author} {\bibfnamefont {S.}~\bibnamefont {Scandolo}},\ }\bibfield  {title}
  {\bibinfo {title} {Nonparabolicity and a sum rule associated with
  bound-to-bound and bound-to-continuum intersubband transitions in quantum
  wells},\ }\href@noop {} {\bibfield  {journal} {\bibinfo  {journal} {Phys.
  Rev. B.}\ }\textbf {\bibinfo {volume} {50}},\ \bibinfo {pages} {8663}
  (\bibinfo {year} {1994})}\BibitemShut {NoStop}%
\bibitem [{\citenamefont {Bahder}(1990)}]{bahder_PRB_1990}%
  \BibitemOpen
  \bibfield  {author} {\bibinfo {author} {\bibfnamefont {T.~B.}\ \bibnamefont
  {Bahder}},\ }\bibfield  {title} {\bibinfo {title} {Eight-band k.p model of
  strained zinc-blende crystals},\ }\href@noop {} {\bibfield  {journal}
  {\bibinfo  {journal} {Phys. Rev. B.}\ }\textbf {\bibinfo {volume} {41}},\
  \bibinfo {pages} {11992} (\bibinfo {year} {1990})}\BibitemShut {NoStop}%
\bibitem [{\citenamefont {Kane}(1966)}]{kane_1966}%
  \BibitemOpen
  \bibfield  {author} {\bibinfo {author} {\bibfnamefont {E.}~\bibnamefont
  {Kane}},\ }\bibfield  {title} {\bibinfo {title} {Semiconductors and
  semimetals},\ }\href@noop {} {\bibfield  {journal} {\bibinfo  {journal}
  {Acad. Press Inc}\ }\textbf {\bibinfo {volume} {1}},\ \bibinfo {pages} {75}
  (\bibinfo {year} {1966})}\BibitemShut {NoStop}%
\bibitem [{\citenamefont {Pidgeon}\ and\ \citenamefont
  {Brown}(1966)}]{pidgeon_PR_1966}%
  \BibitemOpen
  \bibfield  {author} {\bibinfo {author} {\bibfnamefont {C.~R.}\ \bibnamefont
  {Pidgeon}}\ and\ \bibinfo {author} {\bibfnamefont {R.}~\bibnamefont
  {Brown}},\ }\bibfield  {title} {\bibinfo {title} {Interband
  magneto-absorption and faraday rotation in insb},\ }\href@noop {} {\bibfield
  {journal} {\bibinfo  {journal} {Phys. Rev.}\ }\textbf {\bibinfo {volume}
  {146}},\ \bibinfo {pages} {575} (\bibinfo {year} {1966})}\BibitemShut
  {NoStop}%
\bibitem [{\citenamefont {Luttinger}\ and\ \citenamefont
  {Kohn}(1955)}]{luttinger_PR_1957}%
  \BibitemOpen
  \bibfield  {author} {\bibinfo {author} {\bibfnamefont {J.~M.}\ \bibnamefont
  {Luttinger}}\ and\ \bibinfo {author} {\bibfnamefont {W.}~\bibnamefont
  {Kohn}},\ }\bibfield  {title} {\bibinfo {title} {Motion of electrons and
  holes in perturbed periodic fields},\ }\href@noop {} {\bibfield  {journal}
  {\bibinfo  {journal} {Phys. Rev.}\ }\textbf {\bibinfo {volume} {97}},\
  \bibinfo {pages} {869} (\bibinfo {year} {1955})}\BibitemShut {NoStop}%
\bibitem [{\citenamefont {Vurgaftman}\ \emph {et~al.}(2001)\citenamefont
  {Vurgaftman}, \citenamefont {Meyer},\ and\ \citenamefont
  {Ram-Mohan}}]{vurgaftman_JAP_2001}%
  \BibitemOpen
  \bibfield  {author} {\bibinfo {author} {\bibfnamefont {I.}~\bibnamefont
  {Vurgaftman}}, \bibinfo {author} {\bibfnamefont {J.~{\'a}.}\ \bibnamefont
  {Meyer}},\ and\ \bibinfo {author} {\bibfnamefont {L.~{\'a}.}\ \bibnamefont
  {Ram-Mohan}},\ }\bibfield  {title} {\bibinfo {title} {Band parameters for
  iii--v compound semiconductors and their alloys},\ }\href@noop {} {\bibfield
  {journal} {\bibinfo  {journal} {J. Appl. Phys.}\ }\textbf {\bibinfo {volume}
  {89}},\ \bibinfo {pages} {5815} (\bibinfo {year} {2001})}\BibitemShut
  {NoStop}%
\bibitem [{\citenamefont {Bir}\ and\ \citenamefont {Pikus}(1974)}]{bir_1974}%
  \BibitemOpen
  \bibfield  {author} {\bibinfo {author} {\bibfnamefont {G.~L.}\ \bibnamefont
  {Bir}}\ and\ \bibinfo {author} {\bibfnamefont {G.~E.}\ \bibnamefont
  {Pikus}},\ }\bibfield  {title} {\bibinfo {title} {Symmetry and strain-induced
  effects in semiconductors},\ }\href@noop {} {\  (\bibinfo {year}
  {1974})}\BibitemShut {NoStop}%
\bibitem [{\citenamefont {Voon}\ and\ \citenamefont
  {Willatzen}(2009)}]{willatzen_2009_kp}%
  \BibitemOpen
  \bibfield  {author} {\bibinfo {author} {\bibfnamefont {L.~C. L.~Y.}\
  \bibnamefont {Voon}}\ and\ \bibinfo {author} {\bibfnamefont {M.}~\bibnamefont
  {Willatzen}},\ }\href@noop {} {\emph {\bibinfo {title} {The kp method:
  electronic properties of semiconductors}}}\ (\bibinfo  {publisher} {Springer
  Science \& Business Media},\ \bibinfo {year} {2009})\BibitemShut {NoStop}%
\bibitem [{\citenamefont {Fr{\"o}hlich}(1937)}]{frolich_PRSA_1937}%
  \BibitemOpen
  \bibfield  {author} {\bibinfo {author} {\bibfnamefont {H.}~\bibnamefont
  {Fr{\"o}hlich}},\ }\bibfield  {title} {\bibinfo {title} {Theory of electrical
  breakdown in ionic crystals},\ }\href@noop {} {\bibfield  {journal} {\bibinfo
   {journal} {Proc. R. Soc. Lond. A}\ }\textbf {\bibinfo {volume} {160}},\
  \bibinfo {pages} {230} (\bibinfo {year} {1937})}\BibitemShut {NoStop}%
\bibitem [{\citenamefont {Kubis}\ \emph {et~al.}(2008)\citenamefont {Kubis},
  \citenamefont {Yeh},\ and\ \citenamefont {Vogl}}]{kubis_PSSC_2008}%
  \BibitemOpen
  \bibfield  {author} {\bibinfo {author} {\bibfnamefont {T.}~\bibnamefont
  {Kubis}}, \bibinfo {author} {\bibfnamefont {C.}~\bibnamefont {Yeh}},\ and\
  \bibinfo {author} {\bibfnamefont {P.}~\bibnamefont {Vogl}},\ }\bibfield
  {title} {\bibinfo {title} {Quantum theory of transport and optical gain in
  quantum cascade lasers},\ }\href@noop {} {\bibfield  {journal} {\bibinfo
  {journal} {Physica Status Solidi c}\ }\textbf {\bibinfo {volume} {5}},\
  \bibinfo {pages} {232} (\bibinfo {year} {2008})}\BibitemShut {NoStop}%
\bibitem [{\citenamefont {Chiu}\ \emph {et~al.}(2012)\citenamefont {Chiu},
  \citenamefont {Dikmelik}, \citenamefont {Liu}, \citenamefont {Aung},
  \citenamefont {Khurgin},\ and\ \citenamefont {Gmachl}}]{chiu_APL_2012}%
  \BibitemOpen
  \bibfield  {author} {\bibinfo {author} {\bibfnamefont {Y.}~\bibnamefont
  {Chiu}}, \bibinfo {author} {\bibfnamefont {Y.}~\bibnamefont {Dikmelik}},
  \bibinfo {author} {\bibfnamefont {P.~Q.}\ \bibnamefont {Liu}}, \bibinfo
  {author} {\bibfnamefont {N.~L.}\ \bibnamefont {Aung}}, \bibinfo {author}
  {\bibfnamefont {J.~B.}\ \bibnamefont {Khurgin}},\ and\ \bibinfo {author}
  {\bibfnamefont {C.~F.}\ \bibnamefont {Gmachl}},\ }\bibfield  {title}
  {\bibinfo {title} {Importance of interface roughness induced intersubband
  scattering in mid-infrared quantum cascade lasers},\ }\href@noop {}
  {\bibfield  {journal} {\bibinfo  {journal} {Appl. Phys. Lett.}\ }\textbf
  {\bibinfo {volume} {101}},\ \bibinfo {pages} {171117} (\bibinfo {year}
  {2012})}\BibitemShut {NoStop}%
\bibitem [{\citenamefont {Ferry}(2013)}]{ferry2013}%
  \BibitemOpen
  \bibfield  {author} {\bibinfo {author} {\bibfnamefont {D.}~\bibnamefont
  {Ferry}},\ }\href@noop {} {\bibinfo {title} {Semiconductors, 2053-2563}}
  (\bibinfo {year} {2013})\BibitemShut {NoStop}%
\bibitem [{\citenamefont {Roblin}\ \emph {et~al.}(1996)\citenamefont {Roblin},
  \citenamefont {Potter},\ and\ \citenamefont {Fathimulla}}]{roblin_JAP_1996}%
  \BibitemOpen
  \bibfield  {author} {\bibinfo {author} {\bibfnamefont {P.}~\bibnamefont
  {Roblin}}, \bibinfo {author} {\bibfnamefont {R.~C.}\ \bibnamefont {Potter}},\
  and\ \bibinfo {author} {\bibfnamefont {A.}~\bibnamefont {Fathimulla}},\
  }\bibfield  {title} {\bibinfo {title} {Interface roughness scattering in
  alas/ingaas resonant tunneling diodes with an inas subwell},\ }\href@noop {}
  {\bibfield  {journal} {\bibinfo  {journal} {J. Appl. Phys.}\ }\textbf
  {\bibinfo {volume} {79}},\ \bibinfo {pages} {2502} (\bibinfo {year}
  {1996})}\BibitemShut {NoStop}%
\bibitem [{\citenamefont {Ando}(1982)}]{ando_JPSJ_1982}%
  \BibitemOpen
  \bibfield  {author} {\bibinfo {author} {\bibfnamefont {T.}~\bibnamefont
  {Ando}},\ }\bibfield  {title} {\bibinfo {title} {Self-consistent results for
  a gaas/al x ga1-x as heterojunciton. ii. low temperature mobility},\
  }\href@noop {} {\bibfield  {journal} {\bibinfo  {journal} {J. Phys. Soc.
  Jpn.}\ }\textbf {\bibinfo {volume} {51}},\ \bibinfo {pages} {3900} (\bibinfo
  {year} {1982})}\BibitemShut {NoStop}%
\bibitem [{\citenamefont {Gradshteyn}\ and\ \citenamefont
  {Ryzhik}(2007)}]{gradshteyn2007}%
  \BibitemOpen
  \bibfield  {author} {\bibinfo {author} {\bibfnamefont {I.}~\bibnamefont
  {Gradshteyn}}\ and\ \bibinfo {author} {\bibfnamefont {I.}~\bibnamefont
  {Ryzhik}},\ }\bibfield  {title} {\bibinfo {title} {Table of integrals, series
  and products 7th edn, ed a jeffrey and d zwillinger (new york: Academic)},\
  }\href@noop {} {\  (\bibinfo {year} {2007})}\BibitemShut {NoStop}%
\bibitem [{\citenamefont {Varshni}(1967)}]{varshni_physica_1967}%
  \BibitemOpen
  \bibfield  {author} {\bibinfo {author} {\bibfnamefont {Y.~P.}\ \bibnamefont
  {Varshni}},\ }\bibfield  {title} {\bibinfo {title} {Temperature dependence of
  the energy gap in semiconductors},\ }\href@noop {} {\bibfield  {journal}
  {\bibinfo  {journal} {physica}\ }\textbf {\bibinfo {volume} {34}},\ \bibinfo
  {pages} {149} (\bibinfo {year} {1967})}\BibitemShut {NoStop}%
\bibitem [{\citenamefont {Van~Vechten}\ and\ \citenamefont
  {Bergstresser}(1970)}]{vechten_PRB_1970}%
  \BibitemOpen
  \bibfield  {author} {\bibinfo {author} {\bibfnamefont {J.}~\bibnamefont
  {Van~Vechten}}\ and\ \bibinfo {author} {\bibfnamefont {T.}~\bibnamefont
  {Bergstresser}},\ }\bibfield  {title} {\bibinfo {title} {Electronic
  structures of semiconductor alloys},\ }\href@noop {} {\bibfield  {journal}
  {\bibinfo  {journal} {Phys. Rev. B.}\ }\textbf {\bibinfo {volume} {1}},\
  \bibinfo {pages} {3351} (\bibinfo {year} {1970})}\BibitemShut {NoStop}%
\end{thebibliography}

%

\end{document}